\documentclass[11pt]{article}

\usepackage{geometry}
\usepackage{setspace}
\usepackage{indentfirst}
\usepackage{amsmath,amssymb,amsthm,bm}
\newtheorem{assumption}{Assumption}
\newtheorem{theorem}{Theorem}

\usepackage{array}
\usepackage{booktabs,multirow,tabularx,adjustbox}
\usepackage{float}
\usepackage{comment}
\newcolumntype{C}{>{\centering\arraybackslash}X}

\usepackage[authoryear,round]{natbib}
\usepackage[
  bookmarks=false,
  hidelinks
]{hyperref}

\newcounter{algo}
\renewcommand{\thealgo}{\arabic{algo}}

\hypersetup{
  pdftitle={Bayesian Calibration of Computer Experiments and Extrapolation},
  pdfauthor={Jingyi Li and Shifeng Xiong}
}

\begin{document}

\thispagestyle{empty}

\begin{center}
{\fontsize{21pt}{24pt}\selectfont
Bayesian Calibration of Computer Experiments\\[0.25em]
and Extrapolation
\par}
\end{center}

\vspace{0.2em}

\begin{center}
{\fontsize{14pt}{16pt}\selectfont
Jingyi Li\textsuperscript{1,2}, Shifeng Xiong\textsuperscript{2,1*}\par}
\vspace{0.5em}

{\fontsize{11pt}{13pt}\selectfont
1. School of Mathematical Sciences, University of Chinese Academy of Sciences,
Beijing, China\par}
\vspace{0.3em}

{\fontsize{11pt}{13pt}\selectfont
2. SKLMS, Academy of Mathematics and
Systems Science, Chinese Academy of Sciences, Beijing, China\par}
\end{center}
\begingroup
\renewcommand{\thefootnote}{\fnsymbol{footnote}}
\footnotetext[1]{Corresponding author, Email: \href{mailto:xiong@amss.ac.cn}{xiong@amss.ac.cn}}
\endgroup

\vspace{2em}

\begin{center}
{\bfseries\fontsize{15pt}{17pt}\selectfont Abstract\par}
\end{center}

\vspace{-0.3em}

\noindent
Calibration, a fundamental inverse problem in uncertainty quantification, suffers from model identifiability issues. While the discrepancy decomposition model partially addresses this via a first-order Taylor expansion, it struggles when the approximation is biased. We propose a Bayesian modified discrepancy decomposition model that incorporates a residual process to capture higher-order discrepancy structures. By jointly inferring calibration parameters and model discrepancies within a Bayesian framework, our model recovers more physically meaningful parameters and provides reliable uncertainty quantification. Crucially, this accurate parameter recovery enables reliable extrapolative prediction, a challenge current methods largely fail to address. We validate the approach through numerical studies and a real-world industrial application: predicting the thermal stability of multilayer mirrors in extreme ultraviolet lithography. In this application, our method accurately predicts high-temperature responses far beyond the training domain, significantly outperforming existing calibration methods.

\vspace{1em}


\clearpage

\section{Introduction}
\label{sec:introduction}
\noindent
Computer simulation models are widely adopted in engineering, digital twins, environmental science, and
biomedicine to describe, analyze, and predict the behavior of complex physical systems \citep{SantnerWilliamsNotz2018}. The response of a simulation depends deterministically on specified conditions. These conditions are generally divided into two categories: input variables, which are controllable or measurable in the real system, and parameters, which cannot be directly controlled or measured but significantly affect system behavior. A central task in computer experiments is to infer such unknown parameters by combining observations from the physical system with outputs from the computer model. This process is known as \textit{calibration} and is often viewed as an inverse problem; the associated unknown parameters are referred to as \textit{calibration parameters}.

Calibration parameters can be classified into two types: \textit{physical parameters} and \textit{tuning parameters} \citep{HanSantnerRawlinson2009}. Physical parameters have specific scientific meanings within the underlying physical system, and learning their values can enhance our understanding of that system. For example, \citet{SansoForestZantedeschi2008} use climate observations and model simulations to infer climate sensitivity, the rate of deep-ocean heat uptake governed by effective ocean diffusivity, and aerosol forcing. These parameters describe aspects of the climate system's response to external forcing. By contrast, tuning parameters have no direct physical meaning and are used only for model construction. The present work focuses on calibrating physical parameters.

Calibration aims to learn unknown model parameters from physical observations and improve predictions of the physical system \citep{SungTuo2024}. However, even a computer model evaluated at the true physical parameter values may differ from reality because its construction involves idealizations and approximations. This difference is referred to as \textit{model discrepancy}. \citet{BrynjarsdottirOHagan2014} show that ignoring model discrepancy can lead to biased and overconfident parameter estimates and predictions. They also emphasize that recovering physical parameter values and obtaining reliable extrapolations require meaningful information about the discrepancy. Characterizing model discrepancy is therefore a central challenge in computer model calibration.

Among the existing approaches, the Bayesian calibration framework proposed by \citet{KennedyOHagan2001} (abbreviated as KOH) is particularly influential. KOH explicitly accounts for model discrepancy, uncertainty in calibration parameters, and observation error. Gaussian process models are used for the simulator response and the discrepancy function, allowing posterior inference and discrepancy-adjusted predictions of the physical system. \citet{HigdonEtAl2004} develop Bayesian methods for combining field observations and computer simulations for calibration and prediction, while \citet{BayarriEtAl2007} develop a Bayesian and likelihood-based validation framework illustrated with resistance spot welding. Related work includes the probabilistic formulation for linking imperfect simulators to physical systems in \citet{GoldsteinRougier2004} and the sequential empirical Bayes adjustments to engineering models in \citet{JosephMelkote2009}. Applications of Bayesian modeling that explicitly account for model discrepancy or structural model uncertainty include eutrophication modeling \citep{ArhonditsisEtAl2008} and probabilistic climate prediction \citep{MurphyEtAl2007}.

Despite its success, KOH suffers from an identifiability problem. Specifically, for a given computer model, the distribution of the observed physical data generally does not uniquely determine the corresponding calibration parameter, which may lead to unreasonable parameter estimates \citep{BrynjarsdottirOHagan2014}. \citet{Plumlee2017} addresses this problem by imposing orthogonality between the discrepancy prior and the gradient of the computer model. \citet{TuoWu2016} analyze parameter identifiability and the limiting behavior of a simplified KOH calibration procedure, and \citet{TuoWu2015} develop $L_2$ calibration. The $L_2$ target is the parameter that minimizes the $L_2$ distance between the physical response and the computer model. Under suitable regularity and uniqueness assumptions, this definition supports identifiable and semiparametrically efficient estimation \citep{TuoWu2015}.

Methods associated with the $L_2$ calibration target include projected kernel calibration \citep{Tuo2019PKC}, penalized projected kernel calibration \citep{Wang2022PPKC}, and Bayesian projected calibration \citep{XieXu2021}. The scaled Gaussian stochastic process approach connects Gaussian process discrepancy modeling with $L_2$ calibration \citep{GuWang2018SGaSP}; its convergence to the $L_2$ target is established under appropriate conditions in \citet{GuXieWang2022}. Other developments include semiparametric adjustment for predicting the physical response \citep{WangTuo2020} and Sobolev calibration, which uses a Sobolev-space criterion and connects $L_2$ calibration with a frequentist formulation of KOH calibration \citep{ZhangWang2025Sobolev}. In particular, the $L_2$-optimal parameter need not coincide with the true physical parameter when the computer model is imperfect \citep{LiWuXiong2026}. Consequently, identifying a projection-based target does not by itself guarantee recovery of physical parameter values or reliable extrapolation.

To address the identifiability problem and recover physically meaningful calibration parameters, \citet{LiWuXiong2026} propose the discrepancy decomposition model within a frequentist framework. This model decomposes the discrepancy into \textit{intrinsic discrepancy}, which remains when the computer model is evaluated at the true physical parameter, and \textit{parametric discrepancy}, which is linear in the deviation from that parameter. This decomposition can be interpreted as a first-order Taylor expansion of the model discrepancy with respect to the calibration parameter around its true physical value. Under specified constraints, the authors establish identifiability, construct estimators of the physical parameters and discrepancy functions, and derive their asymptotic properties. Their numerical studies demonstrate improved physical parameter estimation relative to $L_2$ calibration.

Nevertheless, the discrepancy decomposition model still has two main limitations. First, it relies on a first-order Taylor approximation and may struggle when this approximation is biased, particularly when higher-order terms in the model discrepancy are non-negligible. Such approximation bias can compromise the recovery of physically meaningful parameters and consequently impair extrapolative prediction beyond the training domain. Second, the original method provides frequentist estimators and asymptotic results, but uncertainty quantification is less straightforward within this framework.

Motivated by these limitations, we propose a Bayesian modified discrepancy decomposition model. Our approach builds on the discrepancy decomposition model and further introduces a residual process to capture higher-order discrepancy structures. We impose appropriate constraints to preserve identifiability and reduce local confounding with the intrinsic and parametric discrepancy components. We establish a Bayesian framework to jointly model the calibration parameter and model discrepancy, and perform posterior inference via a Metropolis-within-Gibbs sampler. This improves the recovery of physically meaningful parameters and provides reliable uncertainty quantification.

In particular, owing to the more accurate estimation of physical parameters, our calibrated model stays closer to the true underlying physical process. Consequently, our approach not only fits the data well within the training domain, but also exhibits superior extrapolative performance, a long-standing challenge that remains largely unaddressed by existing calibration methods. This is fully demonstrated in our extensive numerical experiments and the successful application to the thermal stability of extreme ultraviolet lithography multilayer mirrors.

The rest of the paper is organized as follows. Section \ref{sec:calibration} presents the calibration problem in computer experiments, including classical calibration methods and the discrepancy decomposition model. Section \ref{sec:bayes_bmdd} introduces the Bayesian modified discrepancy decomposition model, including its basic construction, Bayesian inference theory and algorithm, and the prediction method for the physical system. Section \ref{sec:numerical} presents several numerical studies. Section \ref{sec:real_data_application} applies the proposed method to thermal stability analysis of extreme ultraviolet lithography. Section \ref{sec:discussion} concludes the paper and discusses possible directions for future research. Technical details are provided in the separate Supplementary Materials.

\section{The Calibration Problem in Computer Experiments}
\label{sec:calibration}
\noindent
In this section, we aim to provide an introduction to the calibration problem in computer experiments. By reviewing classical calibration methods and outlining the discrepancy decomposition model, we lay the foundation for the improved model proposed in this paper.

\subsection{Classical Calibration Methods}
\noindent
Consider a deterministic computer model $y^{S}(\boldsymbol{x},\boldsymbol{\theta})$ for simulating a physical system $y^{P}$,
where the superscripts $S$ and $P$ stand for ``simulation" and ``physics", respectively. Here, $\boldsymbol{x}\in\Omega$ denotes the controllable inputs, which can be specified in both physical and computer experiments; $\boldsymbol{\theta}\in\Theta$ denotes the vector of calibration parameters, which represents intrinsic attributes of the physical system and is an input vector of the computer model. The physical system relies on the true value of $\boldsymbol{\theta}$, denoted by $\boldsymbol{\theta}^{p}$, where the superscript $p$ stands for ``physical".
In many real problems $\boldsymbol{\theta}^{p}$ cannot be observed in physical experiments, and the task of calibration is to specify it. Throughout this paper, we assume that $\Omega\subset\mathbb{R}^{d}$ and $\Theta\subset\mathbb{R}^{q}$ are bounded closed sets.

Within the above framework, the observation of the physical system is \begin{equation}\label{fp}y^{F}(\boldsymbol{x};\boldsymbol{\theta}^{p})=y^{P}(\boldsymbol{x};\boldsymbol{\theta}^{p})
+\epsilon,\end{equation} where the superscript \(F\) stands for ``field" and $\epsilon$ is the observation noise.
To perform calibration, we need to run the computer model at several different values of the calibration parameter, while also conducting physical experiments and collecting field observations. Given controllable input design points $\{\boldsymbol{x}_{i}\}_{i=1}^{n}\subset\Omega$, one can obtain noisy field data\
$\{y^{F}(\boldsymbol{x}_{i};\boldsymbol{\theta}^{p})\}_{i=1}^{n}$.
The next step is to infer the unknown physical parameters $\boldsymbol{\theta}^{p}$ by combining observations from the physical system and outputs from the computer model.

It is worth noting that conventional calibration methods do not consider $\boldsymbol{\theta}^{p}$ in the physical system/observation, and write $y^{P}$ and $y^{F}$ as $y^{P}(\boldsymbol{x})$ and $y^{F}(\boldsymbol{x})$, respectively.
\citet{KennedyOHagan2001} proposed the classical KOH model under a Bayesian framework. Its basic form can be written as
\[
y^{F}(\boldsymbol{x})
=
\eta\, y^{S}(\boldsymbol{x},\boldsymbol{\theta}_{0})
+
\delta(\boldsymbol{x})
+
\epsilon,
\]
where \(\eta\) is an unknown regression parameter, \(\boldsymbol{\theta}_0\) is interpreted as the calibration parameter in the sense of best fitting and \(\delta(\boldsymbol{x})\) is the discrepancy between the physical system and $\eta\,y^{S}(\boldsymbol{x},\boldsymbol{\theta}_{0})$. In general, \(\boldsymbol{\theta}_0\) need not coincide with \(\boldsymbol{\theta}^p\). The Bayesian calibration framework of the KOH model allows joint inference on the calibration parameter and model discrepancy. To achieve this, they assign prior distributions to \(\delta\) and the computer model \(y^S\), typically using a Gaussian process. Inference is then carried out based on the corresponding posterior distributions. The KOH model yields an estimate of model discrepancy, which is combined with the computer model output for bias correction, leading to an estimate of the true physical system response. This method systematically incorporates model discrepancy into both calibration and prediction. However, the KOH model has an identifiability issue: for a given computer model, the distribution of physical observations does not uniquely determine the calibration parameter, leading to unreasonable parameter estimates. To alleviate this issue within the KOH framework, \citet{Plumlee2017} proposed the orthogonal Gaussian process approach, in which the prior distribution of the discrepancy function is constrained to be orthogonal to the gradient of the computer model with respect to the calibration parameter. This orthogonality constraint reduces the confounding between the calibration parameter and the model discrepancy.

Beyond the KOH framework, \citet{TuoWu2015} addressed the identifiability issue by proposing $L_2$ calibration under a frequentist framework, in which the ``true'' calibration parameter is defined as the projection that makes the computer output closest to the physical process in the $L_2$ sense, i.e.,
\[
{\boldsymbol{\theta }}^{ * }
=
\arg \mathop{\min }\limits_{{\boldsymbol{\theta } \in  \Theta }}
{\begin{Vmatrix}y^{P}(\boldsymbol{x})  - {y}^{S}\left( \boldsymbol{x},\boldsymbol{\theta }\right) \end{Vmatrix}}_{{L}_{2}}
=
\arg \mathop{\min }\limits_{{\boldsymbol{\theta } \in  \Theta }}
{\left( {\int }_{\Omega }{\left( y^{P}(\boldsymbol{x})  - {y}^{S}\left( \boldsymbol{x},\boldsymbol{\theta }\right) \right) }^{2}\mathrm{d}\boldsymbol{x}\right) }^{1/2}.
\]
Under this definition, $\boldsymbol{\theta}^{\ast}$ is strictly identifiable in the frequentist sense.

Overall, the definitions of the ``true" \(\boldsymbol{\theta}\) within the KOH and $L_2$ framework, $\boldsymbol{\theta}_0$ and \(\boldsymbol{\theta}^{\ast}\), are not necessarily equivalent to the physically meaningful parameter \(\boldsymbol{\theta}^{p}\). This may lead to inaccurate estimation of \(\boldsymbol{\theta}^{p}\). Moreover, extrapolative predictions based on these parameters may be unreliable.

\subsection{The Discrepancy Decomposition Model}
\label{subsec:dd_review}
\noindent
In this subsection, we briefly review the discrepancy decomposition (DD) model \citep{LiWuXiong2026}, which provides a straightforward modeling strategy to recognize \(\boldsymbol{\theta}^{p}\), instead of $\boldsymbol{\theta}_0$ and \(\boldsymbol{\theta}^{\ast}\).

Let $\delta(\boldsymbol{x},\boldsymbol{\theta};\boldsymbol{\theta}^{p})=y^{P}(\boldsymbol{x};\boldsymbol{\theta}^{p})-y^{S}(\boldsymbol{x},\boldsymbol{\theta})$ denote the model discrepancy between the physical system and the computer model. By \eqref{fp}, \[
y^{F}(\boldsymbol{x};\boldsymbol{\theta}^{p})
=
y^{S}(\boldsymbol{x},\boldsymbol{\theta})
+
\delta(\boldsymbol{x},\boldsymbol{\theta};\boldsymbol{\theta}^{p})
+
\epsilon.
\]
Without structural constraints on $\delta$, for any given $\boldsymbol{\theta}^p$ one can always construct a discrepancy function such that
$y^{S}(\boldsymbol{x},\boldsymbol{\theta})+\delta(\boldsymbol{x},\boldsymbol{\theta};\boldsymbol{\theta}^{p})$
exactly fits $y^{P}(\boldsymbol{x};\boldsymbol{\theta}^p_0)$, where $\boldsymbol{\theta}^p_0$ represents the true value of $\boldsymbol{\theta}^p$ in the physical system. Therefore, $\boldsymbol{\theta}^p$ cannot be recognized.

\citet{LiWuXiong2026}'s DD model imposes the following structural decomposition on the model discrepancy,
\begin{equation}
\label{eq:dd}
\delta(\boldsymbol{x},\boldsymbol{\theta};\boldsymbol{\theta}^{p})
=
\alpha(\boldsymbol{x})
+
\boldsymbol{\beta}(\boldsymbol{x})^{\mathsf T}
(\boldsymbol{\theta}-\boldsymbol{\theta}^{p}),
\end{equation}
where $\alpha(\boldsymbol{x})$ and
$\boldsymbol{\beta}(\boldsymbol{x})=\big(\beta_{1}(\boldsymbol{x}),\ldots,\beta_{q}(\boldsymbol{x})\big)^{\mathsf T}$
are unknown smooth functions defined on $\Omega$, and the superscript $\mathsf T$ denotes transpose. Here, $\alpha(\boldsymbol{x})$ represents the \textit{intrinsic discrepancy}, namely the discrepancy that
remains even when the computer model is evaluated at the true physical parameter
\(\boldsymbol{\theta}^{p}\). The term
$\boldsymbol{\beta}(\boldsymbol{x})^{\mathsf T}(\boldsymbol{\theta}-\boldsymbol{\theta}^{p})$
represents the \textit{parametric discrepancy} induced by using an incorrect $\boldsymbol{\theta}$, and it vanishes when $\boldsymbol{\theta}=\boldsymbol{\theta}^{p}$. The function $\boldsymbol{\beta}(\boldsymbol{x})$ is referred to as the \textit{discrepancy coefficient function}. Equation \eqref{eq:dd} can also be interpreted as the first-order Taylor expansion of the model discrepancy
$\delta(\boldsymbol{x},\boldsymbol{\theta};\boldsymbol{\theta}^{p})$
with respect to $\boldsymbol{\theta}$ at $\boldsymbol{\theta}=\boldsymbol{\theta}^{p}$, where $\alpha(\boldsymbol{x})$ corresponds to
$\delta(\boldsymbol{x},\boldsymbol{\theta}^{p};\boldsymbol{\theta}^{p})$,
and $\boldsymbol{\beta}(\boldsymbol{x})$ corresponds to
$\left.{\partial\delta(\boldsymbol{x},\boldsymbol{\theta};\boldsymbol{\theta}^{p})}/{\partial\boldsymbol{\theta}}\right|_{\boldsymbol{\theta}=\boldsymbol{\theta}^{p}}$.

The identifiability of DD relies on suitable constraints on the model
discrepancy. For a \(q\)-dimensional \(\boldsymbol{\theta}\), one may choose
\(q\) different subsets \(\Omega_1,\ldots,\Omega_q\subset\Omega\), each with positive
Lebesgue measure, and fix the average level of \(\alpha(\boldsymbol{x})\) on these subsets as
\begin{equation}
\label{eq:dd_alpha_constraint}
\int_{\Omega_t}\alpha(\boldsymbol{x})\,d\boldsymbol{x}
=
c_t,
\qquad
t=1,\ldots,q,
\end{equation}
where \(c_1,\ldots,c_q\) are prespecified constants. These constants reflect
prior information about the average discrepancy between the physical system
and the computer model at \(\boldsymbol{\theta}^{p}\). If such prior
information is unavailable, one may take \(c_t=0\) as a default choice, \(t=1,\dots,q\).
In addition, the average effect of \(\boldsymbol{\beta}(\boldsymbol{x})\) over these subsets is
assumed to be nondegenerate, that is,
\begin{equation}
\label{eq:dd_beta_rank}
\det(\mathbf{M})\neq 0,
\qquad
\mathbf{M}
=
\left(
\int_{\Omega_t}\beta_{\ell}(\boldsymbol{x})\,d\boldsymbol{x}
\right)_{t,\ell=1,\ldots,q}.
\end{equation}
Under \eqref{eq:dd_alpha_constraint} and \eqref{eq:dd_beta_rank}, \citet{LiWuXiong2026} proved that DD is identifiable.

Within a frequentist framework, \citet{LiWuXiong2026} further developed estimation procedures for $\boldsymbol{\theta}^{p}$, $\alpha(\boldsymbol{x})$, and $\boldsymbol{\beta}(\boldsymbol{x})$, and established the corresponding asymptotic properties, such as the convergence rate and asymptotic normality of $\boldsymbol{\theta}^{p}$.
Numerical examples indicate that DD outperforms the classical KOH and $L_2$ methods in many cases. However, there are two limitations in DD. First, the model uses only the
first-order Taylor expansion that is actually a linear relationship. It may perform poorly when this approximation is biased. Second, DD is developed within the frequentist framework and does not directly provide uncertainty quantification. These limitations motivate the Bayesian modified discrepancy decomposition model introduced in the next section.

\section{The Bayesian Modified Discrepancy Decomposition Model}
\label{sec:bayes_bmdd}
\noindent
In this section, we present the Bayesian modified discrepancy decomposition model (BMDD) in detail. Specifically, we describe the model construction, the Bayesian hierarchical structure, the Metropolis-within-Gibbs sampling scheme for posterior inference, and the prediction method for the physical system.

\subsection{Model Construction}
\label{subsec:bmdd_construction}
\noindent
The proposed BMDD is constructed to extend the structure of DD by adding a flexible residual term, while preserving the identifiability mechanism
and embedding the whole model into a Bayesian framework \citep{GhoshDelampadySamanta2006}.

As mentioned in Section \ref{subsec:dd_review}, the discrepancy term \(\delta(\boldsymbol{x},\boldsymbol{\theta};\boldsymbol{\theta}^{p})\) in the DD model \eqref{eq:dd} can be interpreted as a first-order Taylor expansion with respect to \(\boldsymbol{\theta}\) around \(\boldsymbol{\theta}^{p}\), where \(\alpha(\boldsymbol{x})\) and \(\boldsymbol{\beta}(\boldsymbol{x})^{\mathsf T}(\boldsymbol{\theta}-\boldsymbol{\theta}^{p})\) correspond to the zeroth-order and first-order terms, respectively. To further capture the higher-order remainder in the model discrepancy, we
extend the discrepancy decomposition as
\begin{equation}
\label{eq:bmdd_decomposition}
\delta(\boldsymbol{x},\boldsymbol{\theta};\boldsymbol{\theta}^{p})
=
\alpha(\boldsymbol{x})
+
\boldsymbol{\beta}(\boldsymbol{x})^{\mathsf T}
(\boldsymbol{\theta}-\boldsymbol{\theta}^{p})
+
r(\boldsymbol{x},\boldsymbol{\theta};\boldsymbol{\theta}^{p}),
\end{equation}
where \(r(\boldsymbol{x},\boldsymbol{\theta};\boldsymbol{\theta}^{p})\) is a
residual term. This term is introduced to describe higher-order discrepancy structures that cannot be represented by the first-order discrepancy decomposition. A key issue is that \(r(\boldsymbol{x},\boldsymbol{\theta};
\boldsymbol{\theta}^{p})\) should improve flexibility without destroying
identifiability. Thus, we
propose the following assumptions to guarantee the identifiability of \eqref{eq:bmdd_decomposition}. Like in DD, let \(\Omega_1,\ldots,\Omega_q\subset\Omega\) be \(q\) different subsets with positive Lebesgue measures.

\begin{assumption}
\label{ass:bmdd_dd_constraints}
The functions \(\alpha(\boldsymbol{x})\) and
\(\boldsymbol{\beta}(\boldsymbol{x})\) in \eqref{eq:bmdd_decomposition} satisfy the DD identifiability
constraints \eqref{eq:dd_alpha_constraint} and \eqref{eq:dd_beta_rank}.
\end{assumption}

\begin{assumption}
\label{ass:bmdd_local}
The residual term satisfies
\[
r(\boldsymbol{x},\boldsymbol{\theta}^{p};\boldsymbol{\theta}^{p})
=
0,
\qquad
\nabla_{\boldsymbol{\theta}}
r(\boldsymbol{x},\boldsymbol{\theta};\boldsymbol{\theta}^{p})
\big|_{\boldsymbol{\theta}=\boldsymbol{\theta}^{p}}
=
\boldsymbol{0}.
\]
\end{assumption}

\begin{assumption}
\label{ass:bmdd_r_orthogonal}
For any \(\boldsymbol{\theta},\ \boldsymbol{\theta}^{p}\in\Theta\),
\begin{equation}
\label{eq:r_identifiability_constraint}
\int_{\Omega_t}
r(\boldsymbol{x},\boldsymbol{\theta};\boldsymbol{\theta}^{p})
\,d\boldsymbol{x}
=
0, \qquad t=1,\ldots,q.
\end{equation}
\end{assumption}

Assumption~\ref{ass:bmdd_dd_constraints} inherits the conditions of DD. Assumption~\ref{ass:bmdd_local} prevents \(r(\boldsymbol{x},\boldsymbol{\theta};\boldsymbol{\theta}^{p})\) from being locally
confounded with \(\alpha(\boldsymbol{x})\) and \(\boldsymbol{\beta}(\boldsymbol{x})^{\mathsf T}
(\boldsymbol{\theta}-\boldsymbol{\theta}^{p})\) around \(\boldsymbol{\theta}^{p}\).
Thus, \(r(\boldsymbol{x},\boldsymbol{\theta};\boldsymbol{\theta}^{p})\) only captures second-order or higher-order information. Assumption~\ref{ass:bmdd_r_orthogonal}  requires that \(r(\boldsymbol{x},\boldsymbol{\theta};\boldsymbol{\theta}^{p})\) has zero average over \(\Omega_t\) for \(t=1,\dots,q\), in order to preserve the identifiability.

\begin{theorem}
\label{thm:bmdd_identifiability}
Under Assumptions~\ref{ass:bmdd_dd_constraints}--\ref{ass:bmdd_r_orthogonal},
the model \eqref{eq:bmdd_decomposition} is identifiable.
\end{theorem}

The proof of Theorem~\ref{thm:bmdd_identifiability} is given in the Supplementary Materials.
In the following we take \begin{equation}\label{eq:r}
r(\boldsymbol{x},\boldsymbol{\theta};\boldsymbol{\theta}^{p})
=
\|\boldsymbol{\theta}-\boldsymbol{\theta}^{p}\|^2\,
\omega(\boldsymbol{x},\boldsymbol{\theta}),
\end{equation}where $\|\cdot\|$ denotes the Euclidean norm and the sufficiently smooth function $\omega$ defined on $\Omega\times\Theta$ satisfies the following constraint for every \(\boldsymbol{\theta}\in\Theta\),
\begin{equation}
\label{eq:omega_integral_constraint}
\int_{\Omega_t}
\omega(\boldsymbol{x},\boldsymbol{\theta})
\,d\boldsymbol{x}
=
0,
\qquad
t=1,\ldots,q.
\end{equation}It is easy to check that this form satisfies Assumptions~\ref{ass:bmdd_local} and \ref{ass:bmdd_r_orthogonal}.

It is worth noting that the specific form of $r(\boldsymbol{x},\boldsymbol{\theta};\boldsymbol{\theta}^{p})$ in \eqref{eq:r} represents a deliberate compromise between flexibility and tractability. The inherent unidentifiability of the calibration problem dictates that the model discrepancy can only be approximated. Constrained by the typically limited sample sizes in physical experiments, employing an overly complex residual model would introduce excessive variance and computational burden. The proposed parsimonious form effectively addresses higher-order structures while maintaining stability, yielding superior empirical performance over existing calibration methods.

\subsection{Bayesian Hierarchical Structure}
\label{subsec:hierarchy}
\noindent
In this subsection, we introduce the Bayesian hierarchical structure for \eqref{eq:bmdd_decomposition}, which consists of the observation layer, parameter layer, and hyperparameter layer \citep{SantnerWilliamsNotz2018,GhoshDelampadySamanta2006}. The observation layer specifies the conditional distribution of the observed data given the unknown parameters. The parameter layer assigns priors to the calibration parameter and the unknown functions in the model discrepancy. The hyperparameter layer further assigns priors to the hyperparameters in the parameter layer. Together, these three layers define the complete Bayesian hierarchical structure.

We first introduce the observation layer. Let
\(\{\boldsymbol{x}_{i}\}_{i=1}^{n}\) and \(\{\boldsymbol{\theta}_{j}\}_{j=1}^{m}\)
denote the design points on \(\Omega\) and \(\Theta\), respectively. Usually space-filling designs \citep{SantnerWilliamsNotz2018} are adopted here. We conduct physical experiments at \(\{\boldsymbol{x}_{i}\}_{i=1}^{n}\) and obtain the data
\(\{y^{F}(\boldsymbol{x}_{i};\boldsymbol{\theta}^{p})\}_{i=1}^{n}\),
while running the computer model at the design points
\(\{(\boldsymbol{x}_{i},\boldsymbol{\theta}_{j})\}_{i=1,\ldots,n;\,j=1,\ldots,m}\)
to obtain the outputs
\(\{y^{S}(\boldsymbol{x}_{i},\boldsymbol{\theta}_{j})\}_{i=1,\ldots,n;\,j=1,\ldots,m}\).
Based on these data, we construct the difference matrix
\(\boldsymbol{\Delta}=(\delta_{ij})\in\mathbb{R}^{n\times m}\),
where
\[
\delta_{ij}
=
y^{F}(\boldsymbol{x}_{i};\boldsymbol{\theta}^{p})
-
y^{S}(\boldsymbol{x}_{i},\boldsymbol{\theta}_{j}),
\qquad
i=1,\ldots,n,\ j=1,\ldots,m.
\]
Let \(g(\boldsymbol{\theta};\boldsymbol{\theta}^{p})
:=
\|\boldsymbol{\theta}-\boldsymbol{\theta}^{p}\|^2\). By \eqref{eq:bmdd_decomposition} and \eqref{eq:r}, we have
\[
\delta_{ij}
=
\alpha(\boldsymbol{x}_{i})
+
\boldsymbol{\beta}(\boldsymbol{x}_{i})^{\mathsf T}
(\boldsymbol{\theta}_{j}-\boldsymbol{\theta}^{p})
+
g(\boldsymbol{\theta}_{j};\boldsymbol{\theta}^{p})\,
\omega(\boldsymbol{x}_{i},\boldsymbol{\theta}_{j})
+
\epsilon_{i},
\qquad
i=1,\ldots,n,\ j=1,\ldots,m,
\]
where \(\epsilon_i\) denotes the physical observation noise at \(\boldsymbol{x}_i\). In practice,
\(y^{F}(\boldsymbol{x}_{i};\boldsymbol{\theta}^{p})\)
is typically observed only once at each \(\boldsymbol{x}_{i}\). As a result, when constructing the difference matrix
\(\boldsymbol{\Delta}=(\delta_{ij})\in\mathbb{R}^{n\times m}\),
all difference observations in the same row reuse the same physical observation noise, which induces within-row dependence and may lead to numerical degeneracy. To avoid this issue, we further introduce a nugget term \(\xi_{ij}\), yielding
\[
\delta_{ij}
=
\alpha(\boldsymbol{x}_{i})
+\boldsymbol{\beta}(\boldsymbol{x}_{i})^{\mathsf T}
(\boldsymbol{\theta}_{j}-\boldsymbol{\theta}^{p})
+g(\boldsymbol{\theta}_{j};\boldsymbol{\theta}^{p})\,
\omega(\boldsymbol{x}_{i},\boldsymbol{\theta}_{j})
+\epsilon_{i}+\xi_{ij},
\]
where
\(
\epsilon_{i}\stackrel{\text{iid}}{\sim}\mathcal{N}(0,\sigma_{\epsilon}^{2})
\),
\(
\xi_{ij}\stackrel{\text{iid}}{\sim}\mathcal{N}(0,\sigma_{\xi}^{2})
\),
\(\sigma_{\epsilon}^{2}>0\) is the variance of the physical observation noise, and \(\sigma_{\xi}^{2}>0\) is the nugget variance. These two terms are mutually independent. Therefore, the observation layer can be written as
\[
\delta_{ij}
\mid
\alpha,\boldsymbol{\beta},\omega,\boldsymbol{\theta}^{p},\epsilon_i,\sigma_{\xi}^{2}
\ \sim\
\mathcal{N}\!\left(
\alpha(\boldsymbol{x}_{i})
+\boldsymbol{\beta}(\boldsymbol{x}_{i})^{\mathsf T}
(\boldsymbol{\theta}_{j}-\boldsymbol{\theta}^{p})
+g(\boldsymbol{\theta}_{j};\boldsymbol{\theta}^{p})\,
\omega(\boldsymbol{x}_{i},\boldsymbol{\theta}_{j})
+\epsilon_{i},\
\sigma_{\xi}^{2}
\right).
\]
Thus, the nugget term ensures that the conditional covariance of the difference matrix is nondegenerate. This treatment is specific to the difference-matrix construction; for general $(\boldsymbol{x},\boldsymbol{\theta})$ designs with independent observations at each design point, the observation layer can be specified directly without an additional nugget term.

For later derivations, let
\(
\boldsymbol{\alpha}
=
\bigl(\alpha(\boldsymbol{x}_{1}),\ldots,\alpha(\boldsymbol{x}_{n})\bigr)^{\mathsf T}
\in\mathbb{R}^{n}
\),
\(
\mathbf{W}
=
\bigl(\omega(\boldsymbol{x}_{i},\boldsymbol{\theta}_{j})\bigr)_{i,j}
\in\mathbb{R}^{n\times m}
\),
\(
\mathbf{G}(\boldsymbol{\theta}^{p})
=
\mathrm{diag}\!\Bigl(
g(\boldsymbol{\theta}_{1};\boldsymbol{\theta}^{p}),\ldots,
g(\boldsymbol{\theta}_{m};\boldsymbol{\theta}^{p})
\Bigr)
\in\mathbb{R}^{m\times m}
\),
\[
\mathbf{B}_{\beta}
=
\begin{pmatrix}
\boldsymbol{\beta}(\boldsymbol{x}_{1})^{\mathsf T}\\
\vdots\\
\boldsymbol{\beta}(\boldsymbol{x}_{n})^{\mathsf T}
\end{pmatrix}
\in\mathbb{R}^{n\times q},
\qquad
\mathbf{T}(\boldsymbol{\theta}^{p})
=
\begin{pmatrix}
(\boldsymbol{\theta}_{1}-\boldsymbol{\theta}^{p})^{\mathsf T}\\
\vdots\\
(\boldsymbol{\theta}_{m}-\boldsymbol{\theta}^{p})^{\mathsf T}
\end{pmatrix}
\in\mathbb{R}^{m\times q},
\]
and
\(\boldsymbol{\epsilon}=(\epsilon_1,\ldots,\epsilon_n)^{\mathsf T}\in\mathbb{R}^{n}\),
\(\boldsymbol{\Xi}=(\xi_{ij})\in\mathbb{R}^{n\times m}\), and let \(\mathbf{1}_{m}\in\mathbb{R}^{m}\) denote the \(m\)-dimensional vector of ones. Then the observation layer can be written in matrix form as
\begin{equation}
\label{eq:obs_matrix_bmdd}
\boldsymbol{\Delta}
=
\boldsymbol{\alpha}\mathbf{1}_{m}^{\mathsf T}
+\mathbf{B}_{\beta}\mathbf{T}(\boldsymbol{\theta}^{p})^{\mathsf T}
+\mathbf{W}\mathbf{G}(\boldsymbol{\theta}^{p})
+\boldsymbol{\epsilon}\mathbf{1}_{m}^{\mathsf T}
+\boldsymbol{\Xi}.
\end{equation}

Next, we specify the parameter layer. Firstly, we assign a constrained GP prior to
\(\alpha(\boldsymbol{x})\) so that the constraint
\eqref{eq:dd_alpha_constraint} is satisfied almost surely. Let
\(\mu_\alpha(\boldsymbol{x})\) be the mean function and
\(k_\alpha(\boldsymbol{x},\boldsymbol{x}')\) be a positive definite kernel
function on \(\Omega\). For example, one may use the squared exponential kernel
\begin{equation}
\label{eq:kernel}
k(\boldsymbol{x},\boldsymbol{x}')
=
\exp\!\left(
-\sum_{r=1}^{d}\frac{(x_r-x_r')^{2}}{\rho_r^{2}}
\right),
\end{equation}
where \(\boldsymbol{x}=(x_1,\ldots,x_d)^{\mathsf T}\),
\(\boldsymbol{x}'=(x_1',\ldots,x_d')^{\mathsf T}\in\Omega\), and
\(\rho_r>0\) is the correlation length parameter for the \(r\)th
input dimension, which can be prespecified. Let \(\boldsymbol{c}=(c_1,\ldots,c_q)^{\mathsf T}\),
\(\boldsymbol{\mu}_I=
\left(
\int_{\Omega_1}\mu_\alpha(\boldsymbol{x})\,d\boldsymbol{x},
\ldots,
\int_{\Omega_q}\mu_\alpha(\boldsymbol{x})\,d\boldsymbol{x}
\right)^{\mathsf T}\),
\(\boldsymbol{k}_{\alpha I}(\boldsymbol{x})
=
\left(
\int_{\Omega_1} k_\alpha(\boldsymbol{x},\boldsymbol{u})\,d\boldsymbol{u},
\ldots,
\int_{\Omega_q} k_\alpha(\boldsymbol{x},\boldsymbol{u})\,d\boldsymbol{u}
\right)^{\mathsf T}\), and
\(\mathbf K_{II}
=
\left(
\int_{\Omega_s}\int_{\Omega_t}
k_\alpha(\boldsymbol{u},\boldsymbol{v})\,
d\boldsymbol{u}\,d\boldsymbol{v}
\right)_{s,t=1}^{q}\). We assign
\[
\alpha(\boldsymbol{x})
\sim
\mathcal{GP}\big(
\mu_{\alpha,c}(\boldsymbol{x}),
\sigma_\alpha^2 k_{\alpha,c}(\boldsymbol{x},\boldsymbol{x}')
\big),
\]
where \(\mu_{\alpha,c}(\boldsymbol{x})
=
\mu_\alpha(\boldsymbol{x})
+
\boldsymbol{k}_{\alpha I}(\boldsymbol{x})^{\mathsf T}
\mathbf K_{II}^{-1}
(\boldsymbol{c}-\boldsymbol{\mu}_I)\), and
\(k_{\alpha,c}(\boldsymbol{x},\boldsymbol{x}')
=
k_\alpha(\boldsymbol{x},\boldsymbol{x}')
-
\boldsymbol{k}_{\alpha I}(\boldsymbol{x})^{\mathsf T}
\mathbf K_{II}^{-1}
\boldsymbol{k}_{\alpha I}(\boldsymbol{x}')\).
Then, sample paths from this prior satisfy
\eqref{eq:dd_alpha_constraint} almost surely. Let
\(\boldsymbol{\alpha}
=
\big(
\alpha(\boldsymbol{x}_1),\ldots,\alpha(\boldsymbol{x}_n)
\big)^{\mathsf T}\),
\(\boldsymbol{\mu}_{\alpha,c}
=
\big(
\mu_{\alpha,c}(\boldsymbol{x}_1),\ldots,
\mu_{\alpha,c}(\boldsymbol{x}_n)
\big)^{\mathsf T}\), and
\(\mathbf K_{\alpha,c}
=
\big(
k_{\alpha,c}(\boldsymbol{x}_i,\boldsymbol{x}_{i'})
\big)_{i,i'=1}^{n}\). The induced finite-dimensional prior is
\begin{equation}
\label{eq:alpha_prior_constrained}
\boldsymbol{\alpha}\mid\sigma_\alpha^2
\sim
\mathcal{N}\big(
\boldsymbol{\mu}_{\alpha,c},
\sigma_\alpha^2\mathbf K_{\alpha,c}
\big).
\end{equation}

For
\(\boldsymbol{\beta}(\boldsymbol{x})=\bigl(\beta_{1}(\boldsymbol{x}),\ldots,\beta_{q}(\boldsymbol{x})\bigr)^{\mathsf T}\),
we assign mutually independent GP priors to each component,
\[
\beta_{\ell}(\boldsymbol{x})\sim
\mathcal{GP}\bigl(\mu_{\beta_{\ell}}(\boldsymbol{x}),\ \sigma_{\beta_{\ell}}^{2}k_{\beta}(\boldsymbol{x},\boldsymbol{x}')\bigr),
\qquad \ell=1,\ldots,q,
\]
where \(\mu_{\beta_{\ell}}(\boldsymbol{x})\) is the mean function, typically taken to be constant; \(\sigma_{\beta_{\ell}}^{2}>0\) is the process variance; and \(k_{\beta}(\boldsymbol{x},\boldsymbol{x}')\) is a kernel function defined on \(\Omega\). Define
\(
\boldsymbol{\beta}_{\ell}
=
\bigl(\beta_{\ell}(\boldsymbol{x}_{1}),\ldots,\beta_{\ell}(\boldsymbol{x}_{n})\bigr)^{\mathsf T}
\),
\(
\boldsymbol{\mu}_{\beta_{\ell}}
=
\bigl(\mu_{\beta_{\ell}}(\boldsymbol{x}_{1}),\ldots,\mu_{\beta_{\ell}}(\boldsymbol{x}_{n})\bigr)^{\mathsf T},
\)
and \(\mathbf{K}_{\beta}=\big(k_{\beta}(\boldsymbol{x}_{i},\boldsymbol{x}_{i'})\big)_{i,i'=1}^{n}\) .
Then, for each \(\ell=1,\ldots,q\), we have
\begin{equation}
\label{eq:prior_beta_vec_bmdd}
\boldsymbol{\beta}_{\ell}\mid \sigma_{\beta_{\ell}}^{2}
\sim
\mathcal{N}\!\bigl(\boldsymbol{\mu}_{\beta_{\ell}},\ \sigma_{\beta_{\ell}}^{2}\mathbf{K}_{\beta}\bigr).
\end{equation}

For the process \(\omega(\boldsymbol{x},\boldsymbol{\theta})\), we adopt a
constrained GP prior with a separable kernel structure to ensure that
the constraint \eqref{eq:omega_integral_constraint} is satisfied almost surely. Let \(k_x(\boldsymbol{x},\boldsymbol{x}')\) and
\(k_\theta(\boldsymbol{\theta},\boldsymbol{\theta}')\) be positive definite
kernel functions on \(\Omega\) and \(\Theta\), respectively. Define \(\mathbf K_{x,II}
=
\left(
\int_{\Omega_s}\int_{\Omega_t}
k_x(\boldsymbol{u},\boldsymbol{v})\,
d\boldsymbol{u}\,d\boldsymbol{v}
\right)_{s,t=1}^{q}\)
and \(\boldsymbol{k}_{xI}(\boldsymbol{x})
=
\left(
\int_{\Omega_1} k_x(\boldsymbol{x},\boldsymbol{u})\,d\boldsymbol{u},
\ldots,
\int_{\Omega_q} k_x(\boldsymbol{x},\boldsymbol{u})\,d\boldsymbol{u}
\right)^{\mathsf T}\).  We assign
\[
\omega(\boldsymbol{x},\boldsymbol{\theta})
\sim
\mathcal{GP}\Bigl(
0,\ \sigma_{\omega}^{2}k_{x,c}(\boldsymbol{x},\boldsymbol{x}')
k_{\theta}(\boldsymbol{\theta},\boldsymbol{\theta}')
\Bigr),
\]
where \(k_{x,c}(\boldsymbol{x},\boldsymbol{x}')
=
k_x(\boldsymbol{x},\boldsymbol{x}')
-
\boldsymbol{k}_{xI}(\boldsymbol{x})^{\mathsf T}
\mathbf K_{x,II}^{-1}
\boldsymbol{k}_{xI}(\boldsymbol{x}')\).
Then, sample paths from this prior satisfy
\eqref{eq:omega_integral_constraint} almost surely. Therefore,  \(r(\boldsymbol{x},\boldsymbol{\theta};\boldsymbol{\theta}^{p})\) satisfies
Assumption~\ref{ass:bmdd_r_orthogonal}. Let
\(\mathbf{K}_{x,c}
=
\big(
k_{x,c}(\boldsymbol{x}_i,\boldsymbol{x}_{i'})
\big)_{i,i'=1}^{n}\),
\(\mathbf{K}_{\theta}
=
\big(
k_{\theta}(\boldsymbol{\theta}_j,\boldsymbol{\theta}_{j'})
\big)_{j,j'=1}^{m}\), and \(\mathbf{W}
=
\big(\omega(\boldsymbol{x}_i,\boldsymbol{\theta}_j)\big)_{
i=1,\ldots,n;\,j=1,\ldots,m}
\).
Then the induced finite-dimensional prior is
\begin{equation}
\label{eq:prior_W_vec_bmdd}
\operatorname{vec}(\mathbf{W})\mid \sigma_{\omega}^{2}
\sim
\mathcal{N}\!\bigl(
\mathbf{0},\ \sigma_{\omega}^{2}(\mathbf{K}_{\theta}\otimes \mathbf{K}_{x,c})
\bigr),
\end{equation}
where \(\operatorname{vec}(\cdot)\) denotes vectorization, and \(\otimes\) denotes the
Kronecker product.

In addition, we assign a weakly informative prior to the calibration parameter. Specifically, we can take a uniform prior on its support \(\Theta\),
\[
\boldsymbol{\theta}^{p}\sim \mathcal{U}(\Theta).
\]
Alternatively, we may use a multivariate normal prior truncated to \(\Theta\) to reflect prior knowledge about the calibration parameter.

Finally, we specify the hyperparameter layer. The hyperparameters in BMDD consist of the variance \(\sigma_{\alpha}^{2}\) of the intrinsic discrepancy process, the componentwise variances \(\{\sigma_{\beta_{\ell}}^{2}\}_{\ell=1}^{q}\) of the parametric discrepancy coefficient process, the variance \(\sigma_{\omega}^{2}\) of the residual process, the variance \(\sigma_{\epsilon}^{2}\) of the physical observation noise, and the nugget variance \(\sigma_{\xi}^{2}\). We assign inverse-Gamma priors to these variance parameters:
\[
\sigma_{\epsilon}^{2}\sim \mathcal{IG}(a_{\epsilon},b_{\epsilon}),\qquad
\sigma_{\xi}^{2}\sim \mathcal{IG}(a_{\xi},b_{\xi}),\qquad
\sigma_{\omega}^{2}\sim \mathcal{IG}(a_{\omega},b_{\omega}),
\]
\[
\sigma_{\alpha}^{2}\sim \mathcal{IG}(a_{\alpha},b_{\alpha}),\qquad
\sigma_{\beta_{\ell}}^{2}\sim \mathcal{IG}(a_{\beta},b_{\beta}),
\qquad \ell=1,\ldots,q,
\]
where \(a_{\epsilon}, b_{\epsilon}, a_{\xi}, b_{\xi},
a_{\omega}, b_{\omega}, a_{\alpha}, b_{\alpha},
a_{\beta}, b_{\beta}\) are all prespecified positive constants.

The observation, parameter, and hyperparameter layers together constitute the Bayesian hierarchical structure of the BMDD model, which provides the foundation for subsequent posterior inference, physical system prediction, and uncertainty quantification.

\subsection{Posterior Inference and Sampling Algorithm}
\label{subsec:mwG}
\noindent
In this subsection, based on the Bayesian hierarchical structure introduced in Subsection \ref{subsec:hierarchy}, we derive the full conditional posterior distributions of all unknown quantities and present the details of the Metropolis-within-Gibbs sampling algorithm.

Let
\[
\boldsymbol{\Psi}
=
\Big\{
\boldsymbol{\theta}^{p},\ \boldsymbol{\alpha},\ \mathbf{B}_{\beta},\ \mathbf{W},\ \boldsymbol{\epsilon},\ \sigma_{\xi}^{2},\ \sigma_{\epsilon}^{2},\ \sigma_{\omega}^{2},\ \sigma_{\alpha}^{2},\ \{\sigma_{\beta_{\ell}}^{2}\}_{\ell=1}^{q}
\Big\}
\]
denote the collection of all unknown quantities and hyperparameters in BMDD. We use a Metropolis-within-Gibbs sampler to draw posterior samples for the components of \(\boldsymbol{\Psi}\). In this approach, we employ Gibbs sampling as the main framework: for parameters with standard full conditional posteriors (e.g., Gaussian or inverse-gamma distributions), we sample directly from these conditionals. For parameters without standard full conditional forms, we embed a Metropolis--Hastings (MH) step within the Gibbs sampler. This results in a hybrid Metropolis-within-Gibbs sampler \citep{Tierney1994MC}.

Given \(\boldsymbol{\Psi}\), the log-likelihood corresponding to \eqref{eq:obs_matrix_bmdd} is
\begin{equation}
\label{eq:loglik_Delta}
\log p\!\left(\boldsymbol{\Delta}\mid\boldsymbol{\Psi}\right)
=
-\frac{nm}{2}\log\sigma_{\xi}^{2}
-\frac{1}{2\sigma_{\xi}^{2}}
\left\|
\boldsymbol{\Delta}
-\boldsymbol{\alpha}\mathbf{1}_{m}^{\mathsf T}
-\mathbf{B}_{\beta}\mathbf{T}(\boldsymbol{\theta}^{p})^{\mathsf T}
-\mathbf{W}\mathbf{G}(\boldsymbol{\theta}^{p})
-\boldsymbol{\epsilon}\mathbf{1}_{m}^{\mathsf T}
\right\|_{F}^{2}
+C,
\end{equation}
where \(C=-\frac{nm}{2}\log(2\pi)\) is independent of
\(\boldsymbol{\Psi}\), and the Frobenius norm is defined by \(
\|\mathbf{A}\|_{F}^{2}:=\sum_{i,j}a_{ij}^{2}
\), with \(\mathbf{A}=(a_{ij})\) denoting any conformable matrix. By
\[
p(\boldsymbol{\Psi}\mid \boldsymbol{\Delta})
\propto
p(\boldsymbol{\Delta}\mid \boldsymbol{\Psi})\,p(\boldsymbol{\Psi}),
\]
we derive below the full conditional posterior distributions of the components of \(\boldsymbol{\Psi}\). Detailed derivations are provided in the Supplementary Materials.

\textbf{(i) Full conditional posterior distribution of \(\boldsymbol{\alpha}\).}
Let
\(
\widetilde{\boldsymbol{\Delta}}_{\alpha}
=
\boldsymbol{\Delta}
-
\mathbf B_{\beta}\mathbf T(\boldsymbol{\theta}^{p})^{\mathsf T}
-
\mathbf W\mathbf G(\boldsymbol{\theta}^{p})
-
\boldsymbol{\epsilon}\mathbf 1_m^{\mathsf T}.
\)
Then, by \eqref{eq:obs_matrix_bmdd},
\(
\widetilde{\boldsymbol{\Delta}}_{\alpha}
=
\boldsymbol{\alpha}\mathbf 1_m^{\mathsf T}
+
\boldsymbol{\Xi}.
\)
Combining the likelihood in \eqref{eq:loglik_Delta} with the prior in \eqref{eq:alpha_prior_constrained}, we obtain the full conditional posterior distribution
of \(\boldsymbol{\alpha}\) as
\begin{equation}
\label{eq:alpha_full_conditional_constrained}
\boldsymbol{\alpha}
\mid
\big(
\boldsymbol{\Delta},
\mathbf B_{\beta},
\mathbf W,
\boldsymbol{\theta}^{p},
\boldsymbol{\epsilon},
\sigma_{\xi}^{2},
\sigma_{\alpha}^{2}
\big)
\sim
\mathcal{N}\big(
\boldsymbol{\mu}_{\alpha\mid\cdot},
\boldsymbol{\Sigma}_{\alpha\mid\cdot}
\big),
\end{equation}
where
\(
\boldsymbol{\Sigma}_{\alpha\mid\cdot}^{-1}
=
\frac{1}{\sigma_\alpha^2}
\mathbf K_{\alpha,c}^{-1}
+
\frac{m}{\sigma_{\xi}^{2}}\mathbf I_n,
\)
and
\(
\boldsymbol{\mu}_{\alpha\mid\cdot}
=
\boldsymbol{\Sigma}_{\alpha\mid\cdot}
\left(
\frac{1}{\sigma_\alpha^2}
\mathbf K_{\alpha,c}^{-1}\boldsymbol{\mu}_{\alpha,c}
+
\frac{1}{\sigma_{\xi}^{2}}
\widetilde{\boldsymbol{\Delta}}_{\alpha}\mathbf 1_m
\right).
\)
Since the prior in \eqref{eq:alpha_prior_constrained} already incorporates the identifiability constraints  \eqref{eq:dd_alpha_constraint}, the posterior distribution \eqref{eq:alpha_full_conditional_constrained} still satisfies the same constraints.

\textbf{(ii) Full conditional posterior distribution of $\boldsymbol{\epsilon}$.}
Let
\(
\widetilde{\boldsymbol{\Delta}}_{\epsilon}
=
\boldsymbol{\Delta}
-
\boldsymbol{\alpha}\mathbf{1}_{m}^{\mathsf T}
-
\mathbf{B}_{\beta}\mathbf{T}(\boldsymbol{\theta}^{p})^{\mathsf T}
-
\mathbf{W}\mathbf{G}(\boldsymbol{\theta}^{p})
\).
Then, by \eqref{eq:obs_matrix_bmdd},
\(
\widetilde{\boldsymbol{\Delta}}_{\epsilon}
=
\boldsymbol{\epsilon}\mathbf{1}_{m}^{\mathsf T}+\boldsymbol{\Xi}
\).
Combining this with the prior
\(\boldsymbol{\epsilon}\mid\sigma_{\epsilon}^{2}\sim\mathcal{N}(\mathbf{0},\sigma_{\epsilon}^{2}\mathbf{I}_{n})\),
we obtain the full conditional posterior distribution of \(\boldsymbol{\epsilon}\) as
\begin{equation}
\label{eq:eps_full}
\boldsymbol{\epsilon}\mid\big(
\boldsymbol{\Delta},\boldsymbol{\alpha},\mathbf{B}_{\beta},\mathbf{W},\boldsymbol{\theta}^{p},
\sigma_{\xi}^{2},\sigma_{\epsilon}^{2}\big)
\sim
\mathcal{N}\!\big(\boldsymbol{\mu}_{\epsilon\mid\cdot},\ \boldsymbol{\Sigma}_{\epsilon\mid\cdot}\big),
\end{equation}
where
\(
\boldsymbol{\Sigma}_{\epsilon\mid\cdot}^{-1}
=
\left(\frac{1}{\sigma_{\epsilon}^{2}}+\frac{m}{\sigma_{\xi}^{2}}\right)\mathbf{I}_{n}
\),
and
\(
\boldsymbol{\mu}_{\epsilon\mid\cdot}
=
\boldsymbol{\Sigma}_{\epsilon\mid\cdot}
\left(
\frac{1}{\sigma_{\xi}^{2}}\widetilde{\boldsymbol{\Delta}}_{\epsilon}\mathbf{1}_{m}
\right)
\).

\textbf{(iii) Full conditional posterior distribution of $\mathbf{B}_{\beta}$.}
Let
\(
\widetilde{\boldsymbol{\Delta}}_{\beta}
=
\boldsymbol{\Delta}
-
\boldsymbol{\alpha}\mathbf{1}_{m}^{\mathsf T}
-
\mathbf{W}\mathbf{G}(\boldsymbol{\theta}^{p})
-
\boldsymbol{\epsilon}\mathbf{1}_{m}^{\mathsf T}
\).
Then, by \eqref{eq:obs_matrix_bmdd},
\(
\widetilde{\boldsymbol{\Delta}}_{\beta}
=
\mathbf{B}_{\beta}\mathbf{T}(\boldsymbol{\theta}^{p})^{\mathsf T}+\boldsymbol{\Xi}
\).
Under the componentwise independent GP priors in \eqref{eq:prior_beta_vec_bmdd}, define
\(\boldsymbol{\mu}_{b}=\operatorname{vec}\!\big([\boldsymbol{\mu}_{\beta_{1}},\ldots,\boldsymbol{\mu}_{\beta_{q}}]\big)\)
as the prior mean vector of \(\operatorname{vec}(\mathbf{B}_{\beta})\), and define
\[
\mathbf{Q}_{b}
=
\operatorname{blockdiag}\!\left(
\frac{1}{\sigma_{\beta_{1}}^{2}}\mathbf{K}_{\beta}^{-1},\ldots,
\frac{1}{\sigma_{\beta_{q}}^{2}}\mathbf{K}_{\beta}^{-1}
\right)\in\mathbb{R}^{nq\times nq}
\]
as the corresponding prior precision matrix, where \(\operatorname{blockdiag}(\cdot)\) denotes the block diagonal matrix formed from the specified blocks. Then the full conditional posterior distribution of \(\operatorname{vec}(\mathbf{B}_{\beta})\) is
\begin{equation}
\label{eq:beta_full_vec}
\operatorname{vec}(\mathbf{B}_{\beta})\mid \big(
\boldsymbol{\Delta},\boldsymbol{\alpha},\mathbf{W},\boldsymbol{\theta}^{p},\boldsymbol{\epsilon},
\sigma_{\xi}^{2},\{\sigma_{\beta_{\ell}}^{2}\}_{\ell=1}^{q} \big)
\sim
\mathcal{N}\!\big(\boldsymbol{\mu}_{b\mid\cdot},\ \boldsymbol{\Sigma}_{b\mid\cdot}\big),
\end{equation}
where
\(
\boldsymbol{\Sigma}_{b\mid\cdot}^{-1}
=
\mathbf{Q}_{b}
+
\frac{1}{\sigma_{\xi}^{2}}
\big(\mathbf{T}(\boldsymbol{\theta}^{p})^{\mathsf T}\mathbf{T}(\boldsymbol{\theta}^{p})\otimes \mathbf{I}_{n}\big)
\),
and
\(
\boldsymbol{\mu}_{b\mid\cdot}
=
\boldsymbol{\Sigma}_{b\mid\cdot}
\left(
\mathbf{Q}_{b}\boldsymbol{\mu}_{b}
+
\frac{1}{\sigma_{\xi}^{2}}\,
\operatorname{vec}\!\big(\widetilde{\boldsymbol{\Delta}}_{\beta}\mathbf{T}(\boldsymbol{\theta}^{p})\big)
\right)
\).

\textbf{(iv) Full conditional posterior distribution of \(\mathbf{W}\).}
Let
\(
\widetilde{\boldsymbol{\Delta}}_{W}
=
\boldsymbol{\Delta}
-
\boldsymbol{\alpha}\mathbf{1}_{m}^{\mathsf T}
-
\mathbf{B}_{\beta}\mathbf{T}(\boldsymbol{\theta}^{p})^{\mathsf T}
-
\boldsymbol{\epsilon}\mathbf{1}_{m}^{\mathsf T}
\).
Then, by \eqref{eq:obs_matrix_bmdd},
\(
\widetilde{\boldsymbol{\Delta}}_{W}
=
\mathbf{W}\mathbf{G}(\boldsymbol{\theta}^{p})
+
\boldsymbol{\Xi}
\).
Combining this likelihood with the constrained GP prior in
\eqref{eq:prior_W_vec_bmdd}, we obtain the full conditional posterior
distribution of \(\operatorname{vec}(\mathbf{W})\) as
\begin{equation}
\label{eq:W_full_vec}
\operatorname{vec}(\mathbf{W})
\mid
\big(
\boldsymbol{\Delta},
\boldsymbol{\alpha},
\mathbf{B}_{\beta},
\boldsymbol{\theta}^{p},
\boldsymbol{\epsilon},
\sigma_{\xi}^{2},
\sigma_{\omega}^{2}
\big)
\sim
\mathcal{N}\!\big(
\boldsymbol{\mu}_{W\mid\cdot},
\boldsymbol{\Sigma}_{W\mid\cdot}
\big),
\end{equation}
where
\(
\boldsymbol{\Sigma}_{W\mid\cdot}^{-1}
=
\frac{1}{\sigma_{\xi}^{2}}
\big(
\mathbf{G}(\boldsymbol{\theta}^{p})^{2}
\otimes
\mathbf{I}_{n}
\big)
+
\frac{1}{\sigma_{\omega}^{2}}
\big(
\mathbf{K}_{\theta}^{-1}
\otimes
\mathbf{K}_{x,c}^{-1}
\big),
\)
and
\(
\boldsymbol{\mu}_{W\mid\cdot}
=
\boldsymbol{\Sigma}_{W\mid\cdot}
\left[
\frac{1}{\sigma_{\xi}^{2}}
\operatorname{vec}
\left(
\widetilde{\boldsymbol{\Delta}}_{W}
\mathbf{G}(\boldsymbol{\theta}^{p})
\right)
\right].
\)
Since the constrained prior for \(\omega(\boldsymbol{x},\boldsymbol{\theta})\)
incorporates the constraint \eqref{eq:omega_integral_constraint}, the posterior distribution \eqref{eq:W_full_vec} still satisfies the same constraint.

\textbf{(v) Full conditional posterior distributions of the variance parameters.}
By conjugacy, we have
\vspace{-0.3em}
\begin{align}
\label{eq:sig_F_full}
\sigma_{\epsilon}^{2}\mid \boldsymbol{\epsilon}
&\sim
\mathcal{IG}\!\left(
a_{\epsilon}+\frac{n}{2},\
b_{\epsilon}+\frac{1}{2}\boldsymbol{\epsilon}^{\mathsf T}\boldsymbol{\epsilon}
\right),\\[4pt]
\label{eq:sig_alpha_full}
\sigma_{\alpha}^{2}\mid \boldsymbol{\alpha}
&\sim
\mathcal{IG}\!\left(
a_{\alpha}+\frac{n}{2},\
b_{\alpha}+\frac{1}{2}
(\boldsymbol{\alpha}-\boldsymbol{\mu}_{\alpha,c})^{\mathsf T}\mathbf{K}_{\alpha,c}^{-1}(\boldsymbol{\alpha}-\boldsymbol{\mu}_{\alpha,c})
\right),\\[4pt]
\label{eq:sig_beta_full}
\sigma_{\beta_{\ell}}^{2}\mid \boldsymbol{\beta}_{\ell}
&\sim
\mathcal{IG}\!\left(
a_{\beta}+\frac{n}{2},\
b_{\beta}+\frac{1}{2}
(\boldsymbol{\beta}_{\ell}-\boldsymbol{\mu}_{\beta_{\ell}})^{\mathsf T}\mathbf{K}_{\beta}^{-1}(\boldsymbol{\beta}_{\ell}-\boldsymbol{\mu}_{\beta_{\ell}})
\right),\ \ell=1,\ldots,q,\\[4pt]
\label{eq:tau_full}
\sigma_{\omega}^{2}\mid \mathbf{W}
&\sim
\mathcal{IG}\!\left(
a_{\omega}+\frac{nm}{2},\
b_{\omega}+\frac{1}{2}\,
\operatorname{vec}(\mathbf{W})^{\mathsf T}
\bigl(\mathbf{K}_{\theta}^{-1}\otimes \mathbf{K}_{x,c}^{-1}\bigr)
\operatorname{vec}(\mathbf{W})
\right),\\[4pt]
\label{eq:sig_delta_full}
\sigma_{\xi}^{2}\mid \big(
\boldsymbol{\Delta},\boldsymbol{\alpha},\mathbf{B}_{\beta},\mathbf{W},\boldsymbol{\theta}^{p},\boldsymbol{\epsilon}\big)
&\sim
\mathcal{IG}\!\left(
a_{\xi}+\frac{nm}{2},\
b_{\xi}+\frac{1}{2}\left\|
\boldsymbol{\Delta}
-\boldsymbol{\alpha}\mathbf{1}_{m}^{\mathsf T}
-\mathbf{B}_{\beta}\mathbf{T}(\boldsymbol{\theta}^{p})^{\mathsf T}
-\mathbf{W}\mathbf{G}(\boldsymbol{\theta}^{p})
-\boldsymbol{\epsilon}\mathbf{1}_{m}^{\mathsf T}
\right\|_{F}^{2}
\right).
\end{align}
\vspace{-0.3em}

\textbf{(vi) Full conditional form of the calibration parameter $\boldsymbol{\theta}^{p}$ and the MH update.}
Both \(\mathbf{T}(\boldsymbol{\theta}^{p})\) and \(\mathbf{G}(\boldsymbol{\theta}^{p})\) depend on \(\boldsymbol{\theta}^{p}\). As a result, the full conditional posterior distribution of \(\boldsymbol{\theta}^{p}\) generally does not belong to a standard distribution family. Referring to \eqref{eq:loglik_Delta} and using the prior \(\pi(\boldsymbol{\theta}^{p})\), the full conditional posterior density is given by
\[
\pi(\boldsymbol{\theta}^{p}\mid
\boldsymbol{\Delta},\boldsymbol{\alpha},\mathbf{B}_{\beta},\mathbf{W},\boldsymbol{\epsilon},\sigma_{\xi}^{2})
\ \propto\
\pi(\boldsymbol{\theta}^{p})\,
\exp\!\left\{
-\frac{1}{2\sigma_{\xi}^{2}}
\left\|
\boldsymbol{\Delta}
-\boldsymbol{\alpha}\mathbf{1}_{m}^{\mathsf T}
-\mathbf{B}_{\beta}\mathbf{T}(\boldsymbol{\theta}^{p})^{\mathsf T}
-\mathbf{W}\mathbf{G}(\boldsymbol{\theta}^{p})
-\boldsymbol{\epsilon}\mathbf{1}_{m}^{\mathsf T}
\right\|_{F}^{2}
\right\}.
\]
Accordingly, we update \(\boldsymbol{\theta}^{p}\) using an MH step. Specifically, at iteration \(t\), given the current value \(\boldsymbol{\theta}^{p,(t)}\), we propose
\begin{equation}
\label{eq:theta_prop}
\boldsymbol{\theta}^{p,\star}
=
\boldsymbol{\theta}^{p,(t)}+\mathbf{s}_{\theta}\boldsymbol{u},
\qquad
\boldsymbol{u}\sim \mathcal{N}(\mathbf{0},\mathbf{I}_{q}),
\end{equation}
where \(\mathbf{s}_{\theta}\) is a step-size matrix, which may be taken as diagonal to tune the proposal scale separately for each component. If the support of the prior for \(\boldsymbol{\theta}^{p}\) is the bounded set \(\Theta\), then the proposal is rejected immediately whenever \(\boldsymbol{\theta}^{p,\star}\notin\Theta\). The acceptance probability is
\[
\alpha_{\mathrm{MH}}
=
\min\left\{
1,\
\frac{\pi(\boldsymbol{\theta}^{p,\star})}{\pi(\boldsymbol{\theta}^{p,(t)})}
\cdot
\exp\!\left(
-\frac{1}{2\sigma_{\xi}^{2}}
\Big[
\|\mathbf{R}(\boldsymbol{\theta}^{p,\star})\|_{F}^{2}
-
\|\mathbf{R}(\boldsymbol{\theta}^{p,(t)})\|_{F}^{2}
\Big]
\right)
\right\},
\]
where
\[
\mathbf{R}(\boldsymbol{\theta}^{p})=
\boldsymbol{\Delta}
-
\boldsymbol{\alpha}\mathbf{1}_{m}^{\mathsf T}
-
\mathbf{B}_{\beta}\mathbf{T}(\boldsymbol{\theta}^{p})^{\mathsf T}
-
\mathbf{W}\mathbf{G}(\boldsymbol{\theta}^{p})
-
\boldsymbol{\epsilon}\mathbf{1}_{m}^{\mathsf T}
\in\mathbb{R}^{n\times m}.
\]
In practice, we can tune the matrix \(\mathbf{s}_{\theta}\) in \eqref{eq:theta_prop} to achieve a suitable MH acceptance rate and thereby improve the mixing efficiency of the chain.

In summary, posterior sampling for BMDD proceeds by sequentially updating the components within a Gibbs framework, with an embedded MH step for the calibration parameter \(\boldsymbol{\theta}^{p}\). The implementation of the Metropolis-within-Gibbs algorithm is presented in Algorithm \ref{alg:bmdd_mwg_cn}.

\begin{center}
\begin{minipage}{0.98\textwidth}
\refstepcounter{algo}\label{alg:bmdd_mwg_cn}

\small
\setlength{\abovedisplayskip}{4pt}
\setlength{\belowdisplayskip}{4pt}
\setlength{\abovedisplayshortskip}{2pt}
\setlength{\belowdisplayshortskip}{2pt}

\noindent\rule{\linewidth}{0.35pt}
\noindent\textbf{Algorithm \thealgo\ \ Metropolis-within-Gibbs Sampling Algorithm for the BMDD Model}\\[-0.5em]
\noindent\rule{\linewidth}{0.08pt}\\
\vspace{-1,5em}

\noindent\textbf{Input:}
Observed difference matrix $\boldsymbol{\Delta}$; design points
$\{\boldsymbol{x}_i\}_{i=1}^n$ and $\{\boldsymbol{\theta}_j\}_{j=1}^m$; hyperparameters;
total number of iterations $T$; burn-in length $T_0$;
MH step-size matrix $\mathbf{s}_\theta$; initial values
$\boldsymbol{\theta}^{p,(0)},\boldsymbol{\alpha}^{(0)},\mathbf{B}_{\beta}^{(0)},\mathbf{W}^{(0)},\boldsymbol{\epsilon}^{(0)},\sigma_{\xi}^{2,(0)},\sigma_{\epsilon}^{2,(0)},
\sigma_{\omega}^{2,(0)},\sigma_{\alpha}^{2,(0)},\{\sigma_{\beta_\ell}^{2,(0)}\}_{\ell=1}^q$.

\vspace{0.15em}
\noindent\textbf{Output:}
Posterior samples retained after burn-in
$\{\boldsymbol{\Psi}^{(t)}\}_{t=T_0+1}^{T}$ and the associated statistics, where
\[
\boldsymbol{\Psi}^{(t)}
=
\{\boldsymbol{\theta}^{p,(t)},\boldsymbol{\alpha}^{(t)},\mathbf{B}_{\beta}^{(t)},\mathbf{W}^{(t)},\boldsymbol{\epsilon}^{(t)},
\sigma_{\xi}^{2,(t)},\sigma_{\epsilon}^{2,(t)},\sigma_{\omega}^{2,(t)},\sigma_{\alpha}^{2,(t)},\{\sigma_{\beta_\ell}^{2,(t)}\}_{\ell=1}^q\}.
\]

\vspace{-0.4em}
\noindent\textbf{Procedure:}

\noindent\textbf{1.} Initialize all unknown quantities and hyperparameters.

\noindent\textbf{2.} For $t=1,2,\ldots,T$, repeat:

\noindent\hspace*{1.5em}\textbf{(1$^\circ$)} Sample \(\boldsymbol{\alpha}^{(t)}\) from \(p(\boldsymbol{\alpha}\mid\cdots)\)
according to \eqref{eq:alpha_full_conditional_constrained}.

\noindent\hspace*{1.5em}\textbf{(2$^\circ$)} Sample $\mathbf{B}_{\beta}^{(t)}$ from
$p(\mathbf{B}_{\beta}\mid \cdots)$ according to \eqref{eq:beta_full_vec}.

\noindent\hspace*{1.5em}\textbf{(3$^\circ$)} Sample $\mathbf{W}^{(t)}$ from
$p(\mathbf{W}\mid \cdots)$ according to \eqref{eq:W_full_vec}.

\noindent\hspace*{1.5em}\textbf{(4$^\circ$)} Sample $\boldsymbol{\epsilon}^{(t)}$ from
$p(\boldsymbol{\epsilon}\mid \cdots)$ according to \eqref{eq:eps_full}.

\noindent\hspace*{1.5em}\textbf{(5$^\circ$)} Sample $\sigma_{\xi}^{2,(t)}$ from
$p(\sigma_{\xi}^{2}\mid \cdots)$ according to \eqref{eq:sig_delta_full}.

\noindent\hspace*{1.5em}\textbf{(6$^\circ$)} Sample $\sigma_{\epsilon}^{2,(t)}$ from
$p(\sigma_{\epsilon}^{2}\mid \cdots)$ according to \eqref{eq:sig_F_full}.

\noindent\hspace*{1.5em}\textbf{(7$^\circ$)} Sample $\sigma_{\alpha}^{2,(t)}$ from
$p(\sigma_\alpha^2\mid \cdots)$ according to \eqref{eq:sig_alpha_full}.

\noindent\hspace*{1.5em}\textbf{(8$^\circ$)} For each $\ell=1,\ldots,q$, sample
$\sigma_{\beta_\ell}^{2,(t)}$ from
$p(\sigma_{\beta_\ell}^2\mid \cdots)$ according to \eqref{eq:sig_beta_full}.

\noindent\hspace*{1.5em}\textbf{(9$^\circ$)} Sample $\sigma_{\omega}^{2,(t)}$ from
$p(\sigma_{\omega}^{2}\mid \cdots)$ according to \eqref{eq:tau_full}.

\noindent\hspace*{1.5em}\textbf{(10$^\circ$)} Update $\boldsymbol{\theta}^p$ using an MH step:

\noindent\hspace*{3em}\textbf{(a)} Generate
\[
\boldsymbol{\theta}^{p,\star}
=
\boldsymbol{\theta}^{p,(t-1)}+\mathbf{s}_\theta \boldsymbol{u},
\qquad
\boldsymbol{u}\sim \mathcal{N}(\mathbf{0},\mathbf{I}_q),
\]
and reject immediately if $\boldsymbol{\theta}^{p,\star}\notin\Theta$.

\noindent\hspace*{3em}\textbf{(b)} Compute
\[
\alpha_{\mathrm{MH}}
=
\min\left\{
1,\
\frac{\pi(\boldsymbol{\theta}^{p,\star})}{\pi(\boldsymbol{\theta}^{p,(t-1)})}
\exp\!\left[
-\frac{1}{2\sigma_{\xi}^{2,(t)}}
\Bigl(
\|\mathbf{R}(\boldsymbol{\theta}^{p,\star})\|_F^2
-
\|\mathbf{R}(\boldsymbol{\theta}^{p,(t-1)})\|_F^2
\Bigr)
\right]
\right\},
\]
where
\[
\mathbf{R}(\boldsymbol{\theta}^{p})
=
\boldsymbol{\Delta}
-
\boldsymbol{\alpha}^{(t)}\mathbf{1}_m^{\mathsf T}
-
\mathbf{B}_{\beta}^{(t)}\mathbf{T}(\boldsymbol{\theta}^{p})^{\mathsf T}
-
\mathbf{W}^{(t)}\mathbf{G}(\boldsymbol{\theta}^{p})
-
\boldsymbol{\epsilon}^{(t)}\mathbf{1}_m^{\mathsf T}.
\]

\noindent\hspace*{3em}\textbf{(c)} Generate $u^\star\sim \mathrm{Unif}(0,1)$. If
$u^\star<\alpha_{\mathrm{MH}}$, set
$\boldsymbol{\theta}^{p,(t)}=\boldsymbol{\theta}^{p,\star}$; otherwise set
$\boldsymbol{\theta}^{p,(t)}=\boldsymbol{\theta}^{p,(t-1)}$.

\noindent\hspace*{1.5em}\textbf{(11$^\circ$)} If $t>T_0$, save the current sample $\boldsymbol{\Psi}^{(t)}$.

\noindent\textbf{3.} Compute posterior means, credible intervals, and posterior predictive quantities from the retained posterior samples.

\noindent\rule{\linewidth}{0.05pt}
\end{minipage}
\end{center}

\clearpage

\subsection{Prediction}
\label{subsec:prediction}
\noindent
In this subsection, we present posterior predictive inference for the physical-system response \(y^{P}(\boldsymbol{x}^{\ast})\) and a
future physical observation \(y^{F}(\boldsymbol{x}^{\ast})\) at a new input
\(\boldsymbol{x}^{\ast}\in\Omega\). Recall that
\(y^{F}(\boldsymbol{x}^{\ast})
=
y^{P}(\boldsymbol{x}^{\ast})+\epsilon^{\ast}\), where
\(\epsilon^{\ast}\sim\mathcal{N}(0,\sigma_{\epsilon}^{2})\) is independent
of all other quantities.

Let \(\{\boldsymbol{\Psi}^{(s)}\}_{s=1}^{N}\) be the retained posterior
samples from Algorithm~\ref{alg:bmdd_mwg_cn}, where \(N\) denotes the
total number of retained samples. By the BMDD discrepancy
construction in \eqref{eq:bmdd_decomposition} and
Assumption~\ref{ass:bmdd_local},
\(\delta(\boldsymbol{x}^{\ast},\boldsymbol{\theta}^{p};
\boldsymbol{\theta}^{p})=\alpha(\boldsymbol{x}^{\ast})\). Hence, \(
y^{P}(\boldsymbol{x}^{\ast})
=
y^{S}(\boldsymbol{x}^{\ast},\boldsymbol{\theta}^{p})
+
\alpha(\boldsymbol{x}^{\ast})
\). Conditional on \(\boldsymbol{\Psi}\) and \(\boldsymbol{\Delta}\),
\[
\alpha(\boldsymbol{x}^{\ast})
\mid\boldsymbol{\Psi},\boldsymbol{\Delta}
\sim
\mathcal{N}\bigl(
m_{\alpha}(\boldsymbol{x}^{\ast}),
v_{\alpha}(\boldsymbol{x}^{\ast})
\bigr),
\]
where
\(\boldsymbol{k}_{*,c}
=
\bigl(
k_{\alpha,c}(\boldsymbol{x}_{1},\boldsymbol{x}^{\ast}),
\ldots,
k_{\alpha,c}(\boldsymbol{x}_{n},\boldsymbol{x}^{\ast})
\bigr)^{\mathsf T}\),
\(m_{\alpha}(\boldsymbol{x}^{\ast})
=
\mu_{\alpha,c}(\boldsymbol{x}^{\ast})
+
\boldsymbol{k}_{*,c}^{\mathsf T}
\mathbf{K}_{\alpha,c}^{-1}
(\boldsymbol{\alpha}-\boldsymbol{\mu}_{\alpha,c})\), and
\(v_{\alpha}(\boldsymbol{x}^{\ast})
=
\sigma_{\alpha}^{2}
\{k_{\alpha,c}(\boldsymbol{x}^{\ast},\boldsymbol{x}^{\ast})
-
\boldsymbol{k}_{*,c}^{\mathsf T}
\mathbf{K}_{\alpha,c}^{-1}
\boldsymbol{k}_{*,c}\}\).
For the \(s\)-th posterior sample, define
\(\mu_{*}^{(s)}
=
y^{S}(\boldsymbol{x}^{\ast},\boldsymbol{\theta}^{p,(s)})
+
m_{\alpha}^{(s)}(\boldsymbol{x}^{\ast})\),
\(V_{P,*}^{(s)}
=
v_{\alpha}^{(s)}(\boldsymbol{x}^{\ast})\), and
\(V_{F,*}^{(s)}
=
V_{P,*}^{(s)}+\sigma_{\epsilon}^{2,(s)}\). Then,
\[
\begin{aligned}
y^{P}(\boldsymbol{x}^{\ast})
\mid\boldsymbol{\Psi}^{(s)},\boldsymbol{\Delta}
&\sim
\mathcal{N}\bigl(
\mu_{*}^{(s)},V_{P,*}^{(s)}
\bigr),\\
y^{F}(\boldsymbol{x}^{\ast})
\mid\boldsymbol{\Psi}^{(s)},\boldsymbol{\Delta}
&\sim
\mathcal{N}\bigl(
\mu_{*}^{(s)},V_{F,*}^{(s)}
\bigr),
\qquad s=1,\ldots,N.
\end{aligned}
\]
Thus, \(y^{P}(\boldsymbol{x}^{\ast})\) and
\(y^{F}(\boldsymbol{x}^{\ast})\) have the same conditional predictive mean,
whereas the predictive variance for \(y^{F}(\boldsymbol{x}^{\ast})\)
additionally accounts for future observation noise.

The posterior predictive distributions are
approximated by
\[
p\bigl(y^{P}(\boldsymbol{x}^{\ast})\mid\boldsymbol{\Delta}\bigr)
\approx
\frac{1}{N}
\sum_{s=1}^{N}
\phi\bigl(
y^{P}(\boldsymbol{x}^{\ast});
\mu_{*}^{(s)},V_{P,*}^{(s)}
\bigr)\quad\text{and}\quad p\bigl(y^{F}(\boldsymbol{x}^{\ast})\mid\boldsymbol{\Delta}\bigr)
\approx
\frac{1}{N}
\sum_{s=1}^{N}
\phi\bigl(
y^{F}(\boldsymbol{x}^{\ast});
\mu_{*}^{(s)},V_{F,*}^{(s)}
\bigr),
\]
where \(\phi(\,\cdot\,;\mu,V)\) denotes the density function of a normal
distribution with mean \(\mu\) and variance \(V\). Therefore, the posterior
predictive point estimates are
\[
\widehat{y}^{P}(\boldsymbol{x}^{\ast})
=
\widehat{y}^{F}(\boldsymbol{x}^{\ast})
=
\frac{1}{N}
\sum_{s=1}^{N}\mu_{*}^{(s)}.
\]
To construct a \(100(1-\gamma)\%\) interval, define
\[
F_{P,*}(z)
=
\frac{1}{N}
\sum_{s=1}^{N}
\Phi\left(
\frac{z-\mu_{*}^{(s)}}{\sqrt{V_{P,*}^{(s)}}}
\right),
\]
where \(\Phi(\cdot)\) is the cumulative distribution function of the
standard normal distribution. Let
\(q_{\gamma}^{P}(\boldsymbol{x}^{\ast})
=
\inf\{z\in\mathbb{R}:F_{P,*}(z)\ge\gamma\}\). Then, a
\(100(1-\gamma)\%\) credible interval for
\(y^{P}(\boldsymbol{x}^{\ast})\) is
\[
\left[
q_{\gamma/2}^{P}(\boldsymbol{x}^{\ast}),\
q_{1-\gamma/2}^{P}(\boldsymbol{x}^{\ast})
\right].
\]The interval for \(y^{F}(\boldsymbol{x}^{\ast})\) is obtained in the same manner.

\section{Numerical Examples}
\label{sec:numerical}
\noindent
In this section, we evaluate BMDD using several numerical simulations.
The main comparison includes KOH, \(L_2\), DD, BDD, and BMDD, where BDD denotes the Bayesian discrepancy decomposition model and differs from BMDD only by excluding the residual term \(r(\boldsymbol{x},\boldsymbol{\theta};\boldsymbol{\theta}^p)\).

\subsection{A One-Dimensional Example}
\label{subsec:simulation_1d}
\noindent
In this subsection, we consider the case of a one-dimensional calibration parameter. Let the physical system and the corresponding computer model be
\[
\left\{
\begin{aligned}
&y^F(x;\theta^p) = \theta^p x+\sin(2\pi x)+\epsilon,\\
&y^S(x,\theta)= \theta x+\cos(\pi x)\left[\left(\theta-\frac{5}{2}\right)^2-\frac{1}{3}\right],
\end{aligned}
\right.
\qquad x\in[0,1],\ \theta\in[1.5,3.5],
\]
where the true value of \(\theta^p\) is set to 2.5. The resulting model discrepancy is
\[
\delta(x,\theta;\theta^p)
=
\sin(2\pi x)-x(\theta-\theta^p)
-\cos(\pi x)\Bigl[(\theta-\frac{5}{2})^2-\frac{1}{3}\Bigr],
\]
which contains a quadratic term in \(\theta\). Therefore, it no longer satisfies the discrepancy decomposition form \eqref{eq:dd} assumed in DD.

\begin{table}[t]
\centering
\caption{MSEs (\(\times 10^{-4}\)) of \(\theta^p\) for different calibration methods; numbers in parentheses are the standard deviations (\(\times 10^{-4}\)) of the squared errors }
\label{tab:n_compare_all_models_mse_sd}
\renewcommand{\arraystretch}{1.45}

\begin{tabular*}{\linewidth}{@{\hspace{0.8em}\extracolsep{\fill}}lccc@{\hspace{0.8em}}}
\toprule
& $n=11$ & $n=31$ & $n=61$ \\
\midrule
KOH   & 2431.3 (2145.5) & 2526.9 (2216.5) & 2881.4 (2384.6) \\
$L_2$ & 6238.5 (308.45) & 6381.4 (150.02) & 6412.4 (136.80) \\
DD    & 38.343 (37.562) & 16.056 (24.902) & 7.6966 (10.079) \\
BDD   & 37.738 (38.385) & 16.120 (26.021) & 7.8933 (10.617) \\
BMDD  & 1.0150 (1.2216) & 1.2800 (1.5380) & 1.2410 (1.7712) \\
\bottomrule
\end{tabular*}
\end{table}

Let \(x_i=(i-1)/(n-1)\), \(i=1,\ldots,n\), and
\(\theta_j=1.5+2(j-1)/(m-1)\), \(j=1,\ldots,m\). Based on these design
points, we construct the difference matrix design
\({\{(x_i,\theta_j,\delta_{ij})\}}_{i=1,\ldots,n,\;j=1,\ldots,m}\)
introduced in Subsection~\ref{subsec:hierarchy}. The observation noises are generated independently from \(\mathcal{N}(0,0.01)\), and the nugget terms in $\boldsymbol{\Delta}$ are generated independently from \(\mathcal{N}(0,10^{-8})\). Without any prior information, we use the
default constraint \(\int_0^1\alpha(x)\,dx
=0\). The prior distributions of \(\alpha(x)\), \(\beta(x)\), and \(\omega(x,\theta)\) follow \eqref{eq:alpha_prior_constrained}, \eqref{eq:prior_beta_vec_bmdd}, and \eqref{eq:prior_W_vec_bmdd}, respectively, with zero mean functions and the squared exponential kernel in \eqref{eq:kernel}, where \(\rho_x=0.35\) for kernels on \(\Omega\) and \(\rho_\theta=0.2\) for the kernel on \(\Theta\). The prior of \(\theta^p\) is
\(\mathcal{U}(1.5,3.5)\). For the priors of variance parameters, we follow the common practice in Bayesian calibration (e.g., \citet{HigdonEtAl2004}; \citet{HanSantnerRawlinson2009}). Specifically, the GP process variance
parameters \(\sigma_{\alpha}^{2}\), \(\sigma_{\beta}^{2}\),
and \(\sigma_{\omega}^{2}\) are assigned \(\mathcal{IG}(3,0.1)\), while the variance parameters of the observation noise
and the nugget term, \(\sigma_{\epsilon}^{2}\) and \(\sigma_{\xi}^{2}\), are
assigned \(\mathcal{IG}(3,10^{-4})\), yielding weakly informative priors. We
fix \(m=51\), consider three sample sizes \(n=11,31,61\), and repeat the
simulation 50 times for each setting. Posterior samples of \(\theta^p\) are obtained using
Algorithm~\ref{alg:bmdd_mwg_cn}. For comparison, we also implement the
other four calibration methods: KOH, \(L_2\), DD, and BDD.

Table~\ref{tab:n_compare_all_models_mse_sd} summarizes the estimation results
for \(\theta^p\) across the five calibration methods. It shows that BMDD achieves the
smallest MSE and the lowest variability across all settings. DD and BDD
improve substantially on KOH and \(L_2\), but remain less accurate than
BMDD, indicating the benefit of the residual term for nonlinear
discrepancies.

\begin{figure}[t]
\centering
\includegraphics[width=0.9\textwidth]{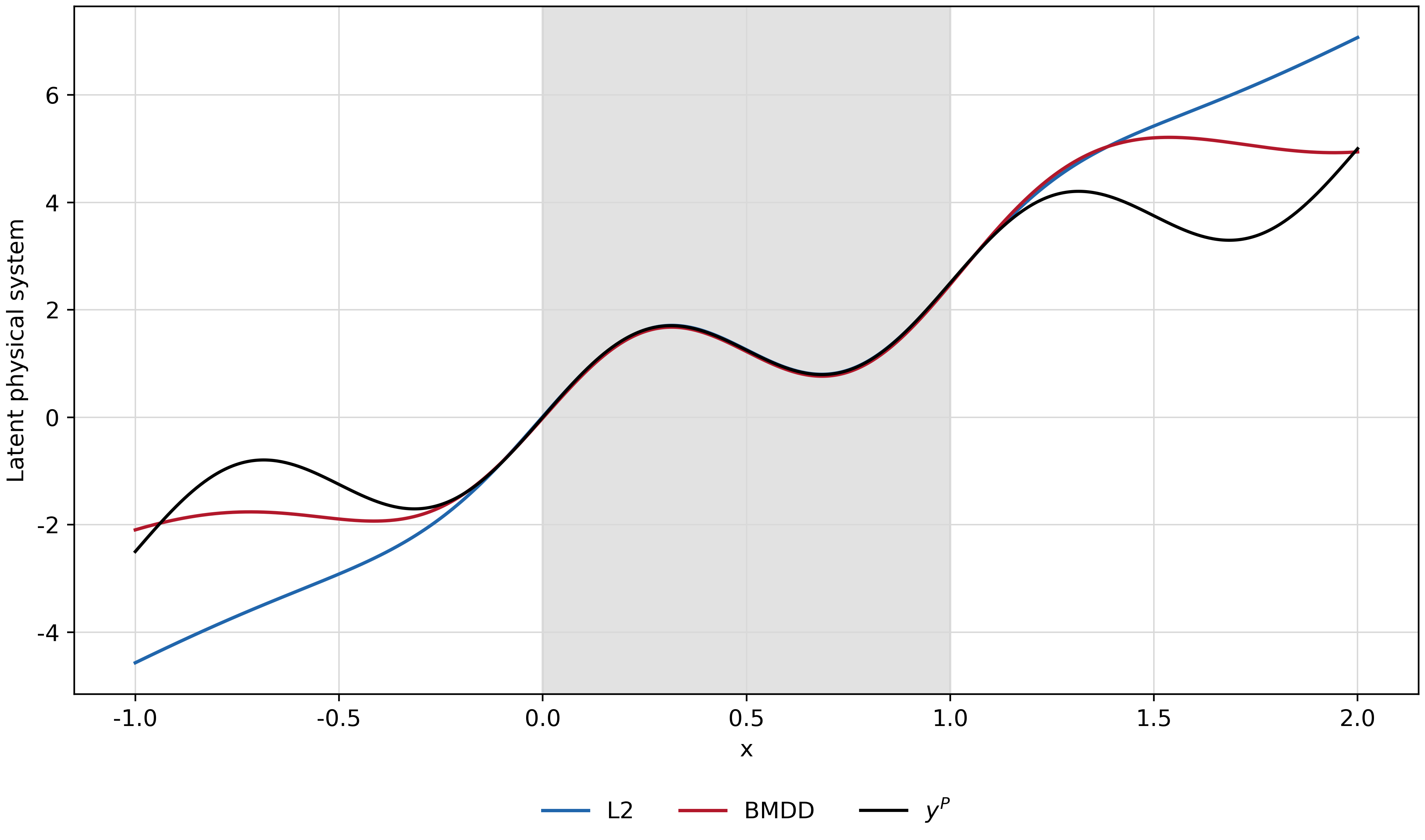}
\caption{Extrapolative prediction in Section \ref{subsec:simulation_1d} ($n=61$).}
\label{fig:simulation_1d_extrapolation}
\end{figure}

To assess extrapolative prediction, we fit the models using field data on
\([0,1]\) and predict the physical system
\(y^P(x;\theta^p)=\theta^p x+\sin(2\pi x)\) over \([-1,2]\). The regions
\([-1,0)\) and \((1,2]\) are extrapolation regions. We compare BMDD with
\(L_2\) calibration to assess whether recovering a physically meaningful calibration parameter improves prediction beyond the field-data range. BMDD predictions follow Subsection~\ref{subsec:prediction}, and the experiment is repeated 50 times.

Figure~\ref{fig:simulation_1d_extrapolation} shows that both methods fit
the calibration domain well, whereas BMDD remains closer to the true physical system outside this domain. This suggests that more accurate recovery of \(\theta^p\) improves extrapolative prediction.

\subsection{An Example with Increasing Nonlinearity}
\noindent
In this subsection, we further investigate the performance of BMDD when the model discrepancy exhibits a nonlinear structure of varying strength. This example was used in \cite{LiWuXiong2026} and shows that DD does not perform well when the nonlinearity is strong.

Consider the Box--Cox transformation
\[
BC(z;r)=
\begin{cases}
(z^r-1)/r, & r>0,\\[2pt]
\log z, & r=0,
\end{cases}
\]
where \(r\) is the power parameter controlling the degree of nonlinearity. Let the physical system and the computer model be
\[
\left\{
\begin{aligned}
&y^F(x;\theta^p)=\exp\!\left\{BC(\theta^p x;r)-\frac{1}{1+x}+\frac{1}{2}\right\}+\epsilon,\\
&y^S(x,\theta)=\exp\!\left\{BC(\theta x;r)\right\},
\end{aligned}
\right.
\qquad x\in[0,2],\ \theta\in[0.5,1.5],
\]
where the true value of \(\theta^p\) is set to 1. The resulting model discrepancy is
\[
\delta(x,\theta;\theta^p)
=
\exp\!\left\{BC(\theta^p x;r)-\frac{1}{1+x}+\frac{1}{2}\right\}
-
\exp\!\left\{BC(\theta x;r)\right\}.
\]
As \(r\) increases, the model discrepancy exhibits increasingly pronounced nonlinear behavior.
Let \(x_i=2(i-1)/(n-1)\), \(i=1,\ldots,n\), and
\(\theta_j=0.5+(j-1)/(m-1)\), \(j=1,\ldots,m\), to construct the
difference-matrix design
\({\{(x_i,\theta_j,\delta_{ij})\}}_{i=1,\ldots,n,\;j=1,\ldots,m}\)
introduced in Subsection~\ref{subsec:hierarchy}. We fix \(n=21\) and
\(m=51\). The observation noises are generated independently from
\(\mathcal{N}(0,0.01)\), and the nugget terms are generated independently from \(\mathcal{N}(0,10^{-8})\). Without any prior information, we use
the default constraint \(\int_0^2 \alpha(x)\,dx=0\). The prior of
\(\theta^p\) is set to \(\mathcal{U}(0.5,1.5)\), while other prior settings are the same as those in Subsection~\ref{subsec:simulation_1d}. We consider six levels of nonlinearity,
\(r\in\{0.5,1.0,1.5,2.0,2.5,3.0\}\), and repeat the simulation 50 times for
each value of \(r\).

Table~\ref{tab:mse_theta_diff_r_bmdd} reports the MSEs of the five methods for estimating \(\theta^p\) under different levels of nonlinearity \(r\). KOH and \(L_2\) have relatively large MSEs when \(r\) is small, while DD and BDD perform well in mildly nonlinear cases but deteriorate rapidly as \(r\) increases. In contrast, BMDD consistently achieves the smallest MSE across all settings, and its estimation error remains at a much lower level even under stronger nonlinearity. This indicates that the additional residual term enables BMDD to better capture nonlinear model discrepancy, thereby improving calibration accuracy and robustness.

\begin{table}[htbp]
\centering
\caption{Comparison of MSEs for estimating \(\theta^p\) under different values of \(r\)}
\label{tab:mse_theta_diff_r_bmdd}
\renewcommand{\arraystretch}{1.35}

\begin{tabular*}{\textwidth}{@{\extracolsep{\fill}}cccccc@{}}
\toprule
\multirow{2}{*}{$r$} & \multicolumn{5}{c}{Method} \\
\cmidrule(lr){2-6}
& KOH & $L_2$ & DD & BDD & BMDD \\
\midrule
0.5 & $1.1895 \times 10^{-2}$ & $8.0322 \times 10^{-3}$ & $4.6696 \times 10^{-4}$ & $5.7846 \times 10^{-4}$ & $7.7372 \times 10^{-6}$ \\
1.0 & $6.4707 \times 10^{-3}$ & $4.9749 \times 10^{-3}$ & $1.2540 \times 10^{-3}$ & $9.7102 \times 10^{-4}$ & $7.8777 \times 10^{-6}$ \\
1.5 & $2.9770 \times 10^{-3}$ & $3.2173 \times 10^{-3}$ & $6.2331 \times 10^{-3}$ & $6.1075 \times 10^{-3}$ & $6.4967 \times 10^{-6}$ \\
2.0 & $1.6681 \times 10^{-3}$ & $1.7544 \times 10^{-3}$ & $1.6993 \times 10^{-2}$ & $1.7764 \times 10^{-2}$ & $1.1587 \times 10^{-5}$ \\
2.5 & $7.6806 \times 10^{-4}$ & $9.5613 \times 10^{-4}$ & $2.9561 \times 10^{-2}$ & $3.3763 \times 10^{-2}$ & $1.0287 \times 10^{-4}$ \\
3.0 & $3.9095 \times 10^{-4}$ & $4.7437 \times 10^{-4}$ & $3.5786 \times 10^{-2}$ & $4.5379 \times 10^{-2}$ & $7.8108 \times 10^{-5}$ \\
\bottomrule
\end{tabular*}
\end{table}

\begin{figure}[htbp]
\centering
\includegraphics[width=1\textwidth]{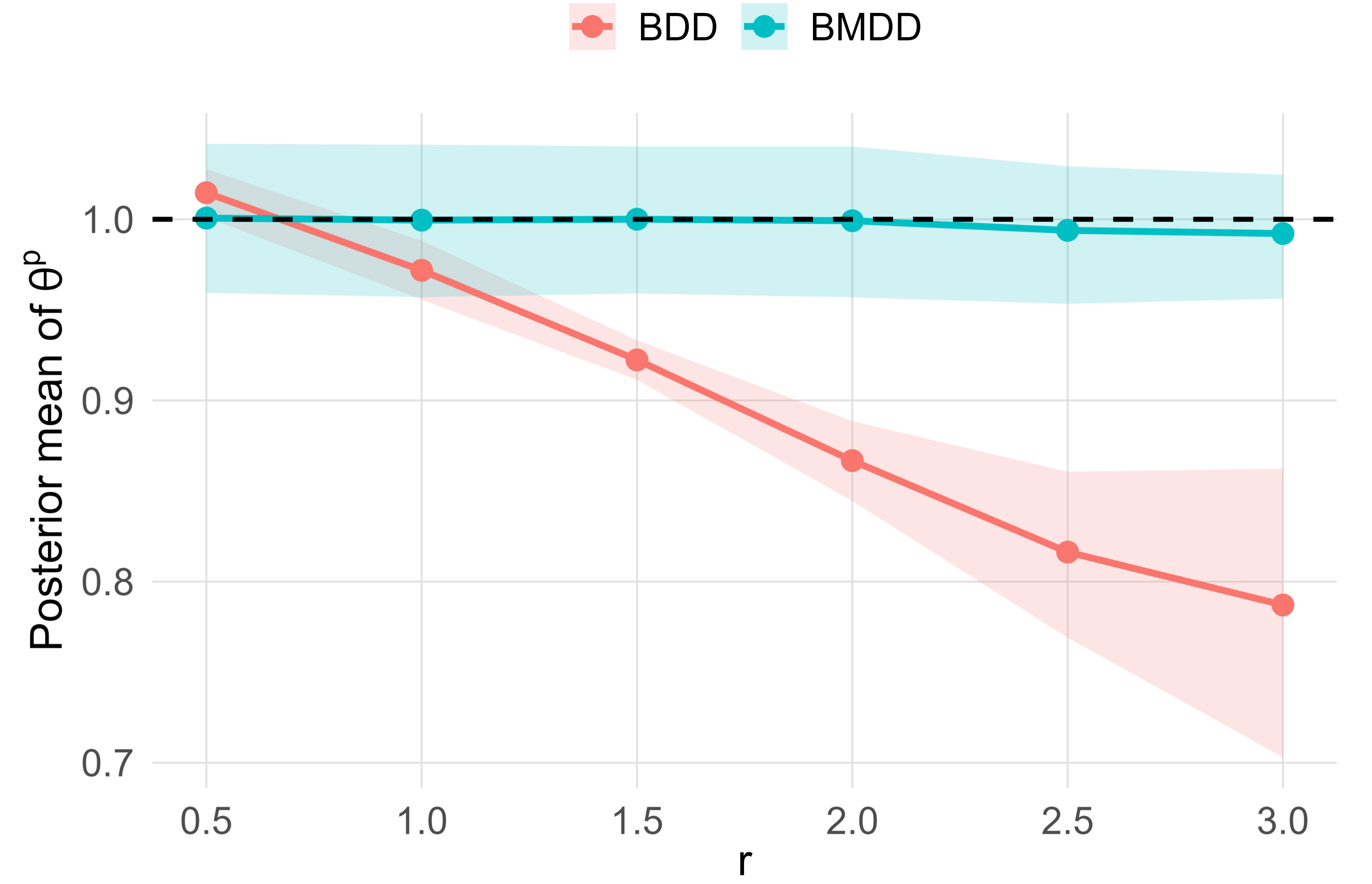}
\caption{Uncertainty quantification performance of BDD and BMDD for \(\theta^p\): the solid lines show the averages of the posterior means over 50 replications, the shaded regions represent 90\% posterior credible intervals, and the black dashed line corresponds to the true value \(\theta^p=1\).}
\label{fig:bdd_bmdd_compare}
\end{figure}

Figure~\ref{fig:bdd_bmdd_compare} displays the 90\% posterior credible intervals of \(\theta^p\) for BDD and BMDD under different values of \(r\). We can see that the credible intervals of BDD gradually fail to cover the true value as the nonlinear level increases, while BMDD maintains coverage across all settings. Meanwhile, the BDD intervals become increasingly wide for large \(r\), indicating unstable uncertainty quantification, whereas BMDD keeps relatively stable interval widths. These results suggest that BMDD provides more reliable uncertainty quantification under stronger nonlinear discrepancy structures.

\subsection{A Two-Dimensional Example}
\label{subsec:simulation_2dtheta}
\noindent
In this subsection, we consider a simulation example with two-dimensional calibration parameters. Let
\(\boldsymbol{\theta}=(\theta_1,\theta_2)^{\mathsf{T}}\in[0.5,1.5]^2\). The physical
system and the corresponding computer model are given by
\[
\left\{
\begin{aligned}
&y^F(x;\boldsymbol{\theta}^p)
=
\theta_1^p x+\theta_2^p x^2
+0.15\sin(4\pi x)+\epsilon,\\
&y^S(x,\boldsymbol{\theta})
=
\theta_1x+\theta_2x^2
+
h(x)
\left\{(\theta_1-1)^2+(\theta_2-1)^2\right\}a(\boldsymbol{\theta})
,
\end{aligned}
\right.
\qquad
x\in[0,1],
\]
where \(
h(x)=0.35\sin(4\pi x)+0.2\sin(8\pi x),
\) and \(
a(\boldsymbol{\theta})
=
0.4+0.2\theta_1+0.15\theta_2+0.1\theta_1\theta_2 .
\)
The true value of \(\boldsymbol{\theta}^p\) is set to \(
(1,1)^{\mathsf{T}}.
\)
The resulting model discrepancy is
\[
\delta(x,\boldsymbol{\theta};\boldsymbol{\theta}^p)=
0.15\sin(4\pi x)
-x(\theta_1-\theta_1^p)
-x^2(\theta_2-\theta_2^p)
-h(x)
\left\{(\theta_1-1)^2+(\theta_2-1)^2\right\}
a(\boldsymbol{\theta}).
\]
This discrepancy contains a second-order term in
\(\boldsymbol{\theta}\).

Let \(x_i=(i-1)/(n-1)\), \(i=1,\ldots,n\),
\(
\theta_{1,k}=0.5+(k-1)/{6}\),
\(\theta_{2,\ell}=0.5+{(\ell-1)}/{6},\
k,\ell=1,\ldots,7.
\)
Thus, the total number of \(\boldsymbol{\theta}\) is
\(m=49\), and we consider three sample sizes
\(n=11,31,61\). Based on these design points, we construct the difference matrix design
\({\{(x_i,\boldsymbol{\theta}_j,\delta_{ij})\}}_{i=1,\ldots,n,\;j=1,\ldots,m}\)
introduced in Subsection~\ref{subsec:hierarchy}. The observation noises are
generated independently from \(\mathcal{N}(0,0.01)\), and the nugget terms in
\(\boldsymbol{\Delta}\) are generated independently from
\(\mathcal{N}(0,10^{-8})\). Since the calibration parameter is two-dimensional, we impose two constraints on
\(\alpha(x)\). Specifically, we use the default constraints
\(\int_0^{0.5}\alpha(x)\,dx=\int_{0.5}^1\alpha(x)\,dx=0\). The prior of \(\boldsymbol{\theta}^p\) is taken as \(\mathcal{U}([0.5,1.5]^2)\). The remaining prior and MCMC settings are the same as those in
Subsection~\ref{subsec:simulation_1d}, with independent GP priors assigned to
\(\beta_1(x)\) and \(\beta_2(x)\). We repeat the simulation 50 times for each setting. Posterior
samples of \(\boldsymbol{\theta}^p\) are obtained using
Algorithm~\ref{alg:bmdd_mwg_cn}. For comparison, we also
implement the other four calibration methods: KOH, \(L_2\), DD, and BDD. Table~\ref{tab:n_compare_all_models_mse_sd_2dtheta} summarizes the estimation
results for \(\boldsymbol{\theta}^p\) across the five calibration methods. The
main entries are the joint mean squared errors over 50 replications, and the values in parentheses represent the standard deviations of the corresponding
squared errors. All results are reported on a \(\times 10^{-4}\) scale for
comparability.

\begin{table}[htbp]
\centering
\caption{Joint MSEs (\(\times 10^{-4}\)) of \(\boldsymbol{\theta}^p\) for different calibration methods; numbers in parentheses are the standard deviations (\(\times 10^{-4}\)) of the squared errors}
\label{tab:n_compare_all_models_mse_sd_2dtheta}
\renewcommand{\arraystretch}{1.45}

\begin{tabular*}{\linewidth}{@{\hspace{0.8em}\extracolsep{\fill}}lccc@{\hspace{0.8em}}}
\toprule
& $n=11$ & $n=31$ & $n=61$ \\
\midrule
KOH   & 1509.0 (1370.4) & 2280.9 (1417.67) & 3006.2 (814.36) \\
$L_2$ & 3274.9 (1117.75) & 3084.6 (774.85) & 3187.7 (620.13) \\
DD    & 1742.0 (1462.5) & 649.37 (860.19) & 440.44 (588.89) \\
BDD   & 619.36 (673.63) & 322.35 (406.55) & 222.46 (256.36) \\
BMDD  & 60.419 (41.958) & 5.9448 (9.9715) & 2.2305 (3.8662) \\
\bottomrule
\end{tabular*}
\end{table}

Table~\ref{tab:n_compare_all_models_mse_sd_2dtheta} shows that KOH and \(L_2\)
still perform poorly in this two-dimensional setting. This may be due to the
identifiability issue in KOH and the fact that the \(L_2\) calibration target does
not necessarily coincide with the true physical parameter. DD and BDD improve as the
sample size increases, but their estimation errors remain relatively large because
the first-order discrepancy decomposition cannot adequately capture the nonlinear term in the discrepancy. BMDD achieves the smallest joint MSE and the smallest standard deviation of squared
errors in all three sample-size settings, indicating that the residual process is useful
for capturing higher-order discrepancy structures and improving the estimation
accuracy of the physical calibration parameters.

\subsection{Sensitivity to the Integral Constraint}
\noindent
In this subsection, we further examine the sensitivity of  BMDD to the integral
constraint imposed on \(\alpha(\boldsymbol{x})\) in \eqref{eq:dd_alpha_constraint}.

First, we use an example with a one-dimensional input and a one-dimensional calibration parameter to study the sensitivity to the specified constant in the integral constraint and to the size of the constraint subdomain. The physical system and the computer model are given by
\[
\left\{
\begin{aligned}
&y^F(x;\theta^p)=
0.2\sin(2\pi x)+0.25x(1-x)+\theta^p x
-1.2\cos(\pi x){\theta^p}^2+\epsilon,\\
&y^S(x,\theta)
=
0.1-0.15\cos(4\pi x)+\theta x-1.2\cos(\pi x)(4\theta-{\theta}^2),
\end{aligned}
\right.
\quad x\in[0,2],\ \theta\in[0.8,3.2],
\]
where the true value of \(\theta^p\) is set to 2. We have
\[
\delta(x,\theta;\theta^p)
=-0.1+0.2\sin(2\pi x)+0.15\cos(4\pi x)+0.25x(1-x)
-x(\theta-\theta^p)
-1.2\cos(\pi x)(\theta-\theta^p)^2
\]
with intrinsic discrepancy \(
\alpha(x)=-0.1+0.2\sin(2\pi x)+0.15\cos(4\pi x)+0.25x(1-x)
\), discrepancy coefficient function \(\beta(x)=-x\), and \(\omega(x,\theta)=-1.2\cos(\pi x)\).

Let \(x_i=2(i-1)/(n-1),i=1,\ldots,n,
\theta_j=0.8+2.4(j-1)/(m-1),j=1,\ldots,m,
\) to construct the difference matrix design
\({\{(x_i,\theta_j,\delta_{ij})\}}_{i=1,\ldots,n,\;j=1,\ldots,m}\)
introduced in Subsection~\ref{subsec:hierarchy}. We fix \(n=m=31\). The observation noises are generated independently from \(\mathcal{N}(0,0.01)\), and the nugget terms are generated independently from \(\mathcal{N}(0,10^{-8})\). The prior of \(\theta^p\) is set to \(\mathcal{U}(0.8,3.2)\), while other settings are the same as those in
Subsection~\ref{subsec:simulation_1d}. Posterior samples of \(\theta^p\) are
obtained using Algorithm~\ref{alg:bmdd_mwg_cn}.

\begin{figure}[htbp]
    \centering
    \begin{minipage}{0.48\textwidth}
        \centering
        \includegraphics[width=\textwidth]{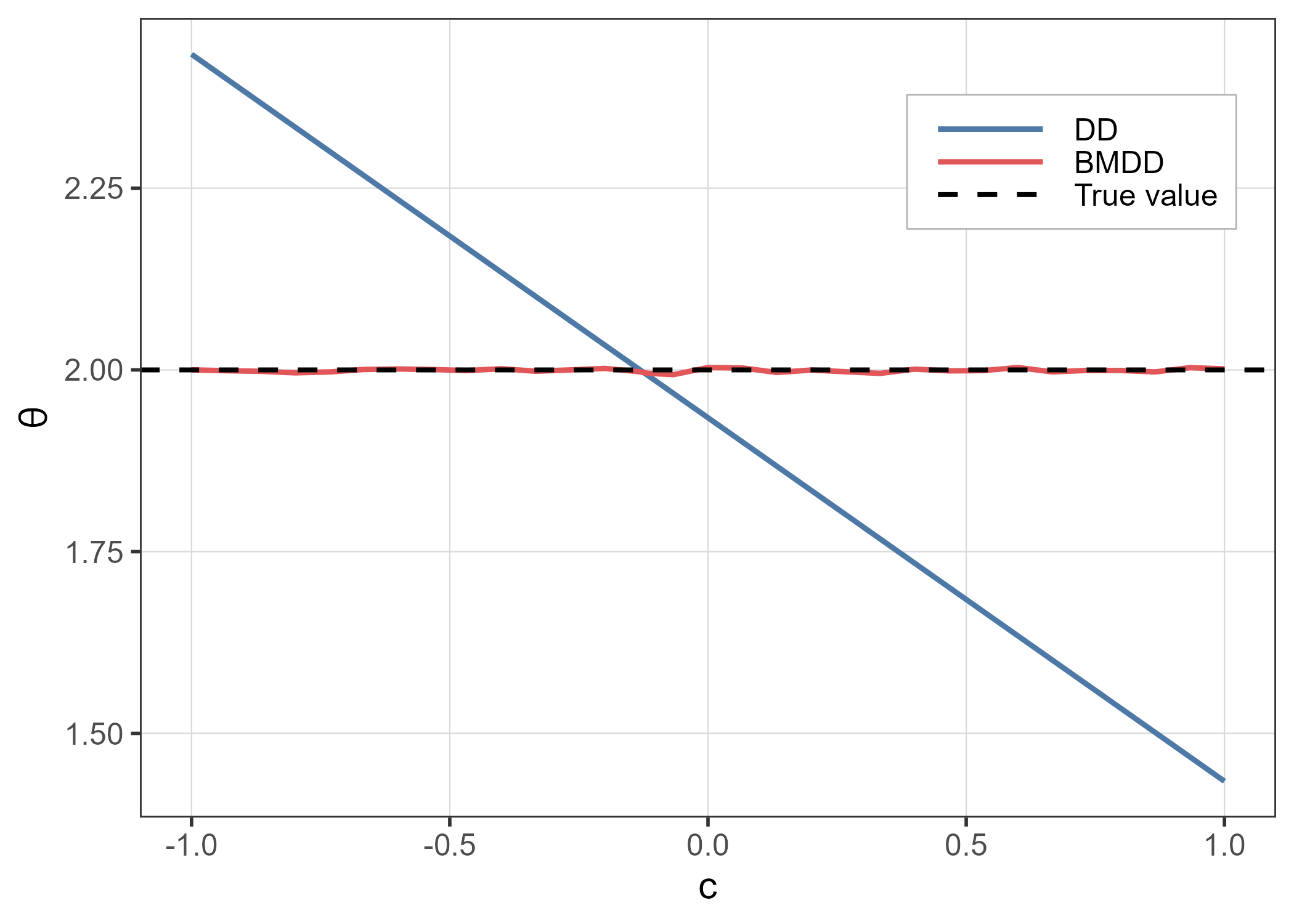}
    \end{minipage}
    \hfill
    \begin{minipage}{0.48\textwidth}
        \centering
        \includegraphics[width=\textwidth]{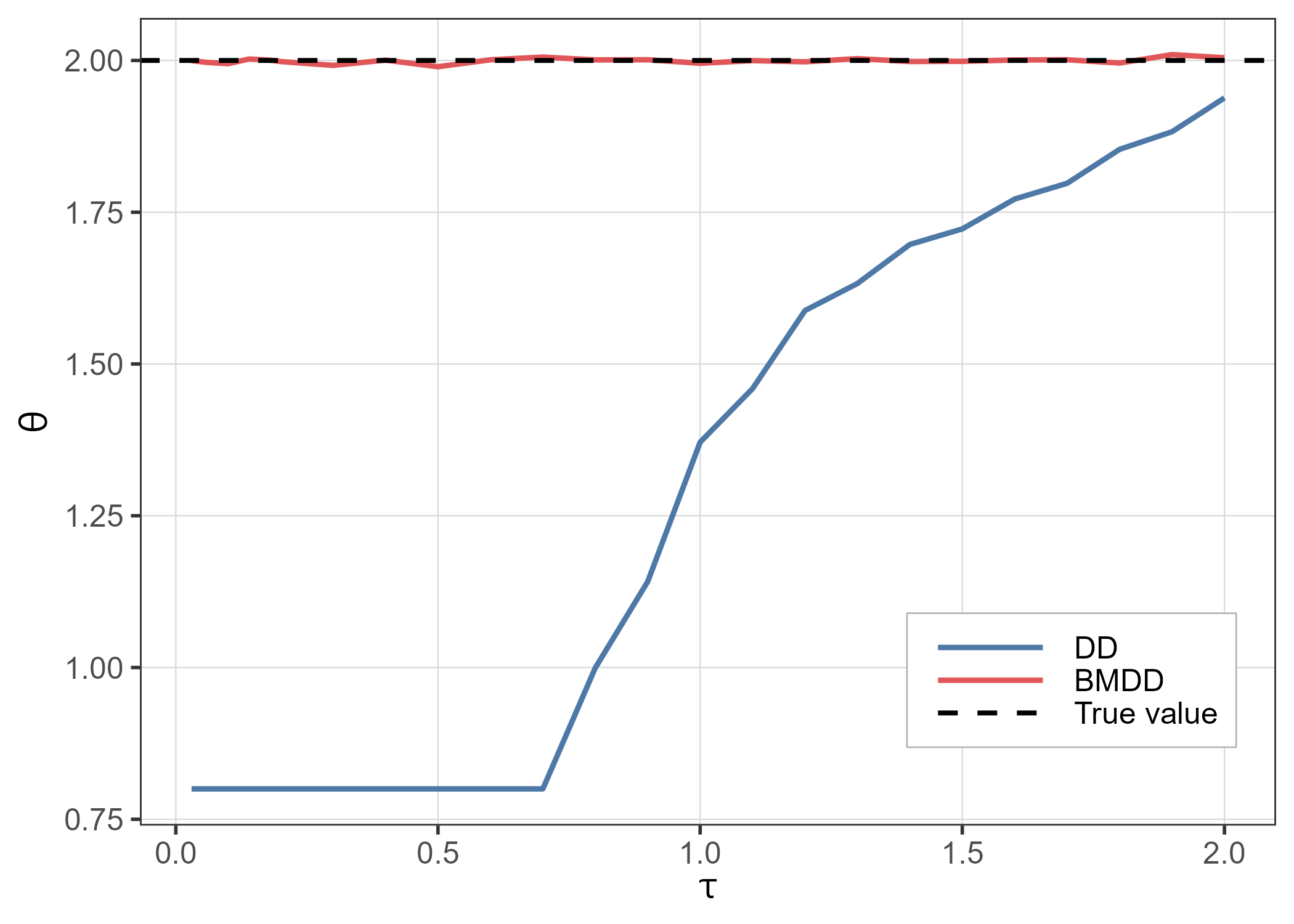}
    \end{minipage}
    \caption{Sensitivity of DD and BMDD to the integral constraint. Left: estimates of
    \(\theta^p\) under \(\int_0^2\alpha(x)\,dx=c\) with \(c\in[-1,1]\).
    Right: estimates of \(\theta^p\) under \(\int_0^\tau\alpha(x)\,dx=0\) with
    \(\tau\in(0,2.0]\).}
    \label{fig:constraint_sensitivity}
\end{figure}

We first vary the constant \(c\) in the full-domain constraint \(
\int_0^2 \alpha(x)\,dx=c
\). It is easy to compute that the oracle constraint is
\(
\int_0^2 \alpha(x)\,dx
=
-11/30.
\)
We also consider the subdomain constraint
\(\int_0^\tau \alpha(x)\,dx=0,
\tau\in(0,2].
\)
Figure \ref{fig:constraint_sensitivity} shows the estimates of \(\theta^p\) in DD and BMDD with varying \(c\) or \(\tau\) over 50 replications. It can be seen that DD is sensitive to the choice of
the integral constraint. Its estimate changes substantially as \(c\) or \(\tau\)
varies. In contrast, BMDD remains close to the true value \(\theta^p=2\), which indicates that the residual process in BMDD makes the estimation of \(\theta^p\) less
sensitive to misspecification of the integral constraint.

We also report repeated simulation results under three representative constraints:
\(
\int_0^2\alpha(x)\,dx=0,
\int_0^1\alpha(x)\,dx=0,
\int_0^2\alpha(x)\,dx=-11/30.
\)
For each constraint, the simulation is repeated 50 times. Table~\ref{tab:constraint_sensitivity}
shows the MSE of \(\theta^p\) and the standard deviation of
squared errors. We can see that BMDD achieves small MSEs under all three constraints and remains more accurate than DD even under the oracle constraint, indicating that BMDD is less sensitive to the choice of the integral constraint.

\begin{table}[htbp]
\centering
\caption{Sensitivity analysis under different integral constraints}
\label{tab:constraint_sensitivity}
\renewcommand{\arraystretch}{1.2}
\setlength{\tabcolsep}{4.2pt}
\begin{tabular}{lcccccc}
\toprule
\multirow{2}{*}{}
& \multicolumn{2}{c}{\(\int_0^2\alpha(x)\,dx=0\)}
& \multicolumn{2}{c}{\(\int_0^1\alpha(x)\,dx=0\)}
& \multicolumn{2}{c}{\(\int_0^2\alpha(x)\,dx=-11/30\)} \\
\cmidrule(lr){2-3}\cmidrule(lr){4-5}\cmidrule(lr){6-7}
& MSE & SD
& MSE & SD
& MSE & SD \\
\midrule
DD
& \(3.825\times10^{-3}\) & \(2.375\times10^{-3}\)
& \(4.050\times10^{-1}\) & \(6.746\times10^{-2}\)
& \(1.690\times10^{-2}\) & \(3.095\times10^{-3}\) \\
BMDD
& \(1.634\times10^{-4}\) & \(2.431\times10^{-4}\)
& \(6.721\times10^{-5}\) & \(1.158\times10^{-4}\)
& \(8.289\times10^{-5}\) & \(5.728\times10^{-5}\) \\
\bottomrule
\end{tabular}
\end{table}

Next, an example with a two-dimensional input and a two-dimensional calibration parameter is used to assess the sensitivity to the partitioning scheme of the constraint regions in multidimensional settings. The physical system and the computer
model are given by
\[
\left\{
\begin{aligned}
&y^F(\boldsymbol{x};\boldsymbol{\theta}^p)
=
\theta_1^p x_1+\theta_2^p x_2
+
0.15\sin(4\pi x_1)\sin(2\pi x_2)
+\epsilon,\\
&y^S(\boldsymbol{x},\boldsymbol{\theta})
=
\theta_1x_1+\theta_2x_2
+
k(\boldsymbol{x})
\left\{
(\theta_1-1)^2+(\theta_2-1)^2
\right\}
a(\boldsymbol{\theta}),
\end{aligned}
\right.
\
\boldsymbol{x}\in[0,1]^2,\
\boldsymbol{\theta}\in[0.5,1.5]^2,
\]
where the true value of \(\boldsymbol{\theta}^p\) is set to
\((1,1)^{\mathsf{T}},\)
and \(
k(\boldsymbol{x})
=
0.35\sin(4\pi x_1)\sin(2\pi x_2)
+
0.2\sin(2\pi x_1)\sin(4\pi x_2)
\), \(
a(\boldsymbol{\theta})
=
0.4+0.2\theta_1+0.15\theta_2+0.1\theta_1\theta_2
\).
We have
\[
\delta(\boldsymbol{x},\boldsymbol{\theta};\boldsymbol{\theta}^p)
=
0.15\sin(4\pi x_1)\sin(2\pi x_2)
-
x_1(\theta_1-\theta_1^p)
-
x_2(\theta_2-\theta_2^p)
-
k(\boldsymbol{x})
\left\{
(\theta_1-1)^2+(\theta_2-1)^2
\right\}
a(\boldsymbol{\theta}).
\]

Let \(x_{1,k}=x_{2,k}=\frac{k-1}{6},k=1,\ldots,7,\) and \(\theta_{1,\ell}=\theta_{2,\ell}=0.5+\frac{\ell-1}{6}, \ell=1,\ldots,7,\)
to construct the difference matrix design
\({\{(\boldsymbol{x}_i,\boldsymbol{\theta}_j,\delta_{ij})\}}_{i=1,\ldots,n;\;j=1,\ldots,m}\)
introduced in Subsection~\ref{subsec:hierarchy}. We fix \(n=m=49\). The observation
noises are generated independently from \(\mathcal{N}(0,0.01)\), and the nugget terms are
generated independently from \(\mathcal{N}(0,10^{-8})\). The prior of
\(\boldsymbol{\theta}^p\) is set to \(\mathcal{U}([0.5,1.5]^2)\). The remaining prior and MCMC settings are the same as those in
Subsection~\ref{subsec:simulation_2dtheta}. Posterior samples of \(\boldsymbol{\theta}^p\) are obtained using
Algorithm~\ref{alg:bmdd_mwg_cn}.

Since the calibration parameter is two-dimensional, two subregions \(\Omega_1,\Omega_2 \subset \Omega\) are required in the
integral constraint on \(\alpha(\boldsymbol{x})\). In a multidimensional input space \(\Omega\), these
subregions may be constructed by splitting \(\Omega\) along different input dimensions. We
therefore compare two natural partitioning schemes. The first one divides
\(\Omega=[0,1]^2\) according to \(x_1\): \(\Omega_1=[0,0.5]\times[0,1],\Omega_2=[0.5,1]\times[0,1].\)
The second one divides \(\Omega\) according to \(x_2\):
\(\Omega_1=[0,1]\times[0,0.5],\Omega_2=[0,1]\times[0.5,1].\)
For both partitioning schemes, we impose
\(\int_{\Omega_1}\alpha(\boldsymbol{x})\,d\boldsymbol{x}=0,
\int_{\Omega_2}\alpha(\boldsymbol{x})\,d\boldsymbol{x}=0
\). The simulation is repeated 50 times. For each replication, we compute the joint squared error. Table~\ref{tab:constraint_split_2d} reports the MSE and the standard deviation of the
squared errors. The results show that DD has relatively large estimation errors under both
partitioning schemes. In contrast, BMDD achieves much
smaller MSEs for both partitioning schemes. Moreover, the BMDD results are
very similar under the two partitions, suggesting that BMDD is not sensitive to whether the
constraint regions are divided along \(x_1\) or \(x_2\).

\begin{table}[htbp]
\centering
\caption{Sensitivity analysis under different partitions of the constraint regions}
\label{tab:constraint_split_2d}
\renewcommand{\arraystretch}{1.2}
\begin{tabular*}{\textwidth}{@{\hspace{0.8em}\extracolsep{\fill}}lcccc@{\hspace{0.8em}}}
\toprule
\multirow{2}{*}{}
& \multicolumn{2}{c}{\(x_1\)-split}
& \multicolumn{2}{c}{\(x_2\)-split} \\
\cmidrule(lr){2-3}\cmidrule(lr){4-5}
& MSE & SD
& MSE & SD \\
\midrule
DD
& \(8.921\times10^{-3}\) & \(1.395\times10^{-2}\)
& \(8.518\times10^{-3}\) & \(9.458\times10^{-3}\) \\
BMDD
& \(2.369\times10^{-6}\) & \(2.617\times10^{-6}\)
& \(3.030\times10^{-6}\) & \(3.538\times10^{-6}\) \\
\bottomrule
\end{tabular*}
\end{table}

\section{Application to Thermal Stability Analysis of EUVL}
\label{sec:real_data_application}

\subsection{Background Introduction}
\noindent
Extreme ultraviolet lithography (EUVL) is the state-of-the-art technology for advanced semiconductor manufacturing. We consider a real data example arising from a study of the thermal stability of multilayer mirrors for EUVL. Multilayer mirrors are core components of extreme ultraviolet optical systems. Their thermal stability and resistance to damage directly affect reflectivity, service life, and operational reliability.

The physical system considered here consists of fabricated Mo/BN/Si/BN multilayer samples that produce a reflectance peak near 13.5 nm under given annealing temperature and time conditions, where 13.5 nm is the standard operating wavelength of EUVL. The physical system involves two physically meaningful parameters: \(\theta_1\), the effective diffusion-blocking strength of the BN barrier layer, and \(\theta_2\), the thermal diffusion sensitivity coefficient. Since both are critical to structural evolution and reflectance retention at elevated temperatures but are difficult to measure directly, we treat them as unknown calibration parameters and denote their true value by \(\boldsymbol{\theta}^p=(\theta_1^p,\theta_2^p)^{\mathsf T}\).

\begin{figure}[htbp]
\centering
\includegraphics[width=0.9\textwidth]{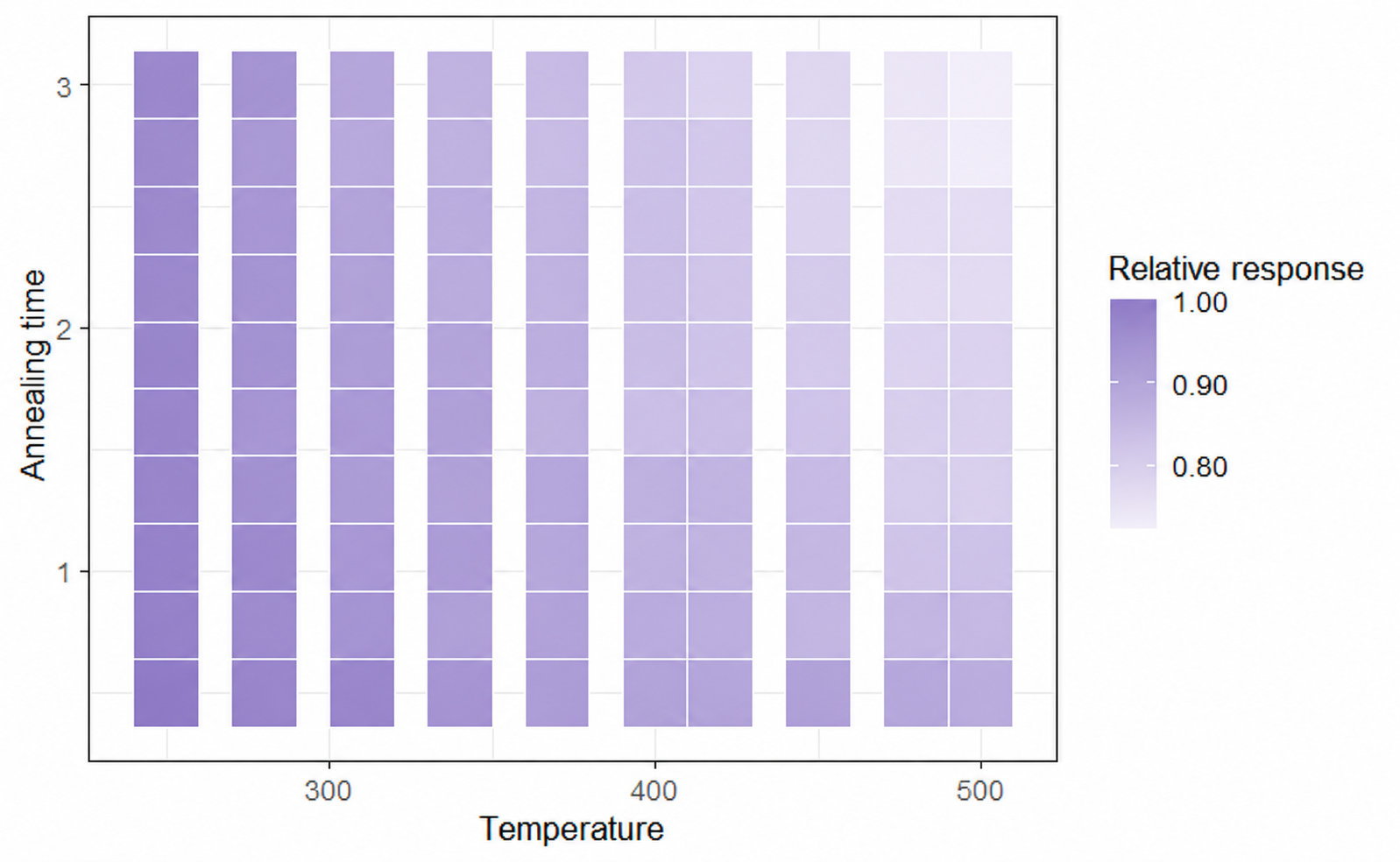}
\caption{Relative variation of the physical system response over the design space.}
\label{fig:real_response_surface}
\end{figure}

Let the controllable input of the physical system be \((T,t)^{\mathsf T}\in\Omega\), where \(T\) denotes annealing temperature and \(t\) denotes annealing time, with input space
\(
\Omega=[250,500]\times[0.5,3.0]
\).
A total of \(n=100\) observations are collected from the physical system over this input region. The original data used in this study were provided by a Chinese lithography company. Due to confidentiality requirements, the raw data have been appropriately modified and processed for publication.
We present the overall pattern of variation in the physical response in the descriptive analysis. Figure~\ref{fig:real_response_surface} illustrates the relative variation of the physical system response across the design space. Reflectance generally decreases as the annealing temperature \(T\) increases, and also shows some decline as the annealing time \(t\) rises. In summary, higher temperatures and longer durations correspond to lower responses, while the low-temperature region displays relatively higher responses.

We introduce an idealized computer model for thermal stability based on interlayer diffusion and thermal degradation mechanisms. The model is defined as
\[
y^S((T,t),\boldsymbol{\theta})
=
0.68\exp\!\left[
-0.28\sqrt{t}\,
\exp\!\left\{
\frac{T-350}{120}
-\theta_1
-\theta_2\frac{T-350}{150}
\right\}
\right],
\]
where \(\boldsymbol{\theta}=(\theta_1,\theta_2)^{\mathsf T}\in[0.4,1.2]\times[0.3,0.9]\). This model is an idealized mechanistic representation. It captures the main trend of diffusion-driven thermal degradation but omits factors such as shifts in the onset temperature, local roughness growth, uneven film deposition, and abrupt structural changes due to stress relief. Thus, model discrepancy with the physical system is inevitable.

We combine physical observations with simulation outputs to infer the calibration parameter \(\boldsymbol{\theta}\), and compare the predictive performance of the Bayesian calibration methods KOH, orthogonal Gaussian process (OGP) \citep{Plumlee2017}, Bayesian projected calibration (BPC) \citep{XieXu2021}, BDD, and BMDD.

\subsection{Calibration Parameter Identification}
\begin{figure}[htbp]
\centering
\includegraphics[width=1\textwidth]{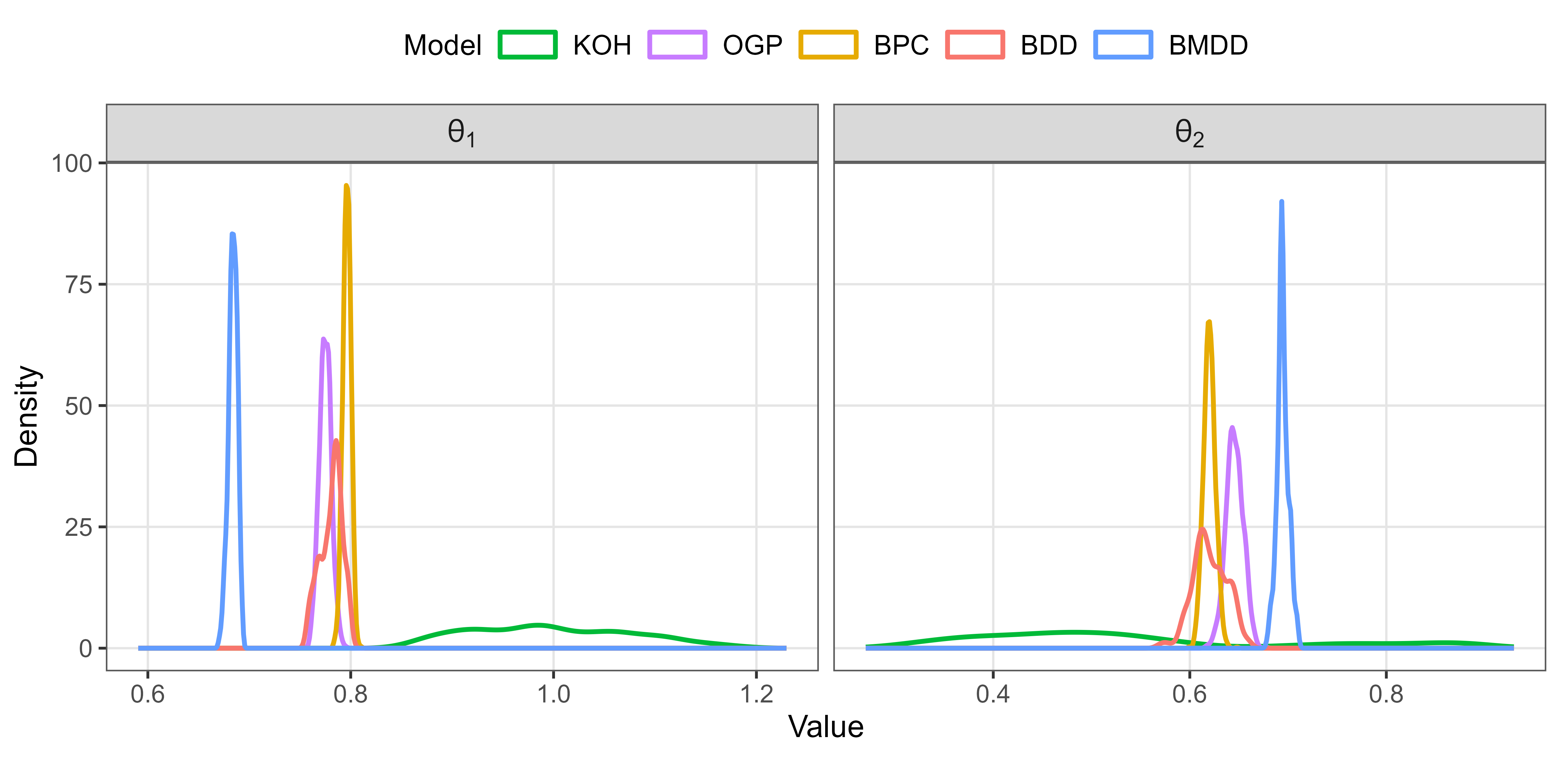}
\caption{Comparison of posterior densities for the calibration parameters.}
\label{fig:real_posterior_density}
\end{figure}

\noindent
We examine how effectively each method identifies \(\boldsymbol{\theta}^p\) in this subsection. Since \(\boldsymbol{\theta}^p\) is not directly observable in this example, the estimation error cannot be evaluated explicitly. We therefore assess the identification performance of different models mainly through the shape of their posterior sampling distributions. Let \(\boldsymbol{x}_i=(T_i,t_i)^{\mathsf T}\), \(i=1,\ldots,n\), denote the input points obtained from the observed EUVL data. Let \(\theta_{1,j_1}=0.4+0.8(j_1-1)/4\), \(j_1=1,\ldots,5\), and \(\theta_{2,j_2}=0.3+0.6(j_2-1)/4\), \(j_2=1,\ldots,5\). Based on these design points, we construct the difference matrix design \({\{(\boldsymbol{x}_i,\boldsymbol{\theta}_j,\delta_{ij})\}}_{i,j}\) introduced in Subsection~\ref{subsec:hierarchy}. The prior of \(\boldsymbol{\theta}^p\) is uniform on \([0.4,1.2]\times[0.3,0.9]\), while other settings are the same as those in
Subsection~\ref{subsec:simulation_1d}. Without any prior information, we apply the default constraints \(\int_{\Omega_1}\alpha(\boldsymbol{x})d\boldsymbol{x}=\int_{\Omega_2}\alpha(\boldsymbol{x})d\boldsymbol{x}=0\), where the two subregions are taken as \(\Omega_1=[250,375]\times[0.5,3.0]\) and \(\Omega_2=(375,400]\times[0.5,3.0]\). For each method, the posterior inference is based on 1000 retained MCMC samples after burn-in. Figure~\ref{fig:real_posterior_density} shows the posterior density estimates of these methods for \(\theta_1\) and \(\theta_2\).

As illustrated in Figure~\ref{fig:real_posterior_density}, KOH yields very flat posterior distributions for both calibration parameters, indicating weak identifiability. BDD exhibits unstable posterior behavior, with irregular and less reliable posterior shapes, suggesting that the discrepancy decomposition in
DD without the additional residual component may not adequately capture the complex model bias. In comparison, OGP, BPC, and BMDD all produce relatively stable and concentrated posterior distributions for \(\theta_1\) and \(\theta_2\). However, since $\boldsymbol{\theta}^p$ is not directly observable,
posterior concentration alone cannot establish that the estimated parameters
correspond to the true physical values. Extrapolative predictive performance
can instead provide indirect evidence regarding whether the calibrated
parameters retain transferable mechanistic information.

\subsection{Extrapolative Prediction Performance}
\noindent
In thermal damage studies of extreme ultraviolet multilayers, the high-temperature annealing regime is particularly important because it involves stronger interlayer diffusion and more complex structural degradation. However, such experiments are costly and may damage samples, making systematic observations difficult. We therefore examine whether low- and medium-temperature data can reliably predict high-temperature responses, thereby assessing both predictive performance and practical applicability.

To compare calibration methods, we adopt a temperature-block prediction design: samples with \(T\le 400\) form the training set, and samples with \(T>400\) the test set. We fit the KOH, OGP, BPC, BDD, and BMDD models using the training set, then predict physical responses in the high-temperature test region. We compare these methods using the following three criteria. Let the test-set size be \(n_{\mathrm{te}}\), let \(y_i^{F}\) denote the observed field response at the \(i\)th test point, and let
\(
\hat y_i = E\!\left(y_i^{F}\mid \mathcal{D}_{\mathrm{tr}}\right)
\)
be the posterior predictive mean, where \(\mathcal{D}_{\mathrm{tr}}\) denotes the training data. Let
\(
[L_i,U_i]
\)
be the corresponding \(90\%\) posterior predictive interval. First, the root mean squared error (RMSE) is defined by
\[
\mathrm{RMSE}
=
\left\{
\frac{1}{n_{\mathrm{te}}}
\sum_{i=1}^{n_{\mathrm{te}}}
(\hat y_i-y_i^{F})^2
\right\}^{1/2},
\]
which measures the overall deviation between the predictive mean and the observed field response. Second, the mean absolute error (MAE) is defined by
\[
\mathrm{MAE}
=
\frac{1}{n_{\mathrm{te}}}
\sum_{i=1}^{n_{\mathrm{te}}}
|\hat y_i-y_i^{F}|,
\]
which measures the average absolute magnitude of the prediction error and is less sensitive to outliers. Smaller values of RMSE and MAE indicate more accurate point prediction. To further assess predictive uncertainty, we consider the coverage of the \(90\%\) predictive interval,
\[
\mathrm{Coverage_{90}}
=
\frac{1}{n_{\mathrm{te}}}
\sum_{i=1}^{n_{\mathrm{te}}}
I\!\left(y_i^{F}\in[L_i,U_i]\right),
\]
where \(I(\cdot)\) is the indicator function, taking the value \(1\) when \(y_i^{F}\in[L_i,U_i]\) and \(0\) otherwise. The predictive results of the calibration methods under the high-temperature block-validation setting are summarized in Table~\ref{tab:block_prediction_results}.

\begin{table}[htbp]
\centering
\caption{Comparison of predictive performance under the high-temperature block-validation setting}
\setlength{\tabcolsep}{12pt}
\renewcommand{\arraystretch}{1.25}
\label{tab:block_prediction_results}
\begin{tabular*}{\textwidth}{@{\hspace{0.8em}\extracolsep{\fill}}lccc@{\hspace{0.8em}}}
\toprule
 & RMSE & MAE & $\mathrm{Coverage}_{90}$ \\
\midrule
KOH  & 0.0303 & 0.0244 & 0.9750 \\
OGP  & 0.0232 & 0.0197 & 0.6000 \\
BPC  & 0.1132 & 0.1075 & 0.0000 \\
BDD  & 0.0276 & 0.0222 & 1.0000 \\
BMDD & 0.0109 & 0.0095 & 1.0000 \\
\bottomrule
\end{tabular*}
\end{table}

Table~\ref{tab:block_prediction_results} shows that BMDD performs best overall under the high-temperature block-validation setting. It achieves the smallest RMSE and MAE among all compared methods, indicating the highest point-prediction accuracy, and it also attains full coverage. BDD also achieves full coverage, but its RMSE and MAE are clearly larger than those of BMDD. KOH and OGP follow the general predictive trend, but their point-prediction errors are higher, and OGP has relatively low coverage. BPC performs worst in this setting, with the largest RMSE and MAE and zero coverage, possibly because it is formulated under the \(L_2\) calibration framework and thus may lead to substantial bias in identifying the physically meaningful calibration parameter.

\begin{figure}[htbp]
\centering
\includegraphics[width=1\textwidth]{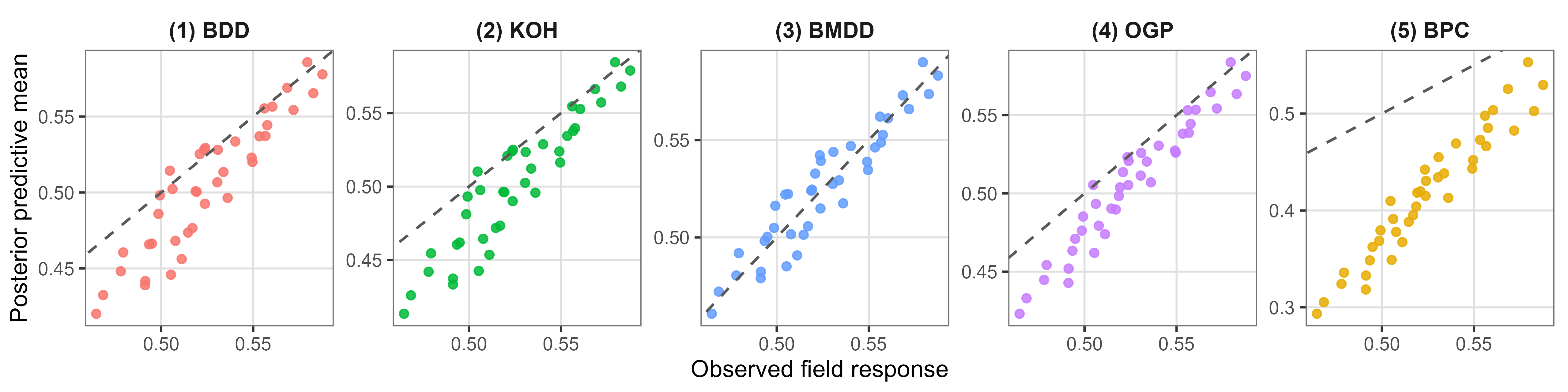}
\caption{Comparison between observed physical responses and posterior predictive means under the high-temperature block-validation setting.}
\label{fig:block_pred_scatter}
\end{figure}

To further clarify the predictive performance, Figure~\ref{fig:block_pred_scatter} plots the observed physical responses against the posterior predictive means, with the dashed line representing the ideal $45^\circ$ reference line. It can be seen that BMDD predictions are closest to the \(45^\circ\) reference line overall, indicating the best agreement with the observed high-temperature responses. BDD, KOH, and OGP generally follow the increasing trend, but their predictions show larger deviations from the reference line than BMDD. BPC predictions are systematically far from the \(45^\circ\) line, which is consistent with its large prediction errors and zero coverage in Table~\ref{tab:block_prediction_results}. These results show that the concentrated posterior distribution of BPC and OGP does not lead to accurate prediction in this extrapolative high-temperature region, whereas BMDD gives more reliable predictive performance.

\section{Discussion}
\label{sec:discussion}
\noindent
This paper proposes a BMDD model for the calibration of imperfect computer experiments. By introducing a residual process into the discrepancy decomposition framework, the model captures higher-order discrepancy components that cannot be represented by the original first-order decomposition. BMDD therefore retains the structural interpretability and parameter identifiability of DD while providing greater flexibility. Within a Bayesian framework, the calibration parameter, discrepancy components, and variance parameters are jointly inferred via a Metropolis-within-Gibbs algorithm, yielding reliable posterior uncertainty quantification and physical-system prediction.

More importantly, this enhanced structural flexibility plays a pivotal role in real-world applications. By recovering a more physically meaningful calibration parameter, our model captures the true physical system more faithfully. Within the training region, a flexible discrepancy can artificially compensate for an inaccurate calibration parameter. Outside that region, however, the discrepancy is no longer directly informed by observations, making the recovery of the physically meaningful parameter absolutely critical. Consequently, our approach yields accurate predictions not only in-sample but also in the extrapolation region, thereby bridging a critical gap left by current calibration methods, which typically struggle with reliable extrapolation.

This advantage is strongly supported by both our simulation studies and a real-world industrial application. While the simulation studies confirm that BMDD accurately recovers physical parameters under complex nonlinear discrepancies, the practical value of this recovery is most evident in the thermal stability analysis of EUVL multilayer mirrors. By utilizing low- and medium-temperature data, BMDD successfully and accurately predicts the physical responses in the high-temperature extrapolation region---an area where experimental data is prohibitively expensive and samples are easily damaged. The high-temperature block-validation results provide compelling empirical evidence that accurate parameter recovery, rather than merely fitting the data, is the key to reliable out-of-domain prediction.

Several directions remain for future research. First, BMDD could be
extended to functional calibration parameters, allowing
$\boldsymbol{\theta}^{p}$ to vary with the controllable input
$\boldsymbol{x}$ and thereby describing operating-condition-dependent
calibration behavior \citep{BrownAtamturktur2018,EzzatPourhabibDing2018}.
Second, the current formulation assumes quantitative inputs and could be
generalized to discrete or mixed input spaces
\citep{KangDeng2020}. Third, extensions to multivariate, functional, and
spatiotemporal responses would broaden the applicability of BMDD to more
complex physical systems \citep{SungTuo2024}. Fourth, the large-sample
properties of the Bayesian procedure, including posterior concentration,
the asymptotic distribution of the calibration parameter, and the
efficiency of the resulting estimator, warrant further theoretical study.
Finally, more efficient posterior-sampling methods are needed for
high-dimensional inputs, large physical datasets, or computationally
intensive discrepancy structures. These developments would improve both
the theoretical foundation and the practical scalability of BMDD.

\section*{Appendix: Proofs}
\phantomsection
\addcontentsline{toc}{section}{Appendix: Proofs}
\label{app:proof}

\renewcommand{\theequation}{A-\arabic{equation}}
\setcounter{equation}{0}

\noindent
This supplementary document provides the proof
 of Theorem~\ref{thm:bmdd_identifiability}  and derivations of equations \eqref{eq:alpha_full_conditional_constrained}--\eqref{eq:sig_delta_full} in Subsection \ref{subsec:mwG} of the main paper.

\begin{proof}[Proof of Theorem~\ref{thm:bmdd_identifiability}]
Suppose that there exist two representations in BMDD, namely
\[(\boldsymbol{\theta}^{p}_{1},\alpha_{1}(\boldsymbol{x}),
\boldsymbol{\beta}_{1}(\boldsymbol{x}),
r_{1}(\boldsymbol{x},\boldsymbol{\theta};\boldsymbol{\theta}^{p}_{1}))\quad \text{and} \quad(\boldsymbol{\theta}^{p}_{2},\alpha_{2}(\boldsymbol{x}),
\boldsymbol{\beta}_{2}(\boldsymbol{x}),
r_{2}(\boldsymbol{x},\boldsymbol{\theta};\boldsymbol{\theta}^{p}_{2})),
\]
where both representations satisfy
Assumptions~\ref{ass:bmdd_dd_constraints}--\ref{ass:bmdd_r_orthogonal}.
Assume that these two representations lead to the same model discrepancy. That
is, for any \((\boldsymbol{x},\boldsymbol{\theta})\in\Omega\times\Theta\),
\begin{equation}
\label{eq:app_bmdd_identical_representation}
\alpha_{1}(\boldsymbol{x})
+
\boldsymbol{\beta}_{1}(\boldsymbol{x})^{\mathsf T}
(\boldsymbol{\theta}-\boldsymbol{\theta}^{p}_{1})
+
r_{1}(\boldsymbol{x},\boldsymbol{\theta};\boldsymbol{\theta}^{p}_{1})
=
\alpha_{2}(\boldsymbol{x})
+
\boldsymbol{\beta}_{2}(\boldsymbol{x})^{\mathsf T}
(\boldsymbol{\theta}-\boldsymbol{\theta}^{p}_{2})
+
r_{2}(\boldsymbol{x},\boldsymbol{\theta};\boldsymbol{\theta}^{p}_{2}).
\end{equation}
We prove that \(\boldsymbol{\theta}^{p}_{1}=\boldsymbol{\theta}^{p}_{2}\).

Taking \(\boldsymbol{\theta}=\boldsymbol{\theta}^{p}_{1}\) in
\eqref{eq:app_bmdd_identical_representation}, we obtain
\[
\alpha_{1}(\boldsymbol{x})
+
\boldsymbol{\beta}_{1}(\boldsymbol{x})^{\mathsf T}
(\boldsymbol{\theta}^{p}_{1}-\boldsymbol{\theta}^{p}_{1})
+
r_{1}(\boldsymbol{x},\boldsymbol{\theta}^{p}_{1};\boldsymbol{\theta}^{p}_{1})
=
\alpha_{2}(\boldsymbol{x})
+
\boldsymbol{\beta}_{2}(\boldsymbol{x})^{\mathsf T}
(\boldsymbol{\theta}^{p}_{1}-\boldsymbol{\theta}^{p}_{2})
+
r_{2}(\boldsymbol{x},\boldsymbol{\theta}^{p}_{1};\boldsymbol{\theta}^{p}_{2}).
\]
Since
\(\boldsymbol{\beta}_{1}(\boldsymbol{x})^{\mathsf T}
(\boldsymbol{\theta}^{p}_{1}-\boldsymbol{\theta}^{p}_{1})=0\)
and, by Assumption~\ref{ass:bmdd_local},
\(r_{1}(\boldsymbol{x},\boldsymbol{\theta}^{p}_{1};
\boldsymbol{\theta}^{p}_{1})=0\), the above equation reduces to
\begin{equation}
\label{eq:app_bmdd_alpha_relation}
\alpha_{1}(\boldsymbol{x})
=
\alpha_{2}(\boldsymbol{x})
+
\boldsymbol{\beta}_{2}(\boldsymbol{x})^{\mathsf T}
(\boldsymbol{\theta}^{p}_{1}-\boldsymbol{\theta}^{p}_{2})
+
r_{2}(\boldsymbol{x},\boldsymbol{\theta}^{p}_{1};\boldsymbol{\theta}^{p}_{2}).
\end{equation}
For each \(t=1,\ldots,q\), integrating both sides of
\eqref{eq:app_bmdd_alpha_relation} over \(\Omega_t\) gives
\begin{equation}
\label{eq:app_bmdd_integral_relation}
\int_{\Omega_t}\alpha_{1}(\boldsymbol{x})\,d\boldsymbol{x}
=
\int_{\Omega_t}\alpha_{2}(\boldsymbol{x})\,d\boldsymbol{x}
+
\int_{\Omega_t}
\boldsymbol{\beta}_{2}(\boldsymbol{x})^{\mathsf T}
(\boldsymbol{\theta}^{p}_{1}-\boldsymbol{\theta}^{p}_{2})
\,d\boldsymbol{x}
+
\int_{\Omega_t}
r_{2}(\boldsymbol{x},\boldsymbol{\theta}^{p}_{1};\boldsymbol{\theta}^{p}_{2})
\,d\boldsymbol{x}.
\end{equation}
By Assumption~\ref{ass:bmdd_dd_constraints}, we have
\(\int_{\Omega_t}\alpha_{1}(\boldsymbol{x})\,d\boldsymbol{x}=\int_{\Omega_t}\alpha_{2}(\boldsymbol{x})\,d\boldsymbol{x}=c_t\)  for
\(t=1,\ldots,q\). Moreover, by
Assumption~\ref{ass:bmdd_r_orthogonal},
\[
\int_{\Omega_t}
r_{2}(\boldsymbol{x},\boldsymbol{\theta}^{p}_{1};\boldsymbol{\theta}^{p}_{2})
\,d\boldsymbol{x}
=
0,
\qquad
t=1,\ldots,q.
\]
Substituting these results into \eqref{eq:app_bmdd_integral_relation}, we have
\begin{equation}
\label{eq:app_bmdd_beta_integral_zero}
\int_{\Omega_t}
\boldsymbol{\beta}_{2}(\boldsymbol{x})^{\mathsf T}
(\boldsymbol{\theta}^{p}_{1}-\boldsymbol{\theta}^{p}_{2})
\,d\boldsymbol{x}
=
0,
\qquad
t=1,\ldots,q.
\end{equation}
Since \(\boldsymbol{\theta}^{p}_{1}-\boldsymbol{\theta}^{p}_{2}\) does not
depend on \(\boldsymbol{x}\), equation
\eqref{eq:app_bmdd_beta_integral_zero} can be written as
\begin{equation}
\label{eq:app_bmdd_matrix_equation}
\mathbf{M}_{2}
(\boldsymbol{\theta}^{p}_{1}-\boldsymbol{\theta}^{p}_{2})
=
\boldsymbol{0},
\end{equation}
where
\(\mathbf{M}_{2}
=
\left(
\int_{\Omega_t}\beta_{2,\ell}(\boldsymbol{x})\,d\boldsymbol{x}
\right)_{t,\ell=1,\ldots,q}\).
By Assumption~\ref{ass:bmdd_dd_constraints}, the DD full-rank condition
\eqref{eq:dd_beta_rank} holds. Therefore, \(\mathbf{M}_{2}\) is nonsingular.
It follows from \eqref{eq:app_bmdd_matrix_equation} that
\(\boldsymbol{\theta}^{p}_{1}-\boldsymbol{\theta}^{p}_{2}=\boldsymbol{0}\).
Thus, \(\boldsymbol{\theta}^{p}_{1}=\boldsymbol{\theta}^{p}_{2}\). Therefore,
the calibration parameter \(\boldsymbol{\theta}^{p}\) in BMDD is identifiable.

It remains to show that the discrepancy components are also uniquely determined.
Since we have proved that
\(\boldsymbol{\theta}^{p}_{1}=\boldsymbol{\theta}^{p}_{2}\), denote the common
value by \(\boldsymbol{\theta}^{p}\). Taking
\(\boldsymbol{\theta}=\boldsymbol{\theta}^{p}\) in
\eqref{eq:app_bmdd_identical_representation}, and using
Assumption~\ref{ass:bmdd_local}, we obtain
\[
\alpha_{1}(\boldsymbol{x})=\alpha_{2}(\boldsymbol{x}).
\]
Thus, the intrinsic discrepancy \(\alpha(\boldsymbol{x})\) is uniquely determined.

Next, differentiating both sides of
\eqref{eq:app_bmdd_identical_representation} with respect to
\(\boldsymbol{\theta}\) and evaluating at
\(\boldsymbol{\theta}=\boldsymbol{\theta}^{p}\), we have
\[
\boldsymbol{\beta}_{1}(\boldsymbol{x})
+
\nabla_{\boldsymbol{\theta}}
r_{1}(\boldsymbol{x},\boldsymbol{\theta};\boldsymbol{\theta}^{p})
\big|_{\boldsymbol{\theta}=\boldsymbol{\theta}^{p}}
=
\boldsymbol{\beta}_{2}(\boldsymbol{x})
+
\nabla_{\boldsymbol{\theta}}
r_{2}(\boldsymbol{x},\boldsymbol{\theta};\boldsymbol{\theta}^{p})
\big|_{\boldsymbol{\theta}=\boldsymbol{\theta}^{p}}.
\]
By Assumption~\ref{ass:bmdd_local}, the two gradient terms are both equal to
\(\boldsymbol{0}\). Hence
\[
\boldsymbol{\beta}_{1}(\boldsymbol{x})
=
\boldsymbol{\beta}_{2}(\boldsymbol{x}).
\]
Therefore, the first-order discrepancy coefficient function \(\boldsymbol{\beta}(\boldsymbol{x})\) is also uniquely
determined.

Finally, substituting
\(\boldsymbol{\theta}^{p}_{1}=\boldsymbol{\theta}^{p}_{2}\),
\(\alpha_{1}(\boldsymbol{x})=\alpha_{2}(\boldsymbol{x})\), and
\(\boldsymbol{\beta}_{1}(\boldsymbol{x})=\boldsymbol{\beta}_{2}(\boldsymbol{x})\)
back into \eqref{eq:app_bmdd_identical_representation}, we obtain
\[
r_{1}(\boldsymbol{x},\boldsymbol{\theta};\boldsymbol{\theta}^{p})
=
r_{2}(\boldsymbol{x},\boldsymbol{\theta};\boldsymbol{\theta}^{p}).
\]
Thus, the residual discrepancy term \(r(\boldsymbol{x},\boldsymbol{\theta};\boldsymbol{\theta}^{p})\) is uniquely determined. Consequently, BMDD is identifiable.
\end{proof}

\begin{proof}[Proof of equation \eqref{eq:alpha_full_conditional_constrained}]
To simplify notation, we write \(\mathbf{T}=\mathbf{T}(\boldsymbol{\theta}^{p})\) and \(\mathbf{G}=\mathbf{G}(\boldsymbol{\theta}^{p})\). The likelihood is
\begin{equation}
\label{eq:app_lik}
p\!\left(
\boldsymbol{\Delta}\mid
\boldsymbol{\alpha},
\mathbf{B}_{\beta},
\mathbf{W},
\boldsymbol{\theta}^{p},
\boldsymbol{\epsilon},
\sigma_{\xi}^{2}
\right)
\propto
(\sigma_{\xi}^{2})^{-nm/2}
\exp\!\left\{
-\frac{1}{2\sigma_{\xi}^{2}}
\left\|
\boldsymbol{\Delta}
-
\boldsymbol{\alpha}\mathbf{1}_m^{\mathsf T}
-
\mathbf{B}_{\beta}\mathbf{T}^{\mathsf T}
-
\mathbf{W}\mathbf{G}
-
\boldsymbol{\epsilon}\mathbf{1}_m^{\mathsf T}
\right\|_F^2
\right\}.
\end{equation}
By Bayes' theorem,
\begin{equation}
\label{eq:app_alpha_bayes}
p\!\left(
\boldsymbol{\alpha}\mid
\boldsymbol{\Delta},
\mathbf{B}_{\beta},
\mathbf{W},
\boldsymbol{\theta}^{p},
\boldsymbol{\epsilon},
\sigma_{\xi}^{2},
\sigma_\alpha^2
\right)
\propto
p\!\left(
\boldsymbol{\Delta}\mid
\boldsymbol{\alpha},
\mathbf{B}_{\beta},
\mathbf{W},
\boldsymbol{\theta}^{p},
\boldsymbol{\epsilon},
\sigma_{\xi}^{2}
\right)
p(\boldsymbol{\alpha}\mid \sigma_\alpha^2).
\end{equation}
Let
\(
\widetilde{\boldsymbol{\Delta}}_{\alpha}
=
\boldsymbol{\Delta}
-
\mathbf{B}_{\beta}\mathbf{T}^{\mathsf T}
-
\mathbf{W}\mathbf{G}
-
\boldsymbol{\epsilon}\mathbf{1}_m^{\mathsf T}
\),
so that
\(
\widetilde{\boldsymbol{\Delta}}_{\alpha}
=
\boldsymbol{\alpha}\mathbf{1}_m^{\mathsf T}
+
\boldsymbol{\Xi}
\).
Hence, the likelihood term in \eqref{eq:app_lik} can be written as
\[
p\!\left(
\boldsymbol{\Delta}\mid
\boldsymbol{\alpha},
\mathbf{B}_{\beta},
\mathbf{W},
\boldsymbol{\theta}^{p},
\boldsymbol{\epsilon},
\sigma_{\xi}^{2}
\right)
\propto
\exp\!\left\{
-\frac{1}{2\sigma_{\xi}^{2}}
\left\|
\widetilde{\boldsymbol{\Delta}}_{\alpha}
-
\boldsymbol{\alpha}\mathbf{1}_m^{\mathsf T}
\right\|_F^2
\right\}.
\]
Expanding the squared Frobenius norm yields
\(
\left\|
\widetilde{\boldsymbol{\Delta}}_{\alpha}
-
\boldsymbol{\alpha}\mathbf{1}_m^{\mathsf T}
\right\|_F^2
=
m\,\boldsymbol{\alpha}^{\mathsf T}\boldsymbol{\alpha}
-
2\boldsymbol{\alpha}^{\mathsf T}\widetilde{\boldsymbol{\Delta}}_{\alpha}\mathbf{1}_m
+
c
\),
where \(c\) is a constant independent of \(\boldsymbol{\alpha}\). Therefore,
\begin{equation}
\label{eq:app_alpha_like_kernel}
p\!\left(
\boldsymbol{\Delta}\mid
\boldsymbol{\alpha},
\mathbf{B}_{\beta},
\mathbf{W},
\boldsymbol{\theta}^{p},
\boldsymbol{\epsilon},
\sigma_{\xi}^{2}
\right)
\propto
\exp\!\left\{
-\frac{1}{2}
\left[
\boldsymbol{\alpha}^{\mathsf T}
\left(\frac{m}{\sigma_{\xi}^{2}}\mathbf{I}_n\right)
\boldsymbol{\alpha}
-
2\boldsymbol{\alpha}^{\mathsf T}
\left(
\frac{1}{\sigma_{\xi}^{2}}
\widetilde{\boldsymbol{\Delta}}_{\alpha}\mathbf{1}_m
\right)
\right]
\right\}.
\end{equation}
On the other hand, from the constrained Gaussian prior
\(
\boldsymbol{\alpha}\mid \sigma_\alpha^2
\sim
\mathcal{N}(\boldsymbol{\mu}_{\alpha,c},\sigma_\alpha^2\mathbf{K}_{\alpha,c})
\),
its kernel, ignoring terms independent of \(\boldsymbol{\alpha}\), is
\begin{equation}
\label{eq:app_alpha_prior_kernel}
p\bigl(\boldsymbol{\alpha}\mid \sigma_{\alpha}^{2}\bigr)
\propto
\exp\!\left\{
-\frac{1}{2}
\left[
\boldsymbol{\alpha}^{\mathsf T}
\frac{1}{\sigma_{\alpha}^{2}}\mathbf{K}_{\alpha,c}^{-1}
\boldsymbol{\alpha}
-
2\boldsymbol{\alpha}^{\mathsf T}
\frac{1}{\sigma_{\alpha}^{2}}\mathbf{K}_{\alpha,c}^{-1}\boldsymbol{\mu}_{\alpha,c}
\right]
\right\}.
\end{equation}
Substituting \eqref{eq:app_alpha_like_kernel} and \eqref{eq:app_alpha_prior_kernel} into \eqref{eq:app_alpha_bayes} and collecting the terms involving \(\boldsymbol{\alpha}\), we obtain
\[
\begin{aligned}
\log p\!\left(
\boldsymbol{\alpha}\mid
\boldsymbol{\Delta},
\mathbf{B}_{\beta},
\mathbf{W},
\boldsymbol{\theta}^{p},
\boldsymbol{\epsilon},
\sigma_{\xi}^{2},
\sigma_\alpha^2
\right)
&=
-\frac{1}{2}
\boldsymbol{\alpha}^{\mathsf T}
\left(
\frac{1}{\sigma_\alpha^2}\mathbf{K}_{\alpha,c}^{-1}
+
\frac{m}{\sigma_{\xi}^{2}}\mathbf{I}_n
\right)
\boldsymbol{\alpha}
\\
&\quad+
\boldsymbol{\alpha}^{\mathsf T}
\left(
\frac{1}{\sigma_\alpha^2}\mathbf{K}_{\alpha,c}^{-1}\boldsymbol{\mu}_{\alpha,c}
+
\frac{1}{\sigma_{\xi}^{2}}\widetilde{\boldsymbol{\Delta}}_{\alpha}\mathbf{1}_m
\right) + C_{\alpha}, 
\end{aligned}
\]
which is precisely the kernel of a multivariate normal distribution. Here \(C_{\alpha}\) is a constant independent of \(\boldsymbol{\alpha}\); it may depend on the quantities being conditioned on. Therefore,
\[
\boldsymbol{\alpha}\mid
(
\boldsymbol{\Delta},
\mathbf{B}_{\beta},
\mathbf{W},
\boldsymbol{\theta}^{p},
\boldsymbol{\epsilon},
\sigma_{\xi}^{2},
\sigma_\alpha^2
)
\sim
\mathcal{N}(\boldsymbol{\mu}_{\alpha\mid\cdot},\boldsymbol{\Sigma}_{\alpha\mid\cdot}),
\]
where
\(
\boldsymbol{\Sigma}_{\alpha\mid\cdot}^{-1}
=
\frac{1}{\sigma_\alpha^2}\mathbf{K}_{\alpha,c}^{-1}
+
\frac{m}{\sigma_{\xi}^{2}}\mathbf{I}_n
\),
and
\(
\boldsymbol{\mu}_{\alpha\mid\cdot}
=
\boldsymbol{\Sigma}_{\alpha\mid\cdot}
\left(
\frac{1}{\sigma_\alpha^2}\mathbf{K}_{\alpha,c}^{-1}\boldsymbol{\mu}_{\alpha,c}
+
\frac{1}{\sigma_{\xi}^{2}}\widetilde{\boldsymbol{\Delta}}_{\alpha}\mathbf{1}_m
\right)
\).
\end{proof}

\begin{proof}[Proof of equation \eqref{eq:eps_full}]
By Bayes' theorem,
\begin{equation}
\label{eq:app_eps_bayes}
p\!\left(
\boldsymbol{\epsilon}\mid
\boldsymbol{\Delta},
\boldsymbol{\alpha},
\mathbf{B}_{\beta},
\mathbf{W},
\boldsymbol{\theta}^{p},
\sigma_{\xi}^{2},
\sigma_{\epsilon}^{2}
\right)
\propto
p\!\left(
\boldsymbol{\Delta}\mid
\boldsymbol{\alpha},
\mathbf{B}_{\beta},
\mathbf{W},
\boldsymbol{\theta}^{p},
\boldsymbol{\epsilon},
\sigma_{\xi}^{2}
\right)
p(\boldsymbol{\epsilon}\mid \sigma_{\epsilon}^{2}).
\end{equation}
Let
\(
\widetilde{\boldsymbol{\Delta}}_{\epsilon}
=
\boldsymbol{\Delta}
-
\boldsymbol{\alpha}\mathbf{1}_m^{\mathsf T}
-
\mathbf{B}_{\beta}\mathbf{T}^{\mathsf T}
-
\mathbf{W}\mathbf{G}
\),
so that
\(
\widetilde{\boldsymbol{\Delta}}_{\epsilon}
=
\boldsymbol{\epsilon}\mathbf{1}_m^{\mathsf T}
+
\boldsymbol{\Xi}
\).
Proceeding as in the proof of equation \eqref{eq:alpha_full_conditional_constrained}, we have
\begin{equation}
\label{eq:app_eps_like_kernel}
p\!\left(
\boldsymbol{\Delta}\mid
\boldsymbol{\alpha},
\mathbf{B}_{\beta},
\mathbf{W},
\boldsymbol{\theta}^{p},
\boldsymbol{\epsilon},
\sigma_{\xi}^{2}
\right) \propto
\exp\!\left\{
-\frac{1}{2}
\left[
\boldsymbol{\epsilon}^{\mathsf T}
\frac{m}{\sigma_{\xi}^{2}}\mathbf{I}_n
\boldsymbol{\epsilon}
-
2\boldsymbol{\epsilon}^{\mathsf T}
\frac{1}{\sigma_{\xi}^{2}}
\widetilde{\boldsymbol{\Delta}}_{\epsilon}\mathbf{1}_m
\right]
\right\}
\end{equation}
Moreover, from the prior
\(
\boldsymbol{\epsilon}\mid \sigma_{\epsilon}^{2}\sim \mathcal{N}(\mathbf{0},\sigma_{\epsilon}^{2}\mathbf{I}_n)
\),
we have
\begin{equation}
\label{eq:app_eps_prior_kernel}
p(\boldsymbol{\epsilon}\mid \sigma_{\epsilon}^{2})
\propto
\exp\!\left(
-\frac{1}{2\sigma_{\epsilon}^{2}}\boldsymbol{\epsilon}^{\mathsf T}\boldsymbol{\epsilon}
\right).
\end{equation}
Substituting \eqref{eq:app_eps_like_kernel} and \eqref{eq:app_eps_prior_kernel} into \eqref{eq:app_eps_bayes} and rearranging terms yields
\[
\log p\!\left(
\boldsymbol{\epsilon}\mid
\boldsymbol{\Delta},
\boldsymbol{\alpha},
\mathbf{B}_{\beta},
\mathbf{W},
\boldsymbol{\theta}^{p},
\sigma_{\xi}^{2},
\sigma_{\epsilon}^{2}
\right)
=
-\frac{1}{2}
\boldsymbol{\epsilon}^{\mathsf T}
\left(
\frac{1}{\sigma_{\epsilon}^{2}}
+
\frac{m}{\sigma_{\xi}^{2}}
\right)\mathbf{I}_n
\boldsymbol{\epsilon}
+
\boldsymbol{\epsilon}^{\mathsf T}
\left(
\frac{1}{\sigma_{\xi}^{2}}\widetilde{\boldsymbol{\Delta}}_{\epsilon}\mathbf{1}_m
\right) + C_{\epsilon}, 
\]
which is again the kernel of a multivariate normal distribution. Here \(C_{\epsilon}\) is a constant independent of \(\boldsymbol{\epsilon}\); it may depend on the quantities being conditioned on. Therefore,
\[
\boldsymbol{\epsilon}\mid \big(
\boldsymbol{\Delta},
\boldsymbol{\alpha},
\mathbf{B}_{\beta},
\mathbf{W},
\boldsymbol{\theta}^{p},
\sigma_{\xi}^{2},
\sigma_{\epsilon}^{2} \big)
\sim
\mathcal{N}(\boldsymbol{\mu}_{\epsilon\mid\cdot},\boldsymbol{\Sigma}_{\epsilon\mid\cdot}),
\]
where
\(
\boldsymbol{\Sigma}_{\epsilon\mid\cdot}^{-1}
=
\left(
\frac{1}{\sigma_{\epsilon}^{2}}
+
\frac{m}{\sigma_{\xi}^{2}}
\right)\mathbf{I}_n
\),
and
\(
\boldsymbol{\mu}_{\epsilon\mid\cdot}
=
\boldsymbol{\Sigma}_{\epsilon\mid\cdot}
\left(
\frac{1}{\sigma_{\xi}^{2}}\widetilde{\boldsymbol{\Delta}}_{\epsilon}\mathbf{1}_m
\right)
\).
\end{proof}

\vspace{5pt}

\begin{proof}[Proof of equation \eqref{eq:beta_full_vec}]
By Bayes' theorem,
\begin{equation}
\label{eq:app_beta_bayes}
p\!\left(
\operatorname{vec}(\mathbf{B}_{\beta})
\mid
\boldsymbol{\Delta},
\boldsymbol{\alpha},
\mathbf{W},
\boldsymbol{\theta}^{p},
\boldsymbol{\epsilon},
\sigma_{\xi}^{2},
\{\sigma_{\beta_\ell}^2\}_{\ell=1}^{q}
\right)
\propto
p\!\left(
\boldsymbol{\Delta}\mid
\boldsymbol{\alpha},
\mathbf{B}_{\beta},
\mathbf{W},
\boldsymbol{\theta}^{p},
\boldsymbol{\epsilon},
\sigma_{\xi}^{2}
\right)
p\!\left(
\operatorname{vec}(\mathbf{B}_{\beta})
\mid
\{\sigma_{\beta_\ell}^2\}_{\ell=1}^{q}
\right).
\end{equation}
Let
\(
\widetilde{\boldsymbol{\Delta}}_{\beta}
=
\boldsymbol{\Delta}
-
\boldsymbol{\alpha}\mathbf{1}_m^{\mathsf T}
-
\mathbf{W}\mathbf{G}
-
\boldsymbol{\epsilon}\mathbf{1}_m^{\mathsf T}
\),
so that
\(
\widetilde{\boldsymbol{\Delta}}_{\beta}
=
\mathbf{B}_{\beta}\mathbf{T}^{\mathsf T}
+
\boldsymbol{\Xi}
\).
Vectorizing both sides and using the identity
\(
\operatorname{vec}(\mathbf{B}_{\beta}\mathbf{T}^{\mathsf T})
=
(\mathbf{T}\otimes \mathbf{I}_n)\operatorname{vec}(\mathbf{B}_{\beta})
\),
we obtain
\(
\operatorname{vec}(\widetilde{\boldsymbol{\Delta}}_{\beta})
=
(\mathbf{T}\otimes \mathbf{I}_n)\operatorname{vec}(\mathbf{B}_{\beta})
+
\operatorname{vec}(\boldsymbol{\Xi})
\).
Hence, the likelihood kernel for \(\mathbf{B}_{\beta}\) can be written as
\begin{equation}
\label{eq:app_beta_like_kernel}
p\!\left(
\boldsymbol{\Delta}\mid
\boldsymbol{\alpha},
\mathbf{B}_{\beta},
\mathbf{W},
\boldsymbol{\theta}^{p},
\boldsymbol{\epsilon},
\sigma_{\xi}^{2}
\right)
\propto
\exp\!\left\{
-\frac{1}{2\sigma_{\xi}^{2}}
\left\|
\operatorname{vec}(\widetilde{\boldsymbol{\Delta}}_{\beta})
-
(\mathbf{T}\otimes \mathbf{I}_n)\operatorname{vec}(\mathbf{B}_{\beta})
\right\|_2^2
\right\}.
\end{equation}
On the other hand, from the componentwise independent GP priors
\(
\boldsymbol{\beta}_{\ell}\mid \sigma_{\beta_\ell}^2
\sim \mathcal{N}(\boldsymbol{\mu}_{\beta_\ell},\sigma_{\beta_\ell}^2\mathbf{K}_{\beta})
\),
\(\ell=1,\dots,q\),
it follows that
\(
\operatorname{vec}(\mathbf{B}_{\beta})
\)
has prior distribution
\[
\operatorname{vec}(\mathbf{B}_{\beta})
\mid
\{\sigma_{\beta_\ell}^2\}_{\ell=1}^{q}
\sim
\mathcal{N}(\boldsymbol{\mu}_{b},\mathbf{Q}_{b}^{-1}),
\]
where
\(
\boldsymbol{\mu}_{b}
=
\operatorname{vec}\!\big([\boldsymbol{\mu}_{\beta_{1}},\ldots,\boldsymbol{\mu}_{\beta_{q}}]\big)
\),
and
\(
\mathbf{Q}_{b}
=
\operatorname{blockdiag}
\left(
\frac{1}{\sigma_{\beta_1}^2}\mathbf{K}_{\beta}^{-1},
\dots,
\frac{1}{\sigma_{\beta_q}^2}\mathbf{K}_{\beta}^{-1}
\right)
\).
Therefore, its prior kernel can be written as
\begin{equation}
\label{eq:app_beta_prior_kernel}
p\!\left(
\operatorname{vec}(\mathbf{B}_{\beta})
\mid
\{\sigma_{\beta_\ell}^2\}_{\ell=1}^{q}
\right)
\propto
\exp\!\left\{
-\frac{1}{2}
\bigl(\operatorname{vec}(\mathbf{B}_{\beta})-\boldsymbol{\mu}_{b}\bigr)^{\mathsf T}
\mathbf{Q}_{b}
\bigl(\operatorname{vec}(\mathbf{B}_{\beta})-\boldsymbol{\mu}_{b}\bigr)
\right\}.
\end{equation}
Substituting \eqref{eq:app_beta_like_kernel} and \eqref{eq:app_beta_prior_kernel} into \eqref{eq:app_beta_bayes}, and ignoring constants independent of \(\operatorname{vec}(\mathbf{B}_{\beta})\), we obtain
\[
\begin{aligned}
&\log p\!\left(
\operatorname{vec}(\mathbf{B}_{\beta})
\mid
\boldsymbol{\Delta},
\boldsymbol{\alpha},
\mathbf{W},
\boldsymbol{\theta}^{p},
\boldsymbol{\epsilon},
\sigma_{\xi}^{2},
\{\sigma_{\beta_\ell}^2\}_{\ell=1}^{q}
\right)
\\
&\quad=
-\frac{1}{2}\,
\operatorname{vec}(\mathbf{B}_{\beta})^{\mathsf T}
\left[
\mathbf{Q}_{b}
+
\frac{1}{\sigma_{\xi}^{2}}
(\mathbf{T}^{\mathsf T}\mathbf{T}\otimes \mathbf{I}_n)
\right]
\operatorname{vec}(\mathbf{B}_{\beta})
\\
&\qquad+ \
\operatorname{vec}(\mathbf{B}_{\beta})^{\mathsf T}
\left[
\mathbf{Q}_{b}\boldsymbol{\mu}_{b}
+
\frac{1}{\sigma_{\xi}^{2}}
(\mathbf{T}^{\mathsf T}\otimes \mathbf{I}_n)\operatorname{vec}(\widetilde{\boldsymbol{\Delta}}_{\beta})
\right] + C_{\beta}, 
\end{aligned}
\]
which shows that the posterior distribution is again multivariate normal. Here \(C_{\beta}\) is a constant independent of \(\operatorname{vec}(\mathbf{B}_{\beta})\); it may depend on the quantities being conditioned on. Therefore,
\[
\operatorname{vec}(\mathbf{B}_{\beta})
\mid \big(
\boldsymbol{\Delta},
\boldsymbol{\alpha},
\mathbf{W},
\boldsymbol{\theta}^{p},
\boldsymbol{\epsilon},
\sigma_{\xi}^{2},
\{\sigma_{\beta_\ell}^2\}_{\ell=1}^{q} \big)
\sim
\mathcal{N}(\boldsymbol{\mu}_{b\mid\cdot},\boldsymbol{\Sigma}_{b\mid\cdot}),
\]
where
\(
\boldsymbol{\Sigma}_{b\mid\cdot}^{-1}
=
\mathbf{Q}_{b}
+
\frac{1}{\sigma_{\xi}^{2}}
(\mathbf{T}^{\mathsf T}\mathbf{T}\otimes \mathbf{I}_n)
\),
and
\(
\boldsymbol{\mu}_{b\mid\cdot}
=
\boldsymbol{\Sigma}_{b\mid\cdot}
\left[
\mathbf{Q}_{b}\boldsymbol{\mu}_{b}
+
\frac{1}{\sigma_{\xi}^{2}}
(\mathbf{T}^{\mathsf T}\otimes \mathbf{I}_n)\operatorname{vec}(\widetilde{\boldsymbol{\Delta}}_{\beta})
\right]
\).
Furthermore, using the identity
\(
(\mathbf{T}^{\mathsf T}\otimes \mathbf{I}_n)\operatorname{vec}(\widetilde{\boldsymbol{\Delta}}_{\beta})
=
\operatorname{vec}(\widetilde{\boldsymbol{\Delta}}_{\beta}\mathbf{T})
\),
the posterior mean can be rewritten as
\(
\boldsymbol{\mu}_{b\mid\cdot}
=
\boldsymbol{\Sigma}_{b\mid\cdot}
\left[
\mathbf{Q}_{b}\boldsymbol{\mu}_{b}
+
\frac{1}{\sigma_{\xi}^{2}}
\operatorname{vec}(\widetilde{\boldsymbol{\Delta}}_{\beta}\mathbf{T})
\right]
\).
\end{proof}

\vspace{5pt}

\begin{proof}[Proof of equation \eqref{eq:W_full_vec}]
By Bayes' theorem,
\begin{equation}
\label{eq:app_W_bayes}
p\!\left(
\operatorname{vec}(\mathbf{W})
\mid
\boldsymbol{\Delta},
\boldsymbol{\alpha},
\mathbf{B}_{\beta},
\boldsymbol{\theta}^{p},
\boldsymbol{\epsilon},
\sigma_{\xi}^{2},
\sigma_{\omega}^{2}
\right)
\propto
p\!\left(
\boldsymbol{\Delta}\mid
\boldsymbol{\alpha},
\mathbf{B}_{\beta},
\mathbf{W},
\boldsymbol{\theta}^{p},
\boldsymbol{\epsilon},
\sigma_{\xi}^{2}
\right)
p\!\left(
\operatorname{vec}(\mathbf{W})\mid \sigma_{\omega}^{2}
\right).
\end{equation}
Let
\(
\widetilde{\boldsymbol{\Delta}}_{W}
=
\boldsymbol{\Delta}
-
\boldsymbol{\alpha}\mathbf{1}_m^{\mathsf T}
-
\mathbf{B}_{\beta}\mathbf{T}^{\mathsf T}
-
\boldsymbol{\epsilon}\mathbf{1}_m^{\mathsf T}
\),
so that
\(
\widetilde{\boldsymbol{\Delta}}_{W}
=
\mathbf{W}\mathbf{G}
+
\boldsymbol{\Xi}
\).
Vectorizing both sides and using the identity
\(
\operatorname{vec}(\mathbf{W}\mathbf{G})
=
(\mathbf{G}\otimes \mathbf{I}_n)\operatorname{vec}(\mathbf{W})
\),
where \(\mathbf{G}\) is diagonal, we obtain
\(
\operatorname{vec}(\widetilde{\boldsymbol{\Delta}}_{W})
=
(\mathbf{G}\otimes \mathbf{I}_n)\operatorname{vec}(\mathbf{W})
+
\operatorname{vec}(\boldsymbol{\Xi})
\).
Accordingly, the likelihood term in \eqref{eq:app_lik} can be written as
\begin{equation}
\label{eq:app_W_like_kernel}
p\!\left(
\boldsymbol{\Delta}\mid
\boldsymbol{\alpha},
\mathbf{B}_{\beta},
\mathbf{W},
\boldsymbol{\theta}^{p},
\boldsymbol{\epsilon},
\sigma_{\xi}^{2}
\right)
\propto
\exp\!\left\{
-\frac{1}{2\sigma_{\xi}^{2}}
\left\|
\operatorname{vec}(\widetilde{\boldsymbol{\Delta}}_{W})
-
(\mathbf{G}\otimes \mathbf{I}_n)\operatorname{vec}(\mathbf{W})
\right\|^2
\right\}.
\end{equation}
On the other hand, from the constrained GP prior for
\(\omega(\boldsymbol{x},\boldsymbol{\theta})\), we have
\(
\operatorname{vec}(\mathbf{W})\mid\sigma_{\omega}^{2}
\sim
\mathcal{N}\!\left(
\mathbf{0},
\sigma_{\omega}^{2}(\mathbf{K}_{\theta}\otimes \mathbf{K}_{x,c})
\right)
\),
where \(\mathbf{K}_{x,c}\) is the covariance matrix induced by the constrained kernel
\(k_{x,c}\). Hence its prior kernel is
\begin{equation}
\label{eq:app_W_prior_kernel}
p\!\left(
\operatorname{vec}(\mathbf{W})\mid \sigma_{\omega}^{2}
\right)
\propto
\exp\!\left\{
-\frac{1}{2\sigma_{\omega}^{2}}
\operatorname{vec}(\mathbf{W})^{\mathsf T}
(\mathbf{K}_{\theta}^{-1}\otimes \mathbf{K}_{x,c}^{-1})
\operatorname{vec}(\mathbf{W})
\right\}.
\end{equation}
Substituting \eqref{eq:app_W_like_kernel} and \eqref{eq:app_W_prior_kernel} into \eqref{eq:app_W_bayes}, we similarly obtain
\[
\begin{aligned}
&\log p\!\left(
\operatorname{vec}(\mathbf{W})
\mid
\boldsymbol{\Delta},
\boldsymbol{\alpha},
\mathbf{B}_{\beta},
\boldsymbol{\theta}^{p},
\boldsymbol{\epsilon},
\sigma_{\xi}^{2},
\sigma_{\omega}^{2}
\right)
\\
&\quad=
-\frac{1}{2}\,
\operatorname{vec}(\mathbf{W})^{\mathsf T}
\left[
\frac{1}{\sigma_{\xi}^{2}}
(\mathbf{G}^{\mathsf T}\mathbf{G}\otimes \mathbf{I}_n)
+
\frac{1}{\sigma_{\omega}^{2}}
(\mathbf{K}_{\theta}^{-1}\otimes \mathbf{K}_{x,c}^{-1})
\right]
\operatorname{vec}(\mathbf{W})
\\
&\qquad+ \
\operatorname{vec}(\mathbf{W})^{\mathsf T}
\left[
\frac{1}{\sigma_{\xi}^{2}}
(\mathbf{G}\otimes \mathbf{I}_n)\operatorname{vec}(\widetilde{\boldsymbol{\Delta}}_{W})
\right] + C_{W}, 
\end{aligned}
\]
which shows that the posterior distribution remains multivariate normal. Here \(C_{W}\) is a constant independent of \(\operatorname{vec}(\mathbf{W})\); it may depend on the quantities being conditioned on.
Since \(\mathbf{G}\) is diagonal, we have \(\mathbf{G}^{\mathsf T}\mathbf{G}=\mathbf{G}^{2}\). Hence,
\[
\operatorname{vec}(\mathbf{W})
\mid
\big(
\boldsymbol{\Delta},
\boldsymbol{\alpha},
\mathbf{B}_{\beta},
\boldsymbol{\theta}^{p},
\boldsymbol{\epsilon},
\sigma_{\xi}^{2},
\sigma_{\omega}^{2}
\big)
\sim
\mathcal{N}(\boldsymbol{\mu}_{W\mid\cdot},\boldsymbol{\Sigma}_{W\mid\cdot}),
\]
where
\(
\boldsymbol{\Sigma}_{W\mid\cdot}^{-1}
=
\frac{1}{\sigma_{\xi}^{2}}
(\mathbf{G}^{2}\otimes \mathbf{I}_n)
+
\frac{1}{\sigma_{\omega}^{2}}
(\mathbf{K}_{\theta}^{-1}\otimes \mathbf{K}_{x,c}^{-1})
\),
and
\(
\boldsymbol{\mu}_{W\mid\cdot}
=
\boldsymbol{\Sigma}_{W\mid\cdot}
\left[
\frac{1}{\sigma_{\xi}^{2}}
\operatorname{vec}(\widetilde{\boldsymbol{\Delta}}_{W}\mathbf{G})
\right]
\).
\end{proof}

\vspace{5pt}

Equations \eqref{eq:sig_F_full}--\eqref{eq:sig_delta_full} all follow directly from normal--inverse-gamma conjugacy. We present them separately below.

\begin{proof}[Proof of equation \eqref{eq:sig_F_full}]
From the prior
\(
\sigma_{\epsilon}^{2}\sim \mathcal{IG}(a_{\epsilon},b_{\epsilon})
\)
and
\(
\boldsymbol{\epsilon}\mid \sigma_{\epsilon}^{2}\sim \mathcal{N}(\mathbf{0},\sigma_{\epsilon}^{2}\mathbf{I}_n)
\),
we have
\[
p(\sigma_{\epsilon}^{2})\propto
(\sigma_{\epsilon}^{2})^{-(a_{\epsilon}+1)}
\exp\!\left(-\frac{b_{\epsilon}}{\sigma_{\epsilon}^{2}}\right),
\qquad
p(\boldsymbol{\epsilon}\mid \sigma_{\epsilon}^{2})
\propto
(\sigma_{\epsilon}^{2})^{-n/2}
\exp\!\left(
-\frac{1}{2\sigma_{\epsilon}^{2}}\boldsymbol{\epsilon}^{\mathsf T}\boldsymbol{\epsilon}
\right).
\]
Multiplying these two expressions and ignoring constants independent of \(\sigma_{\epsilon}^{2}\), we obtain
\[
p(\sigma_{\epsilon}^{2}\mid \boldsymbol{\epsilon})
\propto
(\sigma_{\epsilon}^{2})^{-(a_{\epsilon}+n/2+1)}
\exp\!\left\{
-\frac{1}{\sigma_{\epsilon}^{2}}
\left(
b_{\epsilon}+\frac{1}{2}\boldsymbol{\epsilon}^{\mathsf T}\boldsymbol{\epsilon}
\right)
\right\},
\]
and therefore
\(
\sigma_{\epsilon}^{2}\mid \boldsymbol{\epsilon}
\sim
\mathcal{IG}\!\left(
a_{\epsilon}+\frac{n}{2},
\, b_{\epsilon}+\frac{1}{2}\boldsymbol{\epsilon}^{\mathsf T}\boldsymbol{\epsilon}
\right)
\).
\end{proof}

\vspace{5pt}

\begin{proof}[Proof of equation \eqref{eq:sig_alpha_full}]
Similarly, from
\(
\boldsymbol{\alpha}\mid \sigma_\alpha^2
\sim
\mathcal{N}(\boldsymbol{\mu}_{\alpha,c},
\sigma_\alpha^2\mathbf{K}_{\alpha,c})
\)
and
\(
\sigma_\alpha^2\sim \mathcal{IG}(a_\alpha,b_\alpha)
\),
we obtain
\[
p(\sigma_\alpha^2\mid \boldsymbol{\alpha})
\propto
(\sigma_\alpha^2)^{-(a_\alpha+n/2+1)}
\exp\!\left\{
-\frac{1}{\sigma_\alpha^2}
\left[
b_\alpha
+
\frac{1}{2}
(\boldsymbol{\alpha}-\boldsymbol{\mu}_{\alpha,c})^{\mathsf T}
\mathbf{K}_{\alpha,c}^{-1}
(\boldsymbol{\alpha}-\boldsymbol{\mu}_{\alpha,c})
\right]
\right\},
\]
and hence
\(
\sigma_\alpha^2\mid \boldsymbol{\alpha}
\sim
\mathcal{IG}\!\left(
a_\alpha+\frac{n}{2},
\, b_\alpha+\frac{1}{2}
(\boldsymbol{\alpha}-\boldsymbol{\mu}_{\alpha,c})^{\mathsf T}
\mathbf{K}_{\alpha,c}^{-1}
(\boldsymbol{\alpha}-\boldsymbol{\mu}_{\alpha,c})
\right)
\).
\end{proof}

\vspace{5pt}

\begin{proof}[Proof of equation \eqref{eq:sig_beta_full}]
For any \(\ell=1,\dots,q\), from
\(
\boldsymbol{\beta}_{\ell}\mid \sigma_{\beta_\ell}^2
\sim
\mathcal{N}(\boldsymbol{\mu}_{\beta_\ell},\sigma_{\beta_\ell}^2\mathbf{K}_{\beta})
\)
and
\(
\sigma_{\beta_\ell}^2\sim \mathcal{IG}(a_\beta,b_\beta)
\),
we similarly obtain
\[
p(\sigma_{\beta_\ell}^2\mid \boldsymbol{\beta}_{\ell})
\propto
(\sigma_{\beta_\ell}^2)^{-(a_\beta+n/2+1)}
\exp\!\left\{
-\frac{1}{\sigma_{\beta_\ell}^2}
\left[
b_\beta
+
\frac{1}{2}
(\boldsymbol{\beta}_{\ell}-\boldsymbol{\mu}_{\beta_\ell})^{\mathsf T}
\mathbf{K}_{\beta}^{-1}
(\boldsymbol{\beta}_{\ell}-\boldsymbol{\mu}_{\beta_\ell})
\right]
\right\},
\]
and therefore
\(
\sigma_{\beta_\ell}^2\mid \boldsymbol{\beta}_{\ell}
\sim
\mathcal{IG}\!\left(
a_\beta+\frac{n}{2},
\, b_\beta+\frac{1}{2}
(\boldsymbol{\beta}_{\ell}-\boldsymbol{\mu}_{\beta_\ell})^{\mathsf T}
\mathbf{K}_{\beta}^{-1}
(\boldsymbol{\beta}_{\ell}-\boldsymbol{\mu}_{\beta_\ell})
\right)
\),
\(\ell=1,\dots,q\).
\end{proof}

\vspace{5pt}

\begin{proof}[Proof of equation \eqref{eq:tau_full}]
From
\(
\operatorname{vec}(\mathbf{W})\mid \sigma_{\omega}^{2}
\sim
\mathcal{N}\!\left(
\mathbf{0},
\sigma_{\omega}^{2}
(\mathbf{K}_{\theta}\otimes\mathbf{K}_{x,c})
\right)
\)
and
\(
\sigma_{\omega}^{2}\sim \mathcal{IG}(a_{\omega},b_{\omega})
\),
we similarly obtain
\[
p(\sigma_{\omega}^{2}\mid \mathbf{W})
\propto
(\sigma_{\omega}^{2})^{-(a_{\omega}+nm/2+1)}
\exp\!\left\{
-\frac{1}{\sigma_{\omega}^{2}}
\left[
b_{\omega}
+
\frac{1}{2}
\operatorname{vec}(\mathbf{W})^{\mathsf T}
\bigl(\mathbf{K}_{\theta}^{-1}\otimes\mathbf{K}_{x,c}^{-1}\bigr)
\operatorname{vec}(\mathbf{W})
\right]
\right\},
\]
and hence
\(
\sigma_{\omega}^{2}\mid \mathbf{W}
\sim
\mathcal{IG}\!\left(
a_{\omega}+\frac{nm}{2},
\, b_{\omega}+\frac{1}{2}
\operatorname{vec}(\mathbf{W})^{\mathsf T}
\bigl(\mathbf{K}_{\theta}^{-1}\otimes\mathbf{K}_{x,c}^{-1}\bigr)
\operatorname{vec}(\mathbf{W})
\right)
\).
\end{proof}

\vspace{5pt}

\begin{proof}[Proof of equation \eqref{eq:sig_delta_full}]
Combining the likelihood term in \eqref{eq:app_lik} with the prior
\(
\sigma_{\xi}^{2}\sim \mathcal{IG}(a_{\xi},b_{\xi})
\),
we similarly obtain
\[
\begin{aligned}
p\!\left(\sigma_{\xi}^{2}\mid
\boldsymbol{\Delta},
\boldsymbol{\alpha},
\mathbf{B}_{\beta},
\mathbf{W},
\boldsymbol{\theta}^{p},
\boldsymbol{\epsilon}
\right)
\propto & (\sigma_{\xi}^{2})^{-(a_{\xi}+nm/2+1)}
\\&\exp\!\left\{
-\frac{1}{\sigma_{\xi}^{2}}
\left[
b_{\xi}
+
\frac{1}{2}
\left\|
\boldsymbol{\Delta}
-
\boldsymbol{\alpha}\mathbf{1}_m^{\mathsf T}
-
\mathbf{B}_{\beta}\mathbf{T}^{\mathsf T}
-
\mathbf{W}\mathbf{G}
-
\boldsymbol{\epsilon}\mathbf{1}_m^{\mathsf T}
\right\|_F^2
\right]
\right\},
\end{aligned}
\]
and therefore
\[
\begin{aligned}
&\sigma_{\xi}^{2}\mid \big(
\boldsymbol{\Delta},
\boldsymbol{\alpha},
\mathbf{B}_{\beta},
\mathbf{W},
\boldsymbol{\theta}^{p},
\boldsymbol{\epsilon} \big)
\\
&\quad\sim
\mathcal{IG}\!\left(
a_{\xi}+\frac{nm}{2},
\, b_{\xi}+\frac{1}{2}
\left\|
\boldsymbol{\Delta}
-
\boldsymbol{\alpha}\mathbf{1}_m^{\mathsf T}
-
\mathbf{B}_{\beta}\mathbf{T}^{\mathsf T}
-
\mathbf{W}\mathbf{G}
-
\boldsymbol{\epsilon}\mathbf{1}_m^{\mathsf T}
\right\|_F^2
\right)
.
\end{aligned}
\]
\end{proof}

\section*{Acknowledgments}

This work is partially supported by the National Natural Science Foundation of China (Grant No. 12571276).

\bibliographystyle{apalike}
\bibliography{reference}

@article{ArhonditsisEtAl2008,
  author  = {Arhonditsis, George B. and Perhar, Gurbir and Zhang, Weitao and Massos, Evangelia and Shi, Molu and Das, Argho},
  title   = {Addressing Equifinality and Uncertainty in Eutrophication Models},
  journal = {Water Resources Research},
  year    = {2008},
  volume  = {44},
  number  = {1},
  pages   = {W01420},
  doi     = {10.1029/2007WR005862}
}

@article{BayarriEtAl2007,
  author  = {Bayarri, Maria J. and Berger, James O. and Paulo, Rui and Sacks, Jerry and Cafeo, John A. and Cavendish, James and Lin, Chin-Hsu and Tu, Jian},
  title   = {A Framework for Validation of Computer Models},
  journal = {Technometrics},
  year    = {2007},
  volume  = {49},
  number  = {2},
  pages   = {138--154},
  doi     = {10.1198/004017007000000092}
}

@article{BrownAtamturktur2018,
  author  = {Brown, D. Andrew and Atamturktur, Sez},
  title   = {Nonparametric Functional Calibration of Computer Models},
  journal = {Statistica Sinica},
  year    = {2018},
  volume  = {28},
  number  = {2},
  pages   = {721--742}
}

@article{BrynjarsdottirOHagan2014,
  author  = {Brynjarsd{\'o}ttir, Jenn{\'y} and O'Hagan, Anthony},
  title   = {Learning about Physical Parameters: The Importance of Model Discrepancy},
  journal = {Inverse Problems},
  year    = {2014},
  volume  = {30},
  number  = {11},
  pages   = {114007},
  doi     = {10.1088/0266-5611/30/11/114007}
}

@article{EzzatPourhabibDing2018,
  author  = {Ezzat, Ahmed Aziz and Pourhabib, Arash and Ding, Yu},
  title   = {Sequential Design for Functional Calibration of Computer Models},
  journal = {Technometrics},
  year    = {2018},
  volume  = {60},
  number  = {3},
  pages   = {286--296}
}

@article{SansoForestZantedeschi2008,
  author  = {Sans{\'o}, Bruno and Forest, Chris E. and Zantedeschi, Daniel},
  title   = {Inferring Climate System Properties Using a Computer Model},
  journal = {Bayesian Analysis},
  year    = {2008},
  volume  = {3},
  number  = {1},
  pages   = {1--37},
  doi     = {10.1214/08-BA301}
}

@book{GhoshDelampadySamanta2006,
  author    = {Ghosh, Jayanta K. and Delampady, Mohan and Samanta, Tapas},
  title     = {An Introduction to {Bayesian} Analysis: Theory and Methods},
  publisher = {Springer},
  address   = {New York},
  year      = {2006}
}

@article{GoldsteinRougier2004,
  author  = {Goldstein, M. and Rougier, J.},
  title   = {Probabilistic formulations for transferring inferences from mathematical models to physical systems},
  journal = {SIAM Journal on Scientific Computing},
  year    = {2004},
  volume  = {26},
  number  = {2},
  pages   = {467--487},
  doi     = {10.1137/S106482750342670X}
}

@article{GuWang2018SGaSP,
  author  = {Gu, Mengyang and Wang, Long},
  title   = {Scaled {Gaussian} Stochastic Process for Computer Model Calibration and Prediction},
  journal = {SIAM/ASA Journal on Uncertainty Quantification},
  year    = {2018},
  volume  = {6},
  number  = {4},
  pages   = {1555--1583},
  doi     = {10.1137/17M1159890}
}

@article{GuXieWang2022,
  author  = {Gu, Mengyang and Xie, Fangzheng and Wang, Long},
  title   = {A Theoretical Framework of the Scaled {Gaussian} Stochastic Process in Prediction and Calibration},
  journal = {SIAM/ASA Journal on Uncertainty Quantification},
  year    = {2022},
  volume  = {10},
  number  = {4},
  pages   = {1435--1460},
  doi     = {10.1137/21M1409949}
}

@article{HanSantnerRawlinson2009,
  author  = {Han, Gang and Santner, Thomas J. and Rawlinson, Jeremy J.},
  title   = {Simultaneous Determination of Tuning and Calibration Parameters for Computer Experiments},
  journal = {Technometrics},
  year    = {2009},
  volume  = {51},
  number  = {4},
  pages   = {464--474}
}

@article{HigdonEtAl2004,
  author  = {Higdon, Dave and Kennedy, Marc and Cavendish, James C. and Cafeo, John A. and Ryne, Robert D.},
  title   = {Combining Field Data and Computer Simulations for Calibration and Prediction},
  journal = {SIAM Journal on Scientific Computing},
  year    = {2004},
  volume  = {26},
  number  = {2},
  pages   = {448--466},
  doi     = {10.1137/S1064827503426693}
}

@article{KangDeng2020,
  author  = {Kang, Xiaoning and Deng, Xinwei},
  title   = {Design and Analysis of Computer Experiments with Quantitative and Qualitative Inputs: A Selective Review},
  journal = {WIREs Data Mining and Knowledge Discovery},
  year    = {2020},
  volume  = {10},
  number  = {3},
  pages   = {e1358}
}

@article{KennedyOHagan2001,
  author  = {Kennedy, Marc C. and O'Hagan, Anthony},
  title   = {Bayesian Calibration of Computer Models},
  journal = {Journal of the Royal Statistical Society: Series B (Statistical Methodology)},
  year    = {2001},
  volume  = {63},
  number  = {3},
  pages   = {425--464},
  doi     = {10.1111/1467-9868.00294}
}

@article{LiWuXiong2026,
  author  = {Li, Yang and Wu, C. F. Jeff and Xiong, Shifeng},
  title   = {On the Use of a Discrepancy Decomposition Model for Calibration Experiments},
  journal = {IISE Transactions},
  year    = {2026},
  doi     = {10.1080/24725854.2026.2728122},
  note    = {Published online 24 September 2026}
}

@article{MurphyEtAl2007,
  author  = {Murphy, J. M. and Booth, B. B. B. and Collins, M. and Harris, G. R. and Sexton, D. M. H. and Webb, M. J.},
  title   = {A Methodology for Probabilistic Predictions of Regional Climate Change from Perturbed Physics Ensembles},
  journal = {Philosophical Transactions of the Royal Society A: Mathematical, Physical and Engineering Sciences},
  year    = {2007},
  volume  = {365},
  number  = {1857},
  pages   = {1993--2028},
  doi     = {10.1098/rsta.2007.2077}
}

@article{Plumlee2017,
  author  = {Plumlee, Matthew},
  title   = {Bayesian Calibration of Inexact Computer Models},
  journal = {Journal of the American Statistical Association},
  year    = {2017},
  volume  = {112},
  number  = {519},
  pages   = {1274--1285},
  doi     = {10.1080/01621459.2016.1211016}
}

@book{SantnerWilliamsNotz2018,
  author    = {Santner, Thomas J. and Williams, Brian J. and Notz, William I.},
  title     = {{The Design and Analysis of Computer Experiments}},
  edition   = {2},
  publisher = {Springer},
  address   = {New York},
  year      = {2018},
  doi     = {10.1007/978-1-4939-8847-1}
}

@article{SungTuo2024,
  author  = {Sung, Chih-Li and Tuo, Rui},
  title   = {A Review on Computer Model Calibration},
  journal = {WIREs Computational Statistics},
  year    = {2024},
  volume  = {16},
  number  = {1},
  pages   = {e1645},
  doi     = {10.1002/wics.1645}
}

@article{Tierney1994MC,
  author    = {Tierney, Luke},
  title     = {Markov Chains for Exploring Posterior Distributions},
  journal   = {The Annals of Statistics},
  year      = {1994},
  volume    = {22},
  number    = {4},
  pages     = {1701--1728},
  doi       = {10.1214/aos/1176325750},
  publisher = {Institute of Mathematical Statistics}
}

@article{Tuo2019PKC,
  author  = {Tuo, Rui},
  title   = {Adjustments to Computer Models via Projected Kernel Calibration},
  journal = {SIAM/ASA Journal on Uncertainty Quantification},
  year    = {2019},
  volume  = {7},
  number  = {2},
  pages   = {553--578},
  doi     = {10.1137/17M1128769}
}

@article{TuoWu2015,
  author  = {Tuo, Rui and Wu, C. F. Jeff},
  title   = {Efficient Calibration for Imperfect Computer Models},
  journal = {The Annals of Statistics},
  year    = {2015},
  volume  = {43},
  number  = {6},
  pages   = {2331--2352},
  doi     = {10.1214/15-AOS1314}
}

@article{TuoWu2016,
  author  = {Tuo, Rui and Wu, C. F. Jeff},
  title   = {A Theoretical Framework for Calibration in Computer Models: Parametrization, Estimation and Convergence Properties},
  journal = {SIAM/ASA Journal on Uncertainty Quantification},
  year    = {2016},
  volume  = {4},
  number  = {1},
  pages   = {767--795},
  doi     = {10.1137/151005841}
}

@article{WangTuo2020,
  author  = {Wang, Yan and Tuo, Rui},
  title   = {Semi-parametric adjustment to computer models},
  journal = {Statistics},
  year    = {2020},
  volume  = {54},
  number  = {6},
  pages   = {1255--1275},
  doi     = {10.1080/02331888.2020.1862113}
}

@article{Wang2022PPKC,
  author  = {Wang, Yan},
  title   = {Penalized Projected Kernel Calibration for Computer Models},
  journal = {SIAM/ASA Journal on Uncertainty Quantification},
  year    = {2022},
  volume  = {10},
  number  = {4},
  pages   = {1652--1683},
  doi     = {10.1137/20M1387614}
}

@article{XieXu2021,
  author  = {Xie, Fangzheng and Xu, Yanxun},
  title   = {Bayesian Projected Calibration of Computer Models},
  journal = {Journal of the American Statistical Association},
  year    = {2021},
  volume  = {116},
  number  = {536},
  pages   = {1965--1982},
  doi     = {10.1080/01621459.2020.1753519}
}

@article{ZhangWang2025Sobolev,
  author  = {Zhang, Qingwen and Wang, Wenjia},
  title   = {Sobolev Calibration of Imperfect Computer Models},
  journal = {Journal of the American Statistical Association},
  year    = {2025},
  volume  = {120},
  number  = {549},
  pages   = {408--418},
  doi     = {10.1080/01621459.2024.2340793}
}

@article{JosephMelkote2009,
  author  = {Joseph, V. R. and Melkote, S. N.},
  title   = {Statistical adjustments to engineering models},
  journal = {Journal of Quality Technology},
  year    = {2009},
  volume  = {41},
  number  = {4},
  pages   = {362--375},
  doi     = {10.1080/00224065.2009.11917791}
}

\end{document}